\documentclass[prd,twocolumn,amsmath,amssymb]{revtex4}

\usepackage{epsfig}

\def\al{\alpha}
\def\be{\beta}
\def\ga{\gamma}
\def\de{\delta}
\def\ep{\epsilon}
\def\ve{\varepsilon}
\def\ze{\zeta}
\def\et{\eta}
\def\th{\theta}

\def\ka{\kappa}
\def\la{\lambda}

\def\rh{\rho}

\def\si{\sigma}
\def\vs{\varsigma}

\def\ph{\phi}

\def\ch{\chi}

\def\om{\omega}
\def\Ga{\Gamma}

\def\La{\Lambda}
\def\Si{\Sigma}

\def\Ph{\Phi}

\def\Om{\Omega}
\def\mn{{\mu\nu}}
\def\pt#1{\phantom{#1}}
\def\prt{\partial}
\def\cl{{\cal L}}
\def\vev#1{\langle {#1}\rangle}

\def\fr#1#2{{{#1} \over {#2}}}
\def\frac#1#2{{{#1} \over {#2}}}
\def\Frac#1#2{{\textstyle{{#1}\over {#2}}}}
\def\half{{\textstyle{1\over 2}}}
\def\lsim{\mathrel{\rlap{\lower4pt\hbox{\hskip1pt$\sim$}}
    \raise1pt\hbox{$<$}}}
\def\gsim{\mathrel{\rlap{\lower4pt\hbox{\hskip1pt$\sim$}}
    \raise1pt\hbox{$>$}}}
\def\sqr#1#2{{\vcenter{\vbox{\hrule height.#2pt
         \hbox{\vrule width.#2pt height#1pt \kern#1pt
         \vrule width.#2pt}
         \hrule height.#2pt}}}}
\newcommand{\beq}{\begin{equation}}
\newcommand{\eeq}{\end{equation}}
\newcommand{\bea}{\begin{eqnarray}}
\newcommand{\eea}{\end{eqnarray}}
\newcommand{\rf}[1]{(\ref{#1})}

\def\mbf#1{\mbox{\boldmath$#1$}}
\def\syjm#1#2{\phantom{}_{#1}Y_{#2}}
\def\Wedge#1{\wedge^{#1}}
\def\db{\displaybreak[0]}

\def\etal{{\it et al.}}

\def\kf{\hat k_F}
\def\kaf{\hat k_{AF}}
\def\cf{\hat c_F}
\def\cmf{\hat c^M_F}
\def\kfd#1{k_{F}^{(#1)}}
\def\cfd#1{c_{F}^{(#1)}}
\def\kafd#1{k_{AF}^{(#1)}}
\def\kde{\hat\ka_{DE}}
\def\kdb{\hat\ka_{DB}}
\def\khe{\hat\ka_{HE}}
\def\khb{\hat\ka_{HB}}
\def\kep{\hat\ka_{e+}}
\def\kem{\hat\ka_{e-}}
\def\kop{\hat\ka_{o+}}
\def\kom{\hat\ka_{o-}}
\def\ktp{\hat\ka_{tr+}}
\def\ktm{\hat\ka_{tr-}}
\def\kded#1{\ka_{DE}^{(#1)}}
\def\kdbd#1{\ka_{DB}^{(#1)}}

\def\khbd#1{\ka_{HB}^{(#1)}}
\def\voc{\mathrel{\rlap{\lower0pt\hbox{\hskip1pt{$c$}}}
    \raise3pt\hbox{$\neg$}}}
\def\vok{\mathrel{\rlap{\lower0pt\hbox{\hskip1pt{$k$}}}
    \raise6pt\hbox{$\neg$}}}

\def\sk#1#2#3{#1^{(#2)}_{#3}}

\def\kafoB{\sk{(\kafd{d})}{1B}{njm}}
\def\kafoE{\sk{(\vok_{AF}^{(d)})}{1E}{njm}}
\def\kafzB{\sk{(\kafd{d})}{0B}{njm}}
\def\kafoBdnjm#1#2{\sk{(\kafd{#1})}{1B}{#2}}
\def\kafoEdnjm#1#2{\sk{(\vok_{AF}^{(#1)})}{1E}{#2}}
\def\kafzBdnjm#1#2{\sk{(\kafd{#1})}{0B}{#2}}
\def\kafzBp{\sk{(\kafd{d})}{0B'}{njm}}

\def\cfzE{\sk{(\cfd{d})}{0E}{njm}}
\def\kftE{\sk{(\vok_F^{(d)})}{2E}{njm}}
\def\kfoE{\sk{(\vok_F^{(d)})}{1E}{njm}}
\def\kfzE{\sk{(\kfd{d})}{0E}{njm}}
\def\kftB{\sk{(\vok_F^{(d)})}{2B}{njm}}
\def\kfoB{\sk{(\kfd{d})}{1B}{njm}}
\def\cfzEdnjm#1#2{\sk{(\cfd{#1})}{0E}{#2}}
\def\kftEdnjm#1#2{\sk{(\vok_F^{(#1)})}{2E}{#2}}
\def\kfoEdnjm#1#2{\sk{(\vok_F^{(#1)})}{1E}{#2}}
\def\kfzEdnjm#1#2{\sk{(\kfd{#1})}{0E}{#2}}
\def\kftBdnjm#1#2{\sk{(\vok_F^{(#1)})}{2B}{#2}}
\def\kfoBdnjm#1#2{\sk{(\kfd{#1})}{1B}{#2}}

\def\ring#1{{\mathaccent'27 #1}}
\def\kfdfc#1{\ring k_{F}^{(#1)}}
\def\cfdfc#1{\ring c_{F}^{(#1)}}
\def\kafdfc#1{\ring k_{AF}^{(#1)}}
\def\cffc{(\cfdfc{d})_{n}}
\def\kffc{(\kfdfc{d})_{n}}
\def\kaffc{(\kafdfc{d})_{n}}
\def\cffcdn#1#2{(\cfdfc{#1})_{#2}}
\def\kffcdn#1#2{(\kfdfc{#1})_{#2}}
\def\kaffcdn#1#2{(\kafdfc{#1})_{#2}}

\def\kjm#1#2#3{k^{(#1)}_{(#2)#3}}
\def\cjm#1#2#3{c^{(#1)}_{(#2)#3}}
\def\kI{\cjm{d}{I}{jm}}
\def\koldI{\kjm{d}{I}{jm}}
\def\kE{\kjm{d}{E}{jm}}
\def\kB{\kjm{d}{B}{jm}}
\def\kV{\kjm{d}{V}{jm}}
\def\kIdjm#1#2{\cjm{#1}{I}{#2}}
\def\kEdjm#1#2{\kjm{#1}{E}{#2}}
\def\kBdjm#1#2{\kjm{#1}{B}{#2}}
\def\kVdjm#1#2{\kjm{#1}{V}{#2}}

\def\cftzE{\sk{(\voc_F^{(d)})}{0E}{njm}}
\def\kftzE{\sk{(\vok_F^{(d)})}{0E}{njm}}
\def\kftoB{\sk{(\vok_F^{(d)})}{1B}{njm}}
\def\kaftzB{\sk{(\vok_{AF}^{(d)})}{0B}{njm}}
\def\kaftoB{\sk{(\vok_{AF}^{(d)})}{1B}{njm}}
\def\cftzEdnjm#1#2{\sk{(\voc_F^{(#1)})}{0E}{#2}}
\def\kftzEdnjm#1#2{\sk{(\vok_F^{(#1)})}{0E}{#2}}
\def\kftoBdnjm#1#2{\sk{(\vok_F^{(#1)})}{1B}{#2}}
\def\kaftzBdnjm#1#2{\sk{(\vok_{AF}^{(#1)})}{0B}{#2}}
\def\kaftoBdnjm#1#2{\sk{(\vok_{AF}^{(#1)})}{1B}{#2}}

\def\EtE{\sk{(\sk{\ka}{d}{DE})}{2E}{njm}}
\def\EtB{\sk{(\sk{\ka}{d}{DE})}{2B}{njm}}
\def\EoE{\sk{(\sk{\ka}{d}{DE})}{1E}{njm}}
\def\EoB{\sk{(\sk{\ka}{d}{DE})}{1B}{njm}}
\def\EzE{\sk{(\sk{\ka}{d}{DE})}{0E}{njm}}
\def\ezE{\sk{(\sk{\ka}{d}{DE})}{0E'}{njm}}
\def\DtE{\sk{(\sk{\ka}{d}{HB})}{2E}{njm}}
\def\DtB{\sk{(\sk{\ka}{d}{HB})}{2B}{njm}}
\def\DoE{\sk{(\sk{\ka}{d}{HB})}{1E}{njm}}
\def\DoB{\sk{(\sk{\ka}{d}{HB})}{1B}{njm}}
\def\DzE{\sk{(\sk{\ka}{d}{HB})}{0E}{njm}}
\def\dzE{\sk{(\sk{\ka}{d}{HB})}{0E'}{njm}}
\def\BtB{\sk{(\sk{\ka}{d}{DB})}{2B}{njm}}
\def\BtE{\sk{(\sk{\ka}{d}{DB})}{2E}{njm}}
\def\BoB{\sk{(\sk{\ka}{d}{DB})}{1B}{njm}}
\def\BoE{\sk{(\sk{\ka}{d}{DB})}{1E}{njm}}
\def\BzB{\sk{(\sk{\ka}{d}{DB})}{0B}{njm}}
\def\HoB{\sk{(\sk{\ka}{d}{DB})}{1B'}{njm}}
\def\HoE{\sk{(\sk{\ka}{d}{DB})}{1E'}{njm}}
\def\HzE{\sk{(\sk{\ka}{d}{DB})}{0E'}{njm}}

\def\Mjm#1#2#3{{\cal M}^{(#1)}_{(#2)#3}}
\def\Mc{\Mjm{d}{c_F}{njm}}
\def\Mct{\Mjm{d}{\voc_F}{njm}}

\def\Mcdnjm#1#2{\Mjm{#1}{c_F}{#2}}

\def\MIdjm#1#2{\Mjm{#1}{I}{#2}}
\def\Mccav{{\cal M}^{(d)\, \rm cav}_{(c_F)njm}}
\def\Mctcav{{\cal M}^{(d)\, \rm cav}_{(\voc_F)njm}}

\def\MIcavf{{\cal M}^{(4)\, \rm cav}_{(I)jm}}

\def\Mctcavdnjm#1#2{{\cal M}^{(#1)\, \rm cav}_{(\voc_F)#2}}
\def\MIcavdjm#1#2{{\cal M}^{(#1)\, \rm cav}_{(I)#2}}

\def\Mctlab{{\cal M}^{(d)\, \rm lab}_{(\voc_F)njm}}

\def\MIlabdjm#1#2{{\cal M}^{(#1)\, \rm lab}_{(I)#2}}

\def\sE{\underline E}
\def\sB{\underline B}
\def\sD{\underline D}
\def\sH{\underline H}
\def\sF{\underline F}
\def\ss{\underline s}

\def\cg#1#2{\langle #1 | #2 \rangle}

\def\bc#1#2{\left(\begin{smallmatrix} #1 \\ #2 \end{smallmatrix}\right)}

\begin{document}

\title{Electrodynamics with Lorentz-violating operators 
of arbitrary dimension}

\author{V.\ Alan Kosteleck\'y$^1$ and Matthew Mewes$^2$}
\affiliation{$^1$Physics Department, Indiana University, 
Bloomington, Indiana 47405, USA\\
$^2$Physics Department, Marquette University,
Milwaukee, Wisconsin 53201, USA}

\date{IUHET 527, April 2009; 
accepted in Physical Review D} 

\begin{abstract}
The behavior of photons
in the presence of Lorentz and CPT violation is studied.
Allowing for operators of arbitrary mass dimension, 
we classify all gauge-invariant 
Lorentz- and CPT-violating terms 
in the quadratic Lagrange density 
associated with the effective photon propagator.
The covariant dispersion relation is obtained,
and conditions for birefringence are discussed.
We provide a complete characterization of the
coefficients for Lorentz violation for all mass dimensions
via a decomposition using spin-weighted spherical harmonics.
The resulting nine independent 
sets of spherical coefficients
control birefringence, dispersion, and anisotropy
in the photon propagator. 
We discuss the restriction of the general theory
to various special models,
including among others the minimal Standard-Model Extension,
the isotropic limit,
the case of vacuum propagation,
the nonbirefringent limit,
and the vacuum-orthogonal model.
The transformation 
of the spherical coefficients for Lorentz violation
between the laboratory frame
and the standard Sun-centered frame
is provided.
We apply the results to various astrophysical observations
and laboratory experiments.
Astrophysical searches of relevance include studies 
of birefringence and of dispersion.
We use polarimetric and dispersive data from gamma-ray bursts
to set constraints on coefficients for Lorentz violation
involving operators of dimensions four through nine,
and we describe the mixing of polarizations
induced by Lorentz and CPT violation 
in the cosmic-microwave background.
Laboratory searches of interest include
cavity experiments.
We present the general theory for searches with cavities,
derive the experiment-dependent factors 
for coefficients in the vacuum-orthogonal model,
and predict the corresponding frequency shift
for a circular-cylindrical cavity.
\end{abstract}

\maketitle

\section{Introduction}

The properties of electromagnetic radiation
have proved a fertile testing ground
for relativity since its inception
over a century ago.
Tests such as the classic 
Michelson-Morley, Kennedy-Thorndike,
and Ives-Stilwell experiments
\cite{mm,kt,is,classic}
were key in establishing Lorentz invariance,
the foundational symmetry of relativity. 
The proposal that minuscule deviations from Lorentz symmetry
could emerge from an underlying unified theory 
\cite{ks} 
has rekindled interest in sensitive relativity tests,
and the past decade has seen
a broad variety of searches for Lorentz violation
at impressive sensitivities
\cite{datatables}.

Violations of Lorentz symmetry at attainable energies 
are described by the Standard-Model Extension (SME)
\cite{ck,akgrav}.
The SME is a comprehensive effective field theory
that characterizes general Lorentz and CPT violations.
It contains both General Relativity and the Standard Model,
and so it is a realistic theory that can be applied
to analyze observational and experimental data.
A Lorentz-violating term in the Lagrange density of the SME
is an observer scalar density
formed by contracting a Lorentz-violating operator
with a coefficient that acts to govern the term.
The operator can be characterized
in part by its mass dimension $d$,
which determines the dimensionality of the coefficient
and can be used as a naive guide 
to the size of the associated effects
\cite{kp}.

The focus of the present work
is Lorentz and CPT violation involving photons.
Numerous searches for Lorentz violation in electrodynamics 
have been performed,
yielding some of the best existing constraints
on SME coefficients.
One major class of tests consists of laboratory searches
involving electromagnetic resonators or interferometers
\cite{cav0,km,cav1,cav2,cav3,cav4,cav5,cav6,cav7},
which can be viewed as contemporary versions 
of the classic tests of relativity.
Another major class of tests
consists of astrophysical observations
searching for tiny defects in the propagation of light 
that has traveled over cosmological distances
\cite{ck,km,cfj,km_agn,km_grb,km_cmb,km_apjl}.
A variety of other analyses 
involving photons 
lead to constraints on SME coefficients 
\cite{photonth}.
There is also a substantial literature
on various topics in the photon sector of the SME,
including
renormalization
\cite{renorm},
photon interactions 
\cite{photonint},
vacuum \v Cerenkov radiation
\cite{cerenkov},
the Chern-Simons term 
\cite{kaf},
electromagnetostatics
\cite{emstats},
and related phenomena 
involving photons in other contexts 
\cite{other}.
Outside the photon sector,
the SME serves as the theoretical underpinning 
for studies of Lorentz symmetry involving
electrons \cite{eexpt,eexpt2},
protons and neutrons \cite{ccexpt,spaceexpt,bnsyn},
mesons \cite{hadronexpt},
muons \cite{muexpt},
neutrinos \cite{nuexpt},
the Higgs \cite{higgs},
and gravity \cite{akgrav,gravexpt}.

In this paper, 
we extend the existing treatment
of Lorentz violation in electrodynamics 
to include operators of arbitrary mass dimension $d$.
This further develops and consolidates previous systematic studies
for operators of renormalizable dimension
\cite{ck,km},
and it incorporates many phenomenological models 
for Lorentz violation.
For definiteness,
we focus on an action having the usual U(1) gauge symmetry
and invariance under spacetime translations,
so that charge, energy, and momentum are conserved.
If the Lorentz violation is spontaneous,
then the SME coefficients originate 
as expectation values of operators in an underlying theory,
and the requirement of invariance under spacetime translations
corresponds to disregarding soliton solutions
and any massive or Nambu-Goldstone (NG) modes
\cite{ng}. 
In a more complete treatment including gravity,
the NG modes may play the role of the graviton
\cite{cardinal}.
Alternatively, 
the NG modes may be interpreted 
as the photon for Einstein-Maxwell theory 
embedded in a Lorentz-violating vector model
called bumblebee electrodynamics 
\cite{akgrav,bumblebee}.
The approach discussed here can readily be adapted 
to these and other scenarios,
including applications to the photon sector of the SME 
in the context of the various topics mentioned above.
Our methods are also relevant for other sectors of the SME
\cite{kmfermion}.

The motivation for this work comes in part from current doctrine,
which regards the combination of
General Relativity and the Standard Model 
as the low-energy limit of a unified quantum gravity theory
that holds sway at the Planck scale,
$M_{\mbox{\scriptsize Planck}} \sim 10^{19}$ GeV.
Experience teaches us to expect a smooth transition 
from the known low-energy physics to the underlying theory,
so it is plausible to interpret the low-energy action 
as the zeroth-order term 
in a series approximating the underlying theory.
Dimensional analysis suggests that 
operators with larger $d$
correspond to higher-order corrections.
For physics involving violations of Lorentz and CPT symmetry,
the complete series is given by the SME action,
while the leading corrections
form the action of the minimal SME.
Lorentz-violating operators of larger $d$ 
are therefore likely to be especially relevant 
in searches involving very high energies 
and in theoretical studies of foundational properties
such as causality and stability
\cite{kle}.
Under suitable circumstances,
nonrenormalizable operators may even dominate the physics.
For example,
the action of noncommutative quantum electrodynamics
\cite{hayakawa}
incorporates Lorentz-violating effects
associated with a nontrivial commutator 
for the spacetime coordinates.
When the action is expressed in terms 
of conventional photon fields,
a subset of the SME emerges
in which the lowest-order Lorentz-violating operators
have mass dimension six 
\cite{chklo}.
Similarly,
operators of larger $d$ dominate 
in Lorentz-violating theories with supersymmetry
\cite{susy}.

A comprehensive investigation
of all Lorentz-violating operators 
with arbitrary mass dimensions is a challenging task.
Here,
we concentrate on terms in the action 
that are quadratic in the photon field $A_\mu$
and therefore contribute to the propagator,
which in practice is the quantity of immediate interest
in many searches for Lorentz violation.
Our basic approach consists of
constructing the quadratic action
and developing a scheme to classify the operators.
Rotations are a prominent subgroup of the Lorentz group,
and a spherical decomposition can be performed
on any Lorentz-violating operator.
We use this fact to classify 
all Lorentz-violating terms in the action 
using nine sets of coefficients for Lorentz violation. 
The classification scheme is well matched
to the description of physical Lorentz-violating effects 
in the photon propagator,
including birefringence, dispersion, and anisotropy.

With this classification scheme 
taming the infinite number of operators,
specific analyses of observational and experimental data
become feasible.
We study a variety of methods for seeking Lorentz violation
using predictions from the quadratic action.
The sharpest tests involve astrophysical birefringence,
which involves propagation in the vacuum.
Some Lorentz-violating operators produce
no vacuum birefringence at leading order
but nonetheless cause dispersion in the vacuum,
and these can also be studied using astrophysical observations.
In addition,
there are many other Lorentz-violating operators
that are undetectable at leading order 
via astrophysical observations
and hence are best sought instead in laboratory experiments.
The analyses in this work yield several first measurements
of coefficients for Lorentz violation,
and numerous interesting arenas emerge 
for future exploration.

The structure of this paper is as follows.
The basic theory is discussed in 
Sec.\ \ref{sec_theory},
which contains five subsections.
The construction and counting of Lorentz-violating operators
of all mass dimensions
is presented in 
Sec.\ \ref{sec_construction},
while the Lagrange density and constitutive relations
are obtained in
Sec.\ \ref{sec_lagrangian}.
We derive the covariant dispersion relation
in Sec.\ \ref{sec_cov_disp},
discuss the physics of birefringence 
in Sec.\ \ref{sec_duality},
and offer general comments on Lorentz-violating effects
in Sec.\ \ref{sec_effects}.

The spherical decomposition 
of the coefficients for Lorentz violation
in terms of spin-weighted spherical harmonics
is performed in Sec.\ \ref{sec_gen_coeffs}.
We consider various special limits
in Sec.\ \ref{sec_models},
including the minimal SME,
isotropic models,
the vacuum limit,
nondispersive nonbirefringent `camouflage' models,
the vacuum-orthogonal case,
and some limits providing connections
to other formalisms. 
The rotation properties of the spherical coefficients
for Lorentz violation are discussed 
in Sec.\ \ref{sec_rotations}.
Other key properties of the spin-weighted spherical harmonics
are summarized in Appendix \ref{sec_decomp}.

The remainder of the paper applies
the results to observations and experiments.
Astrophysical observations are studied in
Sec.\ \ref{sec_astro}.
Vacuum dispersive effects are discussed in
Sec.\ \ref{sec_disp},
where new measurements and a summary of existing constraints
are obtained.
Vacuum birefringence is considered in Sec.\ \ref{sec_bire},
which contains three subsections.
Some basic theory for vacuum birefringence is presented
in Sec.\ \ref{bireftheory}.
We apply it to point sources in Sec.\ \ref{sec_point_sources},
using polarimetry from gamma-ray bursts 
to obtain new measurements,
and also to the cosmic microwave background 
in Sec.\ \ref{cmb_sec}.
In both cases,
we tabulate some existing sensitivities.

Laboratory experiments are discussed in
Sec.\ \ref{sec_cavities}.
We construct a general theory for resonant-cavity tests
in Sec.\ \ref{sec_cavity_th}
and apply it in Sec.\ \ref{sec_nb_cav}
to derive cavity factors for nonbirefringent operators 
and the form of the fractional frequency shift  
predicted by Lorentz violation.
Explicit values of some cavity factors
for a circular cylindrical cavity are calculated 
in Sec.\ \ref{sec_cav_example}.
Section \ref{Summary and discussion}
provides a summary and discussion of the results in the paper,
including tables compiling essential properties
of the spherical coefficients and various limiting cases.
Unless otherwise stated,
this paper follows the notation and conventions
of Ref.\ \cite{km}.

\section{Basic theory}
\label{sec_theory}

This section discusses the theory 
and basic features of the quadratic action for electrodynamics 
allowing for arbitrary Lorentz and CPT violation.
We begin with a discussion of the SME procedure 
for constructing the Lagrange density 
associated with the effective photon propagator.
Attention is focused primarily on the case with
conventional U(1) gauge invariance
and translational invariance.
The resulting effective field theory 
conserves charge, energy, and momentum.
It represents an explicit presentation 
of all Lorentz-violating operators for photon propagation
that are consistent with observer covariance.
Following the construction of the theory,
we extract a complete set of 
coefficients for Lorentz and CPT violation 
and discuss some of their basic properties.
A technique for deriving covariant dispersion relations
is presented and used to obtain a general covariant dispersion
relation for the photon in the presence
of arbitrary Lorentz and CPT violation.
The issue of conditions for birefringence
is considered,
and we offer some remarks about the
connections between birefringence,
metricity, and electromagnetic duality.
 
\subsection{Construction} 
\label{sec_construction}

A low-energy Lorentz-violating theory 
that is both coordinate independent
and consistent with current observational data
can be written as a Lagrange density 
containing sums of standard polynomial tensor operators
contracted with coefficients for Lorentz violation
\cite{ck}.
The coefficients may be viewed 
as background fields inducing Lorentz violation,
and they can correspond 
to vacuum expectation values of fundamental tensor fields.
Applying this general idea to source-free linear electrodynamics,
we arrive at an action $S$ 
that is a quadratic functional of the photon field $A_\mu$ 
and its derivatives.
The action $S$ can then be expanded 
in a sum of terms $S_{(d)}$ of the form
\beq
S_{(d)} = \int d^4x\
{\cal K}_{(d)}^{\al_1\al_2\al_3\ldots\al_d}
A_{\al_1}\prt_{\al_3}\ldots\prt_{\al_d}A_{\al_2},
\label{Sd}
\eeq
where $d$ is the dimension of the tensor operator.
Each term $S_{(d)}$ violates CPT if $d$ is odd
or preserves CPT if $d$ is even.
The coefficients
${\cal K}_{(d)}^{\al_1\al_2\al_3\ldots\al_d}$
have mass dimension $4-d$.
In general,
they can be dynamical 
and can depend on spacetime position.
We can ensure invariance of $S$
under spacetime translations,
and hence obtain the usual conservation of energy and momentum,
by restricting attention to 
the case of constant coefficients.
Constant coefficients may arise naturally,
but they may also represent the dominant components
of dynamical background fields
or an averaged effect.

Size estimates for the coefficients for Lorentz violation
depend on the details of their origins.
Since the effects are expected to be small,
it is natural to suppose the Lorentz-violating operators 
are suppressed by some large mass.
The intimate connection between Lorentz symmetry and gravity
suggests the relevant scale is set by Planck-scale physics
and therefore by the Planck mass $M_{\mbox{\scriptsize Planck}}$.
Various scenarios can be imagined,
although the absence of a satisfactory underlying theory
combining gravity and quantum physics,
and hence the lack of specifics 
concerning possible Lorentz violations,
makes such scenarios a matter of surmise and likely naive. 
For example,
one simple estimate has coefficients varying as 
${\cal K}_{(d)} \sim \ze M_{\mbox{\scriptsize Planck}}^{4-d}$,
where $\ze$ is of order 1.
This means operators of renormalizable dimension $d\leq 4$ 
are unsuppressed relative to conventional physics.
Observations then imply only the operators $d\geq 5$
are experimentally viable,
so the dominant new physics is controlled by  
nonrenormalizable terms.
Another class of scenarios has Lorentz-violating effects 
suppressed by a factor involving some power of the ratio 
$m/M_{\mbox{\scriptsize Planck}}$,
where $m$ is an appropriate low-energy scale.
In these cases,
the Lorentz violation may be related to one or more of 
the known hierarchies in nature
\cite{kp}. 
For example,
taking $m\sim 10^2$ GeV as the electroweak scale
gives a dimensionless suppression factor of 
some power of $\sim 10^{-17}$
for Lorentz-violating physics compared to conventional effects.

While a general study of all possible operators
of the form \rf{Sd} would be of interest,
it would be rather cumbersome 
and introduce various features in addition to Lorentz violations, 
thereby complicating 
both theoretical and experimental considerations.
The possibilities are simplified somewhat 
by restricting attention to 
operators that maintain the conservation of 
energy, momentum, and electric charge.
This implies focusing on the case of constant coefficients
and requiring U(1) gauge invariance.
The latter imposes certain symmetries
on the Lorentz-violating operators,
thereby reducing the total number of independent coefficients.

The first step in imposing these symmetries
is to identify properties of the coefficients
${\cal K}_{(d)}^{\al_1\al_2\al_3\ldots\al_d}$
that follow from the intrinsic structure of $S$.
One property of the coefficients
is total symmetry in the $d-2$ indices
$\{\al_3\ldots\al_d\}$.
Another can be displayed 
by integrating Eq.\ \rf{Sd} by parts $d-2$ times.
This reveals that the only coefficients contributing
to CPT-odd terms are antisymmetric 
in the first two indices of 
${\cal K}_{(d)}^{\al_1\al_2\al_3\ldots\al_d}$,
while the only ones contributing to CPT-even terms
are symmetric.
The same conclusion follows by considering the contributions
to the variation of the action:
\begin{align}
\de S_{(d)} =&\ \int d^4x\ 
{\cal K}_{(d)}^{\al_1\al_2\al_3\ldots\al_d}
\big(\prt_{\al_3}\ldots \prt_{\al_d}A_{[\al_2}\big) \de A_{\al_1]_\pm}
\notag \\
&+ \mbox{surface terms} , 
\label{var}
\end{align}
where the minus and plus signs apply for 
CPT-odd and CPT-even terms, respectively,
and where the brackets $[\ ]_\pm$ indicate
symmetrization and antisymmetrization.

\begin{figure}
\begin{center}
\centerline{\psfig{figure=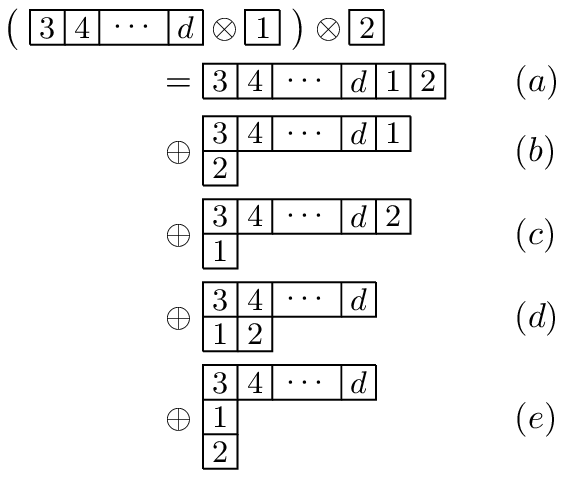,width=0.8\hsize}}
\caption{\label{figyt}
Representation decomposition for the term $S_{(d)}$
using Young tableaux.}
\end{center}
\end{figure}

With these intrinsic symmetries understood,
term-by-term gauge invariance can be imposed 
and the resulting additional symmetries of $S_{(d)}$ identified. 
The usual U(1) gauge invariance,
representing symmetry of the action under the variations 
$\de_g A_\al = \prt_\al\La$,
is achieved by requiring that the variation
\begin{align}
\de_g S_{(d)} =& - \int d^4x\
{\cal K}_{(d)}^{\al_1\al_2\al_3\ldots\al_d}
\notag \\
&\hskip-15pt \times \La \prt_{\al_3}\ldots \prt_{\al_d}
\big( \prt_{[\al_1}A_{\al_2]_\pm}+\half \prt_{[\al_1}\prt_{\al_2]_\pm}\La
\big) \label{gauge}
\end{align}
vanish for an arbitrary scalar function $\La$.
Direct investigation of this equation
is possible but awkward.
To identify the additional symmetries 
and hence the coefficients of interest,
it is more practical first to perform 
representation decompositions
of the associated Lorentz-violating operators.
The intrinsic symmetries drastically
limit the number of representations that can appear.
The total symmetry in the last $d-2$ indices
of ${\cal K}_{(d)}^{\al_1\al_2\al_3\ldots\al_d}$
implies that all representations that are antisymmetric 
in any pair of these indices are absent.
We can therefore construct
the relevant irreducible tensors
from a product of symmetric representations.
It turns out that this limits the possibilities
to only five representations.
In terms of Young tableaux,
these five representations are displayed in Fig.\ \ref{figyt}.

For CPT-odd coefficients,
the antisymmetry condition on the first two indices 
and the gauge variation \rf{gauge}
imply that tableau (a) 
is irrelevant,
since it is symmetric in the indices $\{\al_1\al_2\}$.
Also, 
representation (d) is symmetric 
under the simultaneous permutation 
of $1\leftrightarrow 2$ and $3\leftrightarrow 4$
and so fails to contribute.
The antisymmetry in $\{1,2,3\}$ of representation (e) 
directly implies gauge invariance,
so it satisfies our restrictions.
The remaining two representations (b) and (c)
lead to nonvanishing $\de_g S{(d)}$
and are therefore gauge violating.
We conclude that all gauge-invariant CPT-odd operators
are associated with coefficients 
${\cal K}_{(d)}^{\al_1\al_2\al_3\ldots\al_d}$
belonging to representation (e).
These are antisymmetric in the first three indices 
and symmetric in the last $d-3$.
Since dimension $d\leq 1$ operators are absent 
in a linear theory, 
we take $d\geq 3$ for CPT-odd operators in what follows.

For CPT-even coefficients,
the antisymmetry of tableau (e) in $\{1,2\}$ 
implies it cannot contribute to Eq.\ \rf{var}
and therefore is absent in the theory.
The antisymmetry of tableau (d) 
in \{1,3\} and \{2,4\}
leads to $\de_g S{(d)} = 0$,
so this representation is gauge invariant
and satisfies our restrictions.
The tableau (a) is totally symmetric
and leads to $\de_g S{(d)}\neq 0$,
implying gauge violation.
Similarly, 
tableaux (b) and (c) are also gauge violating.
We thus find that all gauge-invariant CPT-even operators
have coefficients 
${\cal K}_{(d)}^{\al_1\al_2\al_3\ldots\al_d}$
belonging to representation (d).
Note that operators with $d=2$ are gauge-violating,
so in what follows we take $d\geq 4$ for the CPT-even sector.

To simplify handling and to provide a convenient match 
to the usual coefficients for Lorentz violation 
in the minimal SME,
it is convenient to introduce further definitions.
For the CPT-odd case,
the dual coefficients 
\beq
{(\kafd{d})_\ka}^{\al_1\ldots\al_{(d-3)}} 
\equiv\Frac{1}{3!}\ep_{\ka\mu\nu\rh}
{\cal K}_{(d)}^{\mu\nu\rh\al_1\al_2\ldots\al_{d-3}}
\label{kafd}
\eeq
provide a generalization of the usual coefficients $(k_{AF})_\ka$ 
in the minimal SME.
The symmetries of tableau (e)
translate into total symmetry of the coefficients 
${(\kafd{d})_\ka}^{{\al_1}\ldots{\al_{(d-3)}}}$
in the last $d-3$ indices,
along with the trace condition
${(\kafd{d})_{\al_1}}^{\al_1\ldots\al_{(d-3)}} = 0$.
Counting the number of independent
components in representation (e) 
using standard group-theory techniques 
\cite{tableau}
yields for dimension $d$ the result
\beq
N^{(d)}_{AF}=\half(d+1)(d-1)(d-2).
\label{naf}
\eeq
This number can also be obtained
by noting that symmetry in the last $d-3$ indices 
yields $4(d-2)(d-1)d/3!$ components
while the trace condition given above provides 
$(d-3)(d-2)(d-1)/3!$ constraints,
and taking the difference yields $N^{(d)}_{AF}$.
Note that the number of these coefficients 
for CPT-odd Lorentz violation grows rapidly
as the cube of $d$:
the usual 4 for $d=3$, 
then 36 coefficients for $d=5$, 
120 for $d=7$, etc.

For the CPT-even case,
it suffices to define
\beq
(\kfd{d})^{\ka\la\mu\nu\al_1\ldots\al_{(d-4)}} \equiv
{\cal K}_{(d)}^{\ka\mu\la\nu\al_1\al_2\ldots\al_{d-4}}
\label{kfd}
\eeq
to obtain coefficients that mimic 
the definition of $(k_F)^{\ka\la\mu\nu}$ in the minimal SME.
The first four indices of
$(\kfd{d})^{\ka\la\mu\nu\al_1\ldots\al_{(d-4)}}$
have the symmetries of the Riemann tensor,
and there is total symmetry in the 
remaining $d-4$ indices.
Also, 
one can show that antisymmetrization of 
$(\kfd{d})^{\ka\la\mu\nu\al_1\ldots\al_{(d-4)}}$
on any three indices produces zero.
In this case,
counting the number of independent components 
for dimension $d$ gives 
\beq
N^{(d)}_F=(d+1)d(d-3) .
\label{nf}
\eeq
This counting includes the total trace term,
which is Lorentz invariant 
and represents a scaling factor.
Again,
note that the number of these coefficients 
for CPT-even Lorentz violation
grows as the cube of $d$:
the usual 20 (19 plus a Lorentz-invariant trace) for $d=4$,
then 126 coefficients for $d=6$, 
360 for $d=8$, etc.

\subsection{Lagrange density and constitutive relations} 
\label{sec_lagrangian}

The construction outlined in the previous subsection 
leads to a general gauge-invariant Lagrange density 
that can be written in a form similar to the
photon sector of the minimal SME:
\begin{align}
\cl & =  -\Frac 1 4 F_{\mu\nu}F^{\mu\nu}
+\Frac 1 2 \ep^{\ka\la\mu\nu}A_\la (\kaf)_\ka F_{\mu\nu}
\notag \\
&\quad - \Frac 1 4 F_{\ka\la} (\kf)^{\ka\la\mu\nu} F_{\mu\nu} ,
\label{lagrangian}
\end{align}
where the differential operators
$\kaf$ and $\kf$ involve CPT-odd and CPT-even violations, 
respectively.
These operators are given by the expansions 
\begin{align}
(\kaf)_\ka &=\hspace{-3pt}\sum_{d=\mbox{\scriptsize odd}} 
{(\kafd{d})_\ka}^{\al_1\ldots\al_{(d-3)}} 
\prt_{\al_1}\ldots\prt_{\al_{(d-3)}} ,
\label{kafs}\\
(\kf)^{\ka\la\mu\nu} &= \hspace{-5pt}\sum_{d=\mbox{\scriptsize even}} 
(\kfd{d})^{\ka\la\mu\nu\al_1\ldots\al_{(d-4)}} 
\prt_{\al_1}\ldots\prt_{\al_{(d-4)}} ,
\label{kfs}
\end{align}
where the sums range over values $d\geq 3$.
The coefficients
${(\kafd{d})_\ka}^{{\al_1}\ldots{\al_{(d-3)}}}$
are defined in Eq.\ \rf{kafd},
and they have symmetry properties 
yielding $N^{(d)}_{AF}$ independent components
as given in Eq.\ \rf{naf}.
The coefficients
$(\kfd{d})^{\ka\la\mu\nu\al_1\ldots\al_{(d-4)}}$
are defined in Eq.\ \rf{kfd}
and have $N^{(d)}_{F}$ independent components
according to Eq.\ \rf{nf}.
The usual minimal SME terms are 
$k_{AF}\equiv\kafd{3}$ and $k_F\equiv \kfd{4}$,
where in the latter the overall trace is removed
to leave 19 independent coefficients.

In principle,
obtaining and interpreting equations of motion 
for a Lagrange density of the form \rf{lagrangian} 
is problematic due to the infinite sum,
whose action cannot be varied in the usual way,
and also due to Ostrogradski instabilities
\cite{ostrogradski}.
However,
these issues can be circumvented by noting
that Eq.\ \rf{lagrangian} represents 
the low-energy limit of a more fundamental theory,
with each successive term representing
a perturbation on preceding terms.
Truncating the sums at any definite value of $d$ 
and restricting attention to perturbative effects 
therefore can be expected to provide 
a good approximation to the low-energy behavior.
Only at extreme energies can qualitatively new effects 
and late terms in the sum play an important role.
At these energies,
the theory must converge to
the underlying fundamental physics,
which presumably is free of these issues.
Adopting this truncation,
we obtain equations of motion given by
\beq
\big(
\et^{\mu\al}\et^{\nu\be}\prt_\nu
+(\kaf)_\nu \ep^{\mu\nu\al\be}
+(\kf)^{\mu\nu\al\be}\prt_\nu
\big)
F_{\al\be} = 0.
\label{eqnmot}
\eeq
Note the explicit gauge invariance of
these equations.

In the minimal SME,
the coefficients $\kafd{3}$ and $\kfd{4}$
are known to produce photon behavior
analogous to that of conventional electrodynamics 
in anisotropic and gyrotropic materials
\cite{ck,km}.
This analogy can be extended to the present situation
involving Lorentz-violating operators of arbitrary dimension.
The first step is to define a field tensor
\beq
G^{\mu\nu}
= F^{\mu\nu}
-2\ep^{\mu\nu\al\be}(\kaf)_\al A_\be
+(\kf)^{\mu\nu\al\be}F_{\al\be},
\label{macrofs}
\eeq
in terms of which the equations of motion become
\beq
\prt_\nu G^{\mu\nu}= 0 .
\label{geqs}
\eeq
Note that the latter equation is gauge invariant,
even though the definition of $G^{\mu\nu}$ 
depends on the choice of gauge
and is unique only up to a term of the form
$\ep^{\mu\nu\al\be}(\kaf)_\al \prt_\be\La$
for an arbitrary scalar function $\La$.

For conventional electrodynamics in macroscopic media,
a constitutive 4-tensor $\ch$ is typically introduced
that maps the 2-form field strength $F$ 
to the macroscopic 2-tensor field strength $G$,
via $G^{\mu\nu}=\ch^{\mu\nu\rh\si}F_{\rh\si}$.
In the present context,
we can reformulate the situation in terms of
a set of unconventional constitutive relations,
which may explicitly depend on the choice of gauge.
However,
we must generalize the usual notion 
of a constitutive 4-tensor
to encompass more general operator constitutive tensors.
We now require an operator 4-tensor 
$\hat \ch^{\mu\nu\rh\si}$
and an operator 3-tensor
$\hat X^{\mu\nu\rh}$,
defined by
\begin{align}
\hat \ch^{\mu\nu\rh\si} &= 
\half(\et^{\mu\rh}\et^{\nu\si}-\et^{\nu\rh}\et^{\mu\si})
+(\kf)^{\mu\nu\rh\si} ,
\notag \\
\hat X^{\mu\nu\rh} &= \ep^{\mu\nu\rh\si}(\kaf)_\si .    
\label{constit}
\end{align}
Note that the 3-tensor controls CPT violation.
The effective macroscopic field strength $G$
defined in Eq.\ \rf{macrofs}
is then given by 
\beq
G^{\mu\nu}=
\hat \ch^{\mu\nu\rh\si}F_{\rh\si}+2\hat X^{\mu\nu\rh}A_\rh .
\label{fgmap}
\eeq
As in conventional electrodynamics,
the new constitutive relations remain linear.
However,
unlike electrodynamics in linear media,
the relations \rf{fgmap}
inherit a nonlocal aspect due to their differential nature.

Decomposing $G^{\mu\nu}$ into 
an effective vector displacement field $\mbf D$ 
and an effective pseudovector magnetic field $\mbf H$
in the usual way,
the equations of motion \rf{geqs}
take the same form as 
the familiar source-free inhomogeneous Maxwell equations,
\begin{align}
\mbf\nabla\cdot\mbf D & = 0,
\notag\\
\mbf\nabla\times\mbf H - \prt_0\mbf D & = 0 ,
\label{maxeq}
\end{align}
where
\begin{align}
\mbf D =\ & \mbf E +2 \mbf\kaf \times\mbf A 
+\kde \cdot\mbf E + \kdb\cdot\mbf B , 
\notag \\
\mbf H =\ & \mbf B -2(\kaf)_0\mbf A + 2 \mbf\kaf A_0
+ \khb\cdot\mbf B +\khe\cdot\mbf E 
\notag\\
\end{align}
with
\begin{align}
(\kde)^{jk} &= -2(\kf)^{0j0k} , 
\notag\\
(\khb)^{jk} &= \half(\kf)^{lmrs} \ep^{jlm}\ep^{krs} , 
\label{kappas} \\
(\kdb)^{jk} &= -(\khe)^{kj} = (\kf)^{0jlm}\ep^{klm} . 
\notag 
\end{align}
The latter equations are operator generalizations of 
the SO(3) decomposition of the coefficients $\kfd{4}$ 
into $3\times 3$ matrices introduced 
in Ref.\ \cite{km}.

Using the fields $\mbf D$ and $\mbf H$,
the Lagrange density may be written as
\beq
\cl =  -\Frac 1 4 F_{\mu\nu}G^{\mu\nu}
= \half (\mbf E\cdot\mbf D - \mbf B\cdot\mbf H) ,
\eeq
which also parallels conventional electrodynamics
in macroscopic media.
We remark that the analogy with electrodynamics breaks down 
when attempting to construct a conserved energy-momentum tensor.
Standard techniques can be used to build 
a conserved tensor that reduces 
to the conventional symmetrized energy-momentum tensor 
in the limit of vanishing Lorentz violations.
However, 
the resulting tensor takes an unconventional form
in terms of $F_\mn$ and $G^\mn$.
One possibility is the tensor
\beq
{T^\al}_\be =  
-G^{\al\ga}F_{\be\ga}-\de^\al_\be\cl
+\half(\prt_\be A_\ga - A_\ga\prt_\be) G^{\al\ga} .
\eeq
The first two terms parallel those in
conventional electrodynamics,
but the addition of the last term is necessary
for energy-momentum conservation to hold.
This last term is separately conserved 
in conventional electrodynamics and so can be removed there,
but a term of this type must be present 
for ${T^\al}_\be$ to be conserved
in the presence of general Lorentz violation.

\subsection{Covariant dispersion relation} 
\label{sec_cov_disp}

Much of the propagation behavior 
of the photon is encoded in its dispersion relation,
which provides spectral information for the modes.
While standard methods can be used to find 
the dispersion relation from the equations of motion,
at least at leading order 
in the coefficients for Lorentz violation,
handling the gauge freedom typically 
entails the loss of observer Lorentz invariance.
In this subsection,
we present a technique for deriving 
the exact covariant dispersion relation,
based on the rank-nullity properties 
of the equations of motion.
As a concomitant,
the technique provides some insight
into the nature of birefringence,
which arises whenever the dispersion relation 
has non-degenerate physical solutions.

We begin by adopting the ansatz
\beq
A_\mu(x)=A_\mu(p)e^{-ix\cdot p}.
\eeq
This implies the equations of motion \rf{eqnmot}
can be expressed in the matrix form
\beq
M^{\mu\nu}A_\nu = 0
\label{meq}
\eeq
with 
\begin{align}
M^{\mu\nu} =\ & 
(\et^{\mu\nu}\et^{\al\be}-\et^{\mu\al}\et^{\nu\be}
+2(\kf)^{\mu\al\nu\be})p_\al p_\be 
\notag \\
& -2i\ep^{\mu\nu\al\be}(\kaf)_\al p_\be 
\notag\\
=\ & 2\hat \ch^{\mu\al\nu\be} p_\al p_\be
+2i\hat X^{\mu\nu\al}p_\al , 
\label{M} 
\end{align}
where it is understood that each occurrence of $\prt_\al$ 
in the operators $\kaf$ and $\kf$ is replaced with $-ip_\al$.
The matrix $M^{\mu\nu}$ is hermitian, 
$M^{\mu\nu}=(M^{\nu\mu})^*$.
It satisfies the conditions 
for charge conservation, $M^{\mu\nu}p_\mu=0$,
and gauge symmetry, $M^{\mu\nu}p_\nu=0$,
which imply that the determinant
of $M^{\mu\nu}$ vanishes identically.

The standard method to handle the gauge freedom
and the vanishing determinant
involves making a definite gauge choice,
thereby reducing the four-dimensional problem
to a three-dimensional one.
The determinant of the reduced linear equation 
yields the dispersion relation.
Within the particular gauge choice
and for a given solution to the dispersion relation,
one can then solve for the polarization mode $A_\mu$
of the photon.
The general solution for this mode 
is the sum of this solution
and an arbitrary pure gauge term $\propto p_\mu$.
Typically,
the gauge fixing explicitly breaks observer Lorentz invariance.
The alternative method presented below
focuses on the rank-nullity of the linear equation \rf{meq},
which allows us to preserve Lorentz covariance
throughout the calculation.
We take advantage of the exterior product
and its ability to determine linear
independence of a set of vectors.
The reader is reminded that a set of vectors 
$\{A^a\}$, $a=1,2,\ldots$
is linearly independent
if and only if their exterior product is nonzero
\cite{flanders}.

Starting with a set of arbitrary basis vectors
$\{A^1,A^2,A^3,A^4\}$ with
$A^1\wedge A^2\wedge A^3\wedge A^4\neq 0$,
the image space of $M$ is spanned by
the vectors $B^a = M A^a$.
The dimensionality of this space
is equal to the rank of $M$,
and it determines the dimension
of the solution space of Eq.\ \rf{meq}.
In particular,
gauge freedom ensures that $M$ is rank three or lower.
This implies
$B^1\wedge B^2\wedge B^3\wedge B^4=0$,
which is equivalent to the condition
$\det M^{\mu\nu} = 0$.
So to find nontrivial solutions, 
we must impose rank two or less,
which implies $B^a\wedge B^b\wedge B^c =0$
for any $a,b,c = 1,2,3,4$.
This is the minimum requirement,
and it leads to the covariant dispersion relation.

We can translate the above discussion 
into conditions on $M$ 
by noting that 
$M$ generates for each $n$ a linear transformation
between $n$-vectors in its domain and image spaces.
The transformation for a given $n$ 
is denoted $\Wedge{n} M$ and is specified by 
\begin{align}
B^{a_1}\wedge \cdots\wedge B^{a_n}
&= (MA^{a_1})\wedge \cdots\wedge (MA^{a_n})\notag \\
&\equiv (\Wedge{n} M) (A^{a_1}\wedge \cdots\wedge A^{a_n}).
\label{LaM1}
\end{align}
For each $n$,
the linear $\Wedge{n} M$ transformation
takes an arbitrary $n$-vector $\om$
of the domain space 
into an $n$-vector $(\Wedge{n} M) (\om)$
of the image space.
In a coordinate basis,
we get an explicit expression
for the cofactor tensors 
${(\Wedge{n} M)_{\mu_1\mu_2\ldots\mu_n}}^{\nu_1\nu_2\ldots\nu_n}$
controlling the transformation $\Wedge{n} M$:
\beq
{(\Wedge{n} M)_{\mu_1\mu_2\ldots\mu_n}}^{\nu_1\nu_2\ldots\nu_n}
= \frac 1{n!} 
{M_{[\mu_1}}^{\nu_1}{M_{\mu_2}}^{\nu_2}\cdots{M_{\mu_n]}}^{\nu_n}.
\label{LaM2}
\eeq
The key point for our purposes 
is that the rank of $M$ is completely
determined by the set of $\Wedge{n} M$ transformations.
For any given $n$,
the matrix $M$ is rank $n$
if and only if
$\Wedge{n} M \neq 0$ and $\Wedge{n+1} M = 0$.
Note that $\Wedge{n} M = 0$ implies $\Wedge{n+1} M = 0$.

We can now determine the size of the null space of $M$
and hence the nature of the solution space 
for the equations of motion
by using the rank-nullity relation.
Since $M$ is a $4\times4$ matrix,
the rank can in principle range from four down to zero.
The five possibilities can be
summarized as follows.
The case of rank four has $\Wedge{4} M \neq 0$
and is excluded by gauge invariance,
so no solutions exist.
For rank three,
which has $\Wedge{4} M = 0$ and $\Wedge{3} M \neq 0$,
the nullity is one and hence there 
is a one-dimensional solution space.
In the present context, 
it corresponds to pure-gauge solutions.
For rank two
we have $\Wedge{3} M = 0$ and $\Wedge{2} M \neq 0$
with a two-dimensional null space. 
This case yields a one-dimensional non-gauge solution space.
For rank one $\Wedge{2} M = 0$, $\Wedge{1} M \neq 0$,
and the nullity is three,
so there is a two-dimensional non-gauge solution space.
Finally,
for the case of rank 0 we have $\Wedge{1} M = M = 0$, 
which is trivial.

The above discussion reveals that
the covariant dispersion relation
ensuring at least one physical solution is 
\beq
\Wedge{3} M =0,
\label{tensordisprel}
\eeq
in which case $M$ is rank 2 or less.
In a coordinate basis,
this dispersion relation takes the tensor form 
\beq
\frac 1{3!}
{M_{[\mu_1}}^{\nu_1}{M_{\mu_2}}^{\nu_2}{M_{\mu_3]}}^{\nu_3} = 0.
\eeq
Also,
if it so happens that requiring $\Wedge{3} M =0$ 
also leads to $\Wedge{2} M =0$,
then a two-dimensional physical solution space exists.
We interpret this situation as follows.
Suppose we fix the 3-momenta $p_j$,
and find a frequency $p_0$ that solves 
the dispersion relation \rf{tensordisprel}.
If at this frequency $\Wedge{2} M \neq 0$ and hence $M$ 
has rank two,
then there is exactly one non-gauge polarization mode $A_\mu$ 
associated with this solution.
Other frequencies $p_0$ 
that solve the dispersion relation \rf{tensordisprel} 
lead to different polarizations and different phase velocities.
The solution $A_\mu$ is therefore birefringent.
However,
if for a given frequency solution
we find $\Wedge{2} M= 0$ and hence $M$ of rank one, 
then there are two independent polarizations
that propagate with the same phase velocity.
This situation represents nonbirefringence.
We therefore obtain the correspondence 
\beq
\Wedge{2} M = 0 \leftrightarrow \textstyle\rm{no~birefringence},
\label{nobiref}
\eeq
which provides the explicit condition 
for the existence of nonbirefringent modes.
The next subsection explores
the issue of birefringence in more detail.

The covariant tensor dispersion relation \rf{tensordisprel}
can be rewritten as a covariant scalar dispersion relation
by the judicious use of gauge symmetry. 
To see this, 
adopt the special domain basis $\{A^1,A^2,A^3,p\}$.
Since gauge invariance implies $M p =0 $,
the only 3-vectors $\om$ 
that yield nonzero values of $(\Wedge{3} M)(\om)$ 
must be proportional to $A^1\wedge A^2\wedge A^3$.
Consequently,
the transformation $(\Wedge{3} M)(\om)$
is one dimensional,
and therefore its dual $\widetilde{\Wedge{3} M}$
must also be one dimensional.
The hermiticity of $M$
implies that there is a vector $V^\mu$ in terms of which
the dual transformation
$\widetilde{\Wedge{3} M}$
is determined as 
$(\widetilde{\Wedge{3} M})^{\mu\nu}=V^{\mu *}V^\nu$.
Direct calculation with this result shows that
$(\widetilde{\Wedge{3} M})^{\mu\nu}p_\nu 
= {(\widetilde{\Wedge{3} M})^\rh}_\rh p^\mu$,
and from these relations we find 
$p^\rh p_\rh (\widetilde{\Wedge{3} M})^{\mu\nu}
= {(\widetilde{\Wedge{3} M})^\rh}_\rh p^\mu p^\nu$.
Finally, 
this expression implies that 
the tensor dispersion relation \rf{tensordisprel}
is satisfied for any nonzero $p$
if and only if the trace of its dual vanishes.
So we arrive at the covariant dispersion relation
\beq
{(\widetilde{\Wedge{3} M})^\rh}_\rh = 0,
\label{scalardisprel}
\eeq
which is a scalar density.

In terms of the constitutive tensors \rf{constit},
the covariant scalar dispersion relation \rf{scalardisprel}
can be written as
\begin{widetext}
\begin{align}
0 =&\ -\Frac13\hat \ch^{\mu\al\nu\be}
\big((\widetilde{\Wedge{2} M_e})_{\mu\al\nu\be}
-3(\widetilde{\Wedge{2} M_o})_{\mu\al\nu\be}\big) \notag\\
=&\ -\Frac13
\ep_{\mu_1\mu_2\mu_3\mu_4}\ep_{\nu_1\nu_2\nu_3\nu_4}
p_{\rh_1}p_{\rh_2}p_{\rh_3}p_{\rh_4} 
\hat \ch^{\mu_1\mu_2\nu_1\rh_1}
\hat \ch^{\nu_2\rh_2\rh_3\mu_3}
\hat \ch^{\rh_4\mu_4\nu_3\nu_4}
+8p_\al p_\be 
(\kaf)_\mu (\kaf)_\nu \hat \ch^{\al\mu\be\nu}.
\label{dr}
\end{align}
\end{widetext}
In this expression,
the dual of $\Wedge{2} M$ is defined by
\beq
(\widetilde{\Wedge{2} M})_{\mu\al\nu\be}
\equiv \Frac14\ep_{\mu\al\rh\ga}\ep_{\nu\be\si\de}
M^{\rh\si}M^{\ga\de},
\eeq
while 
$M_e$ and $M_o$ are,
respectively,
the CPT-even and CPT-odd parts of $M$.

The covariant scalar dispersion relation \rf{dr} 
is a necessary condition 
for the existence of nontrivial plane-wave solutions.
The first term on the right-hand side of Eq.\ \rf{dr}
is CPT even and matches the result 
found in Ref.\ \cite{ofr} in the appropriate limit.
The last term contains all
CPT-violating contributions.
Note that this covariant scalar dispersion relation
is independent of the spacetime metric $\et_{\mu\nu}$.
Note also that the momentum dependence 
of the constitutive tensor
$\hat \ch^{\mu\nu\rh\si}$ and of $(\kaf)_\mu$
implies that the dispersion relation \rf{dr}
is typically a polynomial of degree greater than four
in the frequency $p_0$.
As a result,
more eigenfrequencies typically exist
in the presence of Lorentz violation
than for the corresponding situation in conventional electrodynamics,
and so more modes can propagate.
However,
following a reasoning similar to that
leading to the equations of motion \rf{eqnmot},
we expect the solutions of interest to be small perturbations
of the limiting physical solutions
in conventional electrodynamics.
It follows that only the corresponding subset of the solutions 
to the dispersion relation \rf{dr}
are relevant to low-energy physics,
while the others represent high-frequency modes
that may play a role as Planck-scale energies are approached.

\subsection{Birefringence}
\label{sec_duality}

Astrophysical searches for vacuum birefringence 
in photon propagation
provide sensitivities to Lorentz violation
that are many orders of magnitude 
beyond those attainable via other techniques.
It is therefore valuable to classify
coefficients for Lorentz violation
according to their birefringence effects.
The analysis in the previous subsection
has already provided some insight
via the condition \rf{nobiref}
for the absence of birefringence.
In this subsection,
we provide a decomposition of 
coefficients for Lorentz violation
that distinguishes birefringent and nonbirefringent cases.
We also offer some remarks 
about generic conditions for birefringence
and their connection to the metric 
and to electromagnetic duality.

\subsubsection{Coefficients for birefringence}

In the minimal SME,
leading-order birefringence is known to be controlled
by the $d=3$ coefficients $(k_{AF})_\mu$ 
and by a subset of the $d=4$ coefficients $(k_F)^{\ka\la\mu\nu}$
\cite{ck,km}.
In the present context with Lorentz-violating operators 
of any dimension,
a similar pattern holds:
leading-order birefringence is associated with 
all the odd-$d$ coefficients $\kaf$
and with some combinations of the even-$d$ coefficients $\kf$.
One way to verify this is via
the covariant scalar dispersion relation \rf{dr}.

To identify the relevant even-$d$ coefficients,
it is useful to introduce the definitions 
\begin{align}
\kep&=\half(\kde+\khb)-\Frac16 \mbox{Tr}(\kde+\khb) , 
\notag\\
\kem&=\half(\kde-\khb)-\Frac16 \mbox{Tr}(\kde-\khb) ,
\notag\\
\kop&=\half(\kdb+\khe) , 
\notag\\
\kom&=\half(\kdb-\khe) , 
\label{kappastwo}\\
\ktp&=\Frac16 \mbox{Tr}(\kde+\khb) , 
\notag\\
\ktm&=\Frac16 \mbox{Tr}(\kde-\khb) ,
\notag
\end{align}
which give an experimentally judicious decomposition 
of the coefficients for Lorentz violation
appearing in Eq.\ \rf{kappas}.
The first four of these are traceless $3\times3$ matrices,
with $\kep$, $\kem$, $\kom$ symmetric
and $\kop$ antisymmetric. 
The last two are SO(3) rotation scalar combinations.
Note that $\ktp$ can be disregarded in the minimal SME
because it represents a simple Lorentz-invariant scaling factor
in that context,
but nonrenormalizable terms of this type 
can violate Lorentz symmetry
and so must be included in the present context.

Among the combinations \rf{kappastwo},
only the matrices $\kep$ and $\kom$
cause vacuum birefringence at leading order.
This result can be understood in terms of a Weyl decomposition
of the constitutive tensor $\hat \ch^{\mu\nu\rh\si}$
analogous to the Weyl decomposition
of the Riemann tensor,
\begin{align}
\hat \ch^{\mu\nu\rh\si} =&\ 
\half(\et^{\mu\rh}\et^{\nu\si}-\et^{\nu\rh}\et^{\mu\si}) 
\notag \\
&\quad 
+\half\big(\et^{\mu\rh}(\cf)^{\nu\si}-\et^{\nu\rh}(\cf)^{\mu\si}
\notag \\
&\quad 
+\et^{\nu\si}(\cf)^{\mu\rh}-\et^{\mu\si}(\cf)^{\nu\rh}\big) 
+ \hat C^{\mu\nu\rh\si} .
\label{decompconstit}
\end{align}
In this equation,
the tensor $\hat C^{\mu\nu\rh\si}$ corresponds
to the Weyl component and is traceless,
$\hat C^{\mu\nu\rh\si}\et_{\nu\si}=0$.
The term corresponding to the Ricci component
involves $(\cf)^{\al\be}$,
which is defined as the symmetric combination 
\beq
(\cf)^{\al\be} \equiv
{(\kf)^{\al\mu\be}}_\mu
-\Frac16 {(\kf)^{\mu\nu}}_{\mu\nu}\et^{\al\be}.
\eeq
The relations between $(\cf)^{\al\be}$, $\hat C^{\mu\nu\rh\si}$
and the $\hat\ka$ matrices are 
\begin{align}
(\cf)^{00} &= \half \big(3\ktm+\ktp\big) , 
\notag \\
(\cf)^{jk} &= -(\kem)^{jk}+\half \big((\ktm)-(\ktp)\big)\de^{jk} , 
\notag \\
(\cf)^{0j} &= -\half (\kop)^{kl} \ep^{jkl}, 
\\
\hat C^{0j0k} &= -\half (\kep)^{jk} , 
\notag \\
\hat C^{jklm} &= \half (\kep)^{np} \ep^{jkn}\ep^{lmp} , 
\notag \\
\hat C^{0jkl} &= \half (\kom)^{jm} \ep^{klm} .
\notag
\end{align}
We see that the ten independent components of $\cf$ are
equivalent to the ten independent components 
of $\kem$, $\kop$, $\ktp$, and $\ktm$,
while the ten independent components of $\hat C$ 
match those of $\kep$ and $\kom$.

The decomposition \rf{decompconstit} reveals that 
the coefficients $\cf$ play the role of
a small distortion of the spacetime metric at leading order.
In this respect,
they are analogous to the $c^{\al\be}$ coefficients 
in the matter sector of the SME,
which motivates the notation.
As further discussed below,
a small metric distortion 
leaves unaffected the usual degeneracy between polarizations
and so cannot cause birefringence at leading order.
In contrast,
the non-metric Weyl piece of Eq.\ \rf{decompconstit} 
breaks the degeneracy and causes birefringence.
Note that the effects of $c$-type coefficients in the minimal SME
are unobservable in experiments involving only one sector,
since they can be removed by a judicious coordinate choice.
However, 
in the present context where $\cf$ depends on energy and momentum,
dispersion effects may arise that are observable.

The above decomposition and results
hold for the vacuum,
where the Weyl decomposition is
performed using the Minkowski metric.
However,
the effective metric $g_{\mu\nu}$
for electrodynamics in a macroscopic medium $M$
is no longer Minkowski,
so the above decomposition must be modified.
The Weyl part $\hat C^M$ 
is required to be traceless with respect to $g_{\mu\nu}$ instead,
so the assignment of $\hat\kappa$ matrices
to $\hat C^M$ and $\hat c^M_F$ differs.
The attribution of birefringence effects to
a coefficient can therefore be medium dependent.
It also follows that an experiment in a suitable medium
can achieve sensitivities to different coefficients
compared to the same experiment performed {\it in vacuo}.

As an example,
consider a uniform isotropic medium $M$ 
with refractive index $n$ and permeability $\mu$.
Performing a decomposition of the constitutive tensor
reveals that the relations between
$\hat C^M$, $\hat c^M_F$ and the $\hat\kappa$ matrices
become
\begin{align}
(\cmf)^{00} &= \half{\sqrt{\mu}(2-n^2)}\ktp
+\half{\sqrt{\mu}(2+n^2)}\ktm , 
\notag \\
(\cmf)^{jk} &=
-\frac{\sqrt{\mu}(n^2+1)}{2n^2}(\kem)^{jk}
+\frac{\sqrt{\mu}(n^2-1)}{2n^2}(\kep)^{jk} 
\notag \\
&\quad
+\half{\sqrt{\mu}} \bigl((\ktm)-(\ktp)\bigr)\de^{jk} , 
\notag \\
(\cmf)^{0j} &= -\half{\sqrt{\mu}} (\kop)^{kl} \ep^{jkl},  
\label{mediumdecomp}
\\
(\hat C^M)^{0j0k} &= -\Frac 1 4 {(n^2+1)} (\kep)^{jk}
+\Frac 1 4 {(n^2-1)} (\kem)^{jk} , 
\notag \\
(\hat C^M)^{jklm} &= 
\frac{(n^2+1)}{4n^2} (\kep)^{np} \ep^{jkn}\ep^{lmp} 
\notag \\
& \qquad 
-\frac{(n^2-1)}{4n^2} (\kem)^{np} \ep^{jkn}\ep^{lmp} , 
\notag \\
(\hat C^M)^{0jkl} &= \half (\kom)^{jm} \ep^{klm} . \notag
\end{align}
For this type of medium,
we see that birefringence is associated with
the matrix $\kem$
as well as the matrices $\kep$ and $\kom$. 
Moreover,
the matrix $\kep$ can now affect nonbirefringent phenomena
at leading order,
unlike the vacuum case.
Additional coefficient mixings
can be expected in anisotropic and gyroscopic media.

\subsubsection{Birefringence, metric, and duality}

An interesting challenge 
is the identification of the minimal set of requirements
leading to birefringence.
For simple local constitutive relations,
nonbirefringence is known to be associated with 
a pure-metric constitutive tensor 
\cite{lhi,gs},
while electromagnetic duality plays a role
\cite{ohr}.
The idea that the essential properties of electrodynamics
rely on constitutive relations 
rather than the underlying metric structure of the spacetime
is a key aspect of the premetric approach to electrodynamics
\cite{ofr,iho}.
In this subsection,
we offer some remarks on the role of
the metric and of electromagnetic duality 
in determining birefringence conditions
within the context 
of the general constitutive relations \rf{constit}.

Our primary conjecture is that the only nonbirefringent terms 
arise from the non-Weyl component of 
$\hat \ch^{\mu\nu\rh\si}$.
We therefore seek a procedure for extracting this component.
The structure of the Weyl decomposition
\rf{decompconstit}
suggests the possibility 
of introducing an effective metric $\hat g_{\mu\nu}$
in terms of which the Ricci component vanishes.
This would imply a natural decomposition of
$\hat \ch^{\mu\nu\rh\si}$
into metric and non-metric components or,
equivalently,
into nonbirefringent and birefringent components.

We therefore postulate the existence of an effective metric 
$\hat g_{\mu\nu}$
satisfying the eigenproblem
\beq
\Frac23\hat \ch^{\mu\nu\rh\si}\hat g_{\nu\si} = 
(\hat g^{-1})^{\mu\rh} ,
\label{metric_def}
\eeq
where the scale factor for $\hat g_{\mu\nu}$
is chosen so that 
the proportionality constant in this equation
matches the value for standard electrodynamics.
Note that in the present context
the effective metric $\hat g_{\mu\rh}$
typically depends on the 4-momentum
and therefore cannot be interpreted 
as a conventional spacetime metric.
The existence and uniqueness of solutions
to Eq.\ \rf{metric_def}
is an interesting open issue,
but here we suppose all physically reasonable constitutive operators
$\hat \ch^{\mu\nu\rh\si}$
lead to a solution that is unique up to a sign.
For practical purposes,
this issue is moot because it suffices to use 
Eqs.\ \rf{constit} and \rf{metric_def}
to find a perturbative expansion for $\hat g_{\mu\nu}$.

The effective metric $\hat g_{\mu\nu}$ provides 
a unique Weyl decomposition of the constitutive tensor
$\hat \ch^{\mu\nu\rh\si}$,
given by
\beq
\hat \ch = \Wedge{2} \hat g^{-1} + \hat \ch_w ,
\eeq
where $\hat \ch_w$ is the trace-free Weyl component,
$(\hat \ch_w)^{\mu\nu\rh\si}\hat g_{\nu\si} = 0$.
Our conjecture now identifies $\hat \ch_w$
as the birefringent component.
We can verify this in special limits.
For vanishing $\hat \ch_w$ and no CPT violation,
the dispersion relation \rf{dr} reduces to 
\beq
-\big((\hat g^{-1})^{\mu\nu}p_\mu p_\nu\big)^2/ \det \hat g=0,
\eeq
as expected.
Also, 
for this case we find that that the cofactor tensor $\Wedge{2} M$ 
is proportional to $(\hat g^{-1})^{\mu\nu}p_\mu p_\nu$,
which demonstrates the absence of birefringence.
Another limit of interest involves small Lorentz violation.
Solving at leading order in $\kf$, 
we find 
\begin{align}
\hat g_{\mu\nu} &\simeq \et_{\mu\nu}-(\cf)_{\mu\nu} ,
\notag\\
(\hat \ch_w)^{\al\be\mu\nu} &\simeq \hat C^{\al\be\mu\nu} .
\end{align}
This is also consistent,
since it demonstrates that $\hat C$ corresponds 
to the leading-order birefringent component
and $\cf$ to the nonbirefringent part.

In conventional electrodynamics,
electromagnetic duality ensures no birefringence occurs
in the vacuum. 
Given the 2-form field strength $F$
and the dual 2-form $H=*F$, 
the Maxwell equations {\it in vacuo} take the compact form 
$dH=0$, $dF=0$.
Here the $*$-operator is defined as
${*_{\mu\nu}}^{\rh\si} = \half{\ep_{\mu\nu}}^{\rh\si}$
and obeys $**=-I$.
In this language,
electromagnetic duality can be understood as the statement 
that if $F$ is a solution then so is 
$F'= \exp(\al *) F = F \cos\al + H \sin\al$. 
This can be viewed as a rotation between $F$ and $H$.
The chiral components $F_\pm=\half(1\pm i*)F$
of the field strength are irreducible representations 
of these duality rotations and also of the Lorentz group.
The relevance to our discussion is that 
duality symmetry excludes birefringence 
because it implies the space of plane-wave solutions 
is two dimensional.
There are two independent polarizations
that mix under duality transformations,
and the duality symmetry ensures that
both polarizations propagate with the same phase velocity,
so no evolution of polarization can occur.

This conventional duality can be generalized 
to the effective metric $\hat g_{\mu\nu}$ 
relevant in our context.
For an arbitrary effective metric, 
we define
\beq
\hat * = \sqrt{- \det \hat g} * (\Wedge{2} \hat g^{-1}),
\eeq
which is normalized to satisfy $\hat * \hat * = -I$ as usual.
This operator induces a chiral structure
with respect to $\hat g_{\mu\nu}$.
The projections of the field strength $F$
onto the chiral subspaces are now 
\beq
F_\pm=\half(1\pm i\hat *)F.
\eeq
Neglecting CPT-odd terms,
the dual field strength 
\beq
H\equiv *G=*\hat \ch F
\eeq
can also be decomposed into chiral components,
\beq
H_\pm = *\hat \ch F_\pm = \half(1\pm i\hat *)H.
\eeq

According to our primary conjecture,
the constitutive tensor $\ch$ 
provides a natural decomposition of the solution space.
It is therefore plausible that  
the absence of birefringence 
is associated with a duality symmetry
generated by $\hat *$,
with any nonzero Weyl piece $\hat \ch_w$
or nonzero CPT violation
breaking this symmetry and so causing birefringence.
We can explore this idea for the case of 
constitutive relations of the metric type,
$\hat \ch=\Wedge{2}\hat g^{-1}$.
This form of the constitutive tensor 
yields the closure relation 
$*\hat \ch*\hat \ch= 1/\det \hat g$.
Under these circumstances,
if $F$ represents a solution to
the Maxwell equations $dF=dH=0$,
then it can be shown that 
$F\to *\hat \ch F$ is also a solution.
This generalizes the result for conventional duality rotations 
to the $\hat *$-chiral subspaces.
It follows that every polarization is associated
with a second polarization that is also a solution
and hence that birefringence is absent,
as expected.

A rigorous derivation of the above results
is an open problem of definite interest,
although it lies beyond our present scope.
Note that our arguments hold inside any simply connected
source-free region.
We anticipate that they can also be applied 
to more general scenarios 
with inhomogeneous constituent tensors
and curved spacetimes,
in which duality would be defined locally.
It is also plausible that duality breaking
could arise via boundary conditions.
This could lead,
for example,
to a foundational understanding of degeneracy splitting 
in resonant cavities.
Generalizations of the framework may also merit investigation. 
For instance,
more complicated closure relations
of the form $*\hat \ch*\hat \ch = a+b*\hat \ch$
for scalar $a$ and $b$
also result in duality symmetries,
and they may be related to improved decompositions
of the constitutive relations.
One could also consider a duality at the potential level, 
involving mixing of two 1-forms $A$ and $B$
obeying $F=dA$ and $H=dB$,
which can be found for any vacuum solution.
Such approaches may make it possible to incorporate
CPT-violating effects in this picture.

\subsection{Effects of Lorentz violation}
\label{sec_effects}

It is of interest to categorize 
the types of effects produced by Lorentz violation
in various physical situations.
A number of schemes are possible.
In this subsection,
we offer some remarks about the categorization 
used in later sections,
which is based on identifying Lorentz-violating effects
in terms of birefringence, dispersion, and anisotropy.

In the previous subsections,
we have defined birefringence 
as the existence of only one low-energy eigenmode
for a particular solution 
to the covariant dispersion relation \rf{dr},
and we have conjectured its connection 
to a breakdown of duality. 
This definition contains more than 
the notion of rotation of the polarization of light
in the vacuum.
The breaking of eigenmode degeneracy 
can cause effects in circumstances other than vacuum propagation,
such as the splitting of resonant frequencies in cavities.
In the remainder of this work,
the term birefringence is used 
in the sense of eigenmode nondegeneracy.
In vacuum propagation,
this reduces to the usual notion of rotation of polarization 
and can be termed vacuum birefringence.

A similar dichotomy appears in the definition of dispersion.
In what follows,
we adopt the term dispersive to refer 
to Lorentz-violating operators 
in the Lagrange density that appear with other than two derivatives.
Only operators with mass dimension $d=4$ are nondispersive 
in this sense.
In the momentum-space covariant dispersion relation,
these operators contribute terms 
that are non-quadratic in the momentum.
However,
this definition implies more than merely
a nonlinear relationship between the frequency $p_0$ 
and the momentum $\mbf p$ for vacuum propagation.
As shown explicitly in Sec.\ \ref{vacorthog},
some operators involving non-quadratic derivative terms
and hence labeled as dispersive according to our usage 
in fact produce no leading-order dispersion in vacuum propagation,
although they can produce 
analogous effects in other situations
such as cavity resonators.
Throughout this work,
we use the term dispersive to mean $d\neq 4$,
and we reserve the term vacuum dispersion
for modifications to the usual relation 
$p_0 = |\mbf p|$ for vacuum propagation. 

In any physical situation,
the properties of electromagnetic waves
are determined by the coefficients for Lorentz violation,
the medium,
and the boundary conditions. 
Some useful intuition can be gained by considering
the interplay between these
and the resulting birefringent, 
dispersive, 
and anisotropic effects.

Consider first the presence of a macroscopic medium.
We have seen in the previous subsection 
that a coefficient controls birefringence
if it is associated with CPT violation
or if it contributes to the Weyl piece
of the constitutive tensor
$\hat \ch^{\mu\nu\rh\si}$.
The presence of a medium 
can affect the Weyl decomposition,
as shown in Eq.\ \rf{mediumdecomp},
so the manifestation of Lorentz-violating birefringence
depends on the medium.
In contrast,
a coefficient controls Lorentz-violating dispersion
according to the derivative structure
of the corresponding operator in the Lagrange density,
which is unaffected by the medium.
Similarly,
Lorentz-violating anisotropic effects are 
determined by the rotation properties
of the relevant operators,
and so they too are independent of the medium.

The role of the boundary conditions is different.
A given choice of boundary conditions
determines which coefficients for Lorentz violation 
are measurable.
This feature is similar to a property 
of conventional electrodynamics.
A given solution,
such as a vacuum plane wave
or a cavity eigenmode,
is determined by both the Maxwell equations
and by the boundary conditions.
Changing the boundary conditions
reveals distinct sets of eigensolutions
with different physical properties.
One cannot, for instance,
expand vacuum plane waves in terms of cavity eigenmodes.
Similarly,
in the presence of Lorentz violation,
a specific choice of boundary conditions
fixes certain eigenmodes as solutions.
However,
only a suitable subset of the coefficients for Lorentz violation
affects a given set of eigenmodes.
For example,
far-field boundary conditions 
suitable for vacuum radiation 
yield solutions that depend on 
a much reduced subset of coefficients,
as is explicitly identified in Sec.\ \ref{sec_vac}.

Despite their role in determining 
the observability of coefficients for Lorentz violation,
the boundary conditions
have no impact on the associated physical effects.
For example,
the birefringence properties of a given operator are unaffected,
essentially because the duality symmetry
is a local property and therefore
is independent of the boundary conditions. 
Dispersive properties are also unaffected
because they are associated with position-space derivatives
on the Lorentz-violating operators
and therefore are basically a local feature in position space.
The associated modification of the momentum-space relation 
between $p_0$ and $\mbf p$ reflects
an impact on the eigensolution space as described above,
rather than a change in the underlying dispersive properties
of the operators. 
Similarly,
the anisotropy properties of Lorentz-violating operators
are fixed by the Lagrange density 
and are independent of the boundary conditions. 

This feature of the boundary conditions
has the interesting implication
that a comprehensive search for Lorentz violation
requires multiple observational and experimental methods,
since any one method typically applies one type of boundary condition
and so cannot be expected to access the whole coefficient space.
It also means there are multiple approaches 
for categorizing the coefficients.
For example,
one can split the coefficient space
into a subset selected by boundary conditions
appropriate for vacuum propagation
and its complement.
This turns out to be an apposite splitting
for several reasons,
and we develop it in some detail 
beginning in Sec.\ \ref{sec_models}.
However,
one could in principle consider alternative 
splittings of the coefficient space using other 
boundary conditions,
such as ones for resonant cavities.

\section{Spherical decomposition}
\label{sec_gen_coeffs}

In the previous section,
we have demonstrated
that the effects on photon propagation 
of Lorentz-violating operators of arbitrary dimension 
are specified by the Lagrange density \rf{lagrangian}
and are determined 
by the coefficients for Lorentz violation 
${(\kafd{d})_\ka}^{\al_1\ldots\al_{(d-3)}}$ and
$(\kfd{d})^{\ka\la\mu\nu\al_1\ldots\al_{(d-4)}}$.
A classification of these coefficients
that characterizes the key effects
of the corresponding Lorentz-violating operators
is both useful and convenient for more detailed investigations.
The ideal scenario is to establish
a minimal collection of independent coefficients
associated with operators having physical properties
of direct relevance to observation and experiment.

One natural classification scheme 
takes advantage of the role of spatial rotations
to perform an SO(3) decomposition
of the coefficients for Lorentz violation.
This technique has been applied
to obtain first measurements of certain coefficients 
relevant to vacuum photon propagation 
from astrophysical observations 
of active galaxies, gamma-ray bursts, 
and the cosmic microwave background
\cite{km_cmb,km_apjl}.
The SO(3) decomposition uses 
spin-weighted spherical harmonics
\cite{sYjm1,sYjm2},
which are angular-momentum eigenstates
and so obey relatively simple
transformation rules under rotations.
The method has the advantage that
spin-weighted spherical harmonics
are commonly used in some areas of astrophysics 
\cite{zs}
and are well understood.
A summary of some properties
of spin-weighted spherical harmonics
is provided in Appendix A,
which also derives several mathematical relations
used in what follows.

In this section, 
we discuss the decomposition
of the coefficients $\kaf$ and $\kf$ 
into spin-weighted components.
For the analysis,
the coefficient $\kaf$  
is separated into the pseudoscalar $(\kaf)_0$
and the pseudovector $\mbf\kaf$,
while $\kf$ 
is separated into the tensors $\kde$ and $\khb$
and the pseudotensor $\kdb$
defined in Eq.\ \rf{kappas}.
Each of these five components 
is expanded in a helicity basis 
and decomposed into spin-weighted spherical harmonics.
The symmetries of $\kaf$ and $\kf$
then permit extraction of
a minimal collection of spherical coefficients
for Lorentz violation.

The results in this section demonstrate
that the minimal collection of spherical coefficients 
includes nine sets of coefficients. 
For convenience,
we summarize the notation here. 
Three sets are extracted from $\kaf$
and are denoted 
$\kafzB$,
$\kafoB$,
and $\kafoE$.
Six sets emerge from $\kf$,
denoted 
$\cfzE$,
$\kfzE$, 
$\kfoE$,
$\kftE$,
$\kfoB$,
and $\kftB$.
The symbol $c$ specifies coefficients 
associated with nonbirefringent operators,
while $k$ specifies birefringent ones.
A negation diacritic $\neg$ denotes coefficients that have 
no leading-order effects
on the vacuum propagation of light,
a property derived in Sec.\ \ref{vacorthog}.
The subscripts $n$, $j$, and $m$
determine the frequency or wavelength dependence,
the total angular momentum, 
and the $z$-component of the angular momentum,
respectively.
The allowed ranges of these three indices 
and the counting of the coefficients
are given in Tables 
\ref{kaf_ranges},
\ref{kf_B_ranges},
\ref{cf_ranges},
and \ref{kf_E_ranges}.
The label $d$ gives the mass dimension
of the corresponding operator for Lorentz violation,
while the numerals 0,1, or 2 
preceding either $E$ or $B$
refer to the spin weight of the operator.
The superscripts $E$ and $B$ refer
to the parity of the operator,
with parity $(-1)^j$ labeled as $E$ 
and parity $(-1)^{j+1}$ as $B$.
The phases are chosen so
that each spherical coefficient ${\cal K}_{jm}$ 
for Lorentz violation
obeys the complex-conjugation relation
\beq
({\cal K}_{jm})^*=(-1)^m{\cal K}_{j(-m)}.
\label{ccrel}
\eeq
Note that this implies that
the pair of complex coefficients
${\cal K}_{jm}$ and ${\cal K}_{j(-m)}$
are codependent
but represent two real degrees of freedom.

\subsection{General CPT-odd coefficients}
\label{sec_odd_coeffs}

\begin{table*}
\tabcolsep 3pt
\renewcommand{\arraystretch}{0.5}
\begin{tabular}{c||*{7}{c}|*{7}{c}|*{6}{c}}
\multicolumn{1}{c||}{}&
\multicolumn{7}{c|}{$\kafzB$}&
\multicolumn{7}{c|}{$\kafoB$}&
\multicolumn{6}{c}{$\kafoE$}\\[4pt]
\hline
\hline
\multicolumn{1}{c||}{$n$}&
\multicolumn{7}{c|}{$j$}&
\multicolumn{7}{c|}{$j$}&
\multicolumn{6}{c}{$j$}\\[4pt]
\hline
0  &0&&&&&&   &1&&&&&&   &&&&&&\\
1  &&1&&&&&   &&2&&&&&   &1&&&&&\\
2  &0&&2&&&&  &1&&3&&&&  &&2&&&&\\
3  &&1&&3&&&  &&2&&4&&&  &1&&3&&&\\
4  &0&&2&&4&& &1&&3&&5&& &&2&&4&&\\
$\vdots$  &$\vdots$&&&&&$\ddots$&  
   &$\vdots$&&&&&$\ddots$&  &$\vdots$&&&&$\ddots$&\\
$d-3$&0&&2&&$\cdots$&&$d-3$  &1&&3&&$\cdots$&&$d-2$  
   &&2&&$\cdots$&&$d-3$ \\[4pt]
\hline
\hline
\multicolumn{1}{c||}{total}&
\multicolumn{7}{c|}{$\Frac16 d(d-1)(d-2)$}&
\multicolumn{7}{c|}{$\Frac16 (d-1)(d^2+d-3)$}&
\multicolumn{6}{c}{$\Frac16 (d+1)(d-1)(d-3)$}
\end{tabular}
\caption{\label{kaf_ranges}
Summary of the allowed ranges of indices $n$ and $j$
for the independent spherical coefficients
associated with CPT-odd operators.
The dimension $d$ is odd with $d\geq3$,
while $n\leq d-3$.
The index $m$ satisfies the usual restrictions
$-j\leq m\leq j$,
so there are $2j+1$ coefficients for each $j$.
For a given dimension $d$,
the number of coefficients of each type
is given in the last row. 
Adding these gives the expected total of 
$\half (d+1)(d-1)(d-2)$.}
\end{table*}

We begin the decomposition of $(\kaf)_\ka$
by performing an expansion of the pseudoscalar component $(\kaf)_0$
in spherical harmonics.
The first step is to separate $(\kaf)_0$ 
into pieces with definite frequency and 3-momentum dependence. 
Denoting the frequency by $\om=p^0$
and the components of the 3-momentum $\mbf p$ by $p^k$,
we obtain the expression
\begin{align}
(\kaf)_0 &= \sum_d\sum_{n=0}^{d-3} (-1)^{n+(d+1)/2} 
\left(\begin{smallmatrix}d-3\\n\end{smallmatrix}\right) 
\notag \\
&\qquad 
\times
{(\kafd{d})_0}^{0\ldots 0k_1\ldots k_n}
\om^{d-3-n} p^{k_1}\ldots p^{k_n} .
\end{align}
Writing the magnitude of the 3-momentum as $p=|\mbf p|$,
we see that
each term in the sum involves a factor of $\om^{d-3-n}p^n$.
The frequency and wavelength dependence of each term
is therefore controlled by the new index $n$.
The direction dependence introduced by the components of 
$\mbf p$ is characterized 
through the expansion in spherical harmonics below.

To determine the relevant angular-momentum eigenvalues $j$ 
for the spherical-harmonic expansion,
we can break the coefficients for each $n$ 
into a series of three-dimensional traceless symmetric tensors 
of rank $n$, $n-2$, $n-4$, $\ldots$.
This implies the range of eigenvalues 
$l=n$, $n-2$, $n-4$, $\ldots$
for the orbital angular momentum.
Since the spin of $(\kaf)_0$ is zero, 
the eigenvalue $j$ must also span this range.
An alternative and more elegant approach makes use of parity.
For a given $n$, 
the orbital angular momentum is limited by $n$.
Since the spin is zero, 
we must have $j\leq n$.
Given that $(\kaf)^0$ is a pseudoscalar
and $\mbf p$ is a vector,
each term in the expansion must have parity $(-1)^{n+1}$
with only $B$-type parity occurring.
This imposes the relation $(-1)^{n+1}=(-1)^{j+1}$,
which implies that $j-n$ is even.
The conclusion is that $j=n$, $n-2$, $n-4$, $\ldots \geq 0$,
as before.

The resulting expansion in spherical harmonics is given by
\beq
(\kaf)_0 = \sum_{dnjm} \om^{d-3-n}p^n\,
\syjm{0}{jm}(\mbf{\hat p})\ \kafzB .
\label{kaf0}
\eeq
The spherical coefficients for Lorentz violation 
$\kafzB$
are nonzero for the $n$ and $j$
values listed in Table \ref{kaf_ranges}.
As an example, 
consider $d=5$.
In this case,
there are two $j=0$ singlets, 
$\kafzBdnjm 5 {000}$ and $\kafzBdnjm 5 {200}$.
There is also one $j=1$ triplet
$\kafzBdnjm 5 {11m}$,
with $m = -1$, $0$, $1$.
Finally,
there is one $j=2$ quintuplet
$\kafzBdnjm 5 {22m}$,
with $m = -2$, $-1$, $0$, $1$, $2$.
The total number of coefficients is $1+1+3+5=10$.

Next,
we perform a spherical-harmonic expansion of the radial component
$(\kaf)_r = \mbf{\hat p}\cdot \mbf{\kaf}$ 
$= \mbf{\hat e}_r\cdot \mbf{\kaf}$.
This component is also a pseudoscalar,
and the decomposition follows the same
basic steps as above.
However, 
the additional $\mbf{\hat p}$ factor
implies that the total angular momentum
is now limited by $j\leq n+1$.
Also, 
the parity of each term is given by $(-1)^n=(-1)^{j+1}$, 
yielding the range of $j$ as
$j=n+1$, $n-1$, $n-3$, $\ldots\geq 0$.
We therefore obtain the expansion
\beq
(\kaf)_r = \sum_{dnjm} \om^{d-3-n}p^n\,
\syjm{0}{jm}(\mbf{\hat p})\ \kafzBp ,
\label{kafr}
\eeq
involving another set of spherical coefficients 
$\kafzBp$.
However,
it turns out that the symmetries of the tensors
${(\kafd{d})_\ka}^{\al_1\ldots\al_{(d-3)}}$
imply that these new coefficients 
can all be expressed as combinations of
the other spherical coefficients 
occurring in the expansion of $(\kaf)_\ka$.
Before demonstrating this,
we first complete the expansion of $\mbf{\kaf}$.

The remaining components of $\mbf{\kaf}$
have spin weight $\pm 1$,
$(\kaf)_\pm = \mbf{\hat e}_\pm\cdot \mbf{\kaf}$.
For these cases,
we again find $j\leq n+1$,
but now coefficients with both $E$- and $B$-type parities occur.
The parity is $(-1)^n$,
which implies $j=n+1$, $n-1$, $\ldots \geq 1$ 
for the $B$-type components 
and $j=n$, $n-2$, $\ldots \geq 1$ 
for the $E$-type components.
Since the spin weight is $\pm1$, 
the index $j$ for the total angular momentum 
is limited from below by 1.
The expansions of the components $(\kaf)_\pm$
in terms of spin-weighted spherical harmonics
take the forms
\begin{align}
(\kaf)_\pm &= \sum_{dnjm} \om^{d-3-n}p^n\,
\syjm{\pm1}{jm}(\mbf{\hat p})\
\notag\\
&\vspace{-10pt} \times\Frac{1}{\sqrt{2j(j+1)}} 
\big( \pm\kafoB + i \kafoE \big),
\label{kafpm}
\end{align}
where we introduce a factor of $1/\sqrt{2j(j+1)}$
for later convenience.
The spherical coefficients for Lorentz violation 
$\kafoB$ and $\kafoE$
represent the remaining two independent sets.
Their ranges are summarized in Table \ref{kaf_ranges}.

The symmetries of 
${(\kafd{d})_\ka}^{\al_1\ldots\al_{(d-3)}}$
described in Sec.\ \ref{sec_lagrangian}
imply certain constraints
on the four sets of coefficients 
$\kafzB$, $\kafzBp$, $\kafoB$, and $\kafoE$.
To determine these constraints,
the symmetries must be formulated 
in the momentum-space helicity basis.
Recall that the tensors 
${(\kafd{d})_\ka}^{\al_1\ldots\al_{(d-3)}}$
are totally symmetric in the last $d-3$ indices
and that any trace involving the first index vanishes.
The total-symmetry condition is implicit
in the above spherical-harmonic decompositions,
but the trace condition provides 
a nontrivial constraint.

In momentum space,
the trace condition can be written 
as the differential equation
\begin{align}
0 &= \frac{\prt\phantom{p_\al}}{\prt p_\ka} (\kaf)_\ka = 
\frac{\prt\phantom{\om}}{\prt \om} (\kaf)_0 
+\mbf\nabla\cdot\mbf\kaf  \notag \\
&= \frac{\prt\phantom{\om}}{\prt \om} (\kaf)_0 
+\nabla_r (\kaf)_r 
+\nabla_+ (\kaf)_- 
+\nabla_- (\kaf)_+ \notag \\
&=\frac{\prt\phantom{\om}}{\prt \om} (\kaf)_0 
+\big(\frac{\prt\phantom{p}}{\prt p} 
+\frac{2}{p}\big) (\kaf)_r  \notag \\
& \qquad 
+\frac{1}{p}\big(J_+ (\kaf)_- - J_- (\kaf)_+ \big) , 
\label{bconst}
\end{align}
where we have used the identities \rf{J-ident}.
The combination $J_+ (\kaf)_- - J_- (\kaf)_+$
is a pseudoscalar that is generated
by the $B$ component of $\mbf\kaf$
and that contains no $E$ component.
Consequently, 
Eq.\ \rf{bconst} provides a symmetry constraint 
involving only the $B$-type coefficients
$\kafzB$, $\kafzBp$, and $\kafoB$.
Inserting the spherical-harmonic expansions
into Eq.\ \rf{bconst} 
and making use of the identity \rf{J-ident3}
yields an explicit relation 
involving these three coefficient sets.
Careful consideration of the index ranges 
then reveals that $\kafzBp$ may be written as 
\begin{align}
\kafzBp &= \frac{-1\ }{n+2} \big( \kafoB\,
\notag \\
& \qquad \qquad 
+(d-2-n) \kafzBdnjm{d}{(n-1)jm} \big) . 
\label{kafconst}
\end{align}
We conclude that the auxiliary coefficients $\kafzBp$
are completely determined as linear combinations of 
the independent coefficients $\kafzB$ and $\kafoB$.

The net yield of the helicity decomposition of $(\kaf)_\ka$
is therefore three independent sets 
of spherical coefficients for Lorentz violation,
which are the $B$-type coefficients $\kafzB$ and $\kafoB$,
and the $E$-type coefficients $\kafoE$.
These coefficients completely characterize
the CPT-odd Lorentz-violating operators 
of arbitrary mass dimension 
associated with the quadratic action for electrodynamics.
All the operators are birefringent.
On each set of coefficients,
the label $d$ specifies the (odd) mass dimension
of the operator.
The three indices $n$, $j$, and $m$
determine the frequency or wavelength dependence,
the total angular momentum, 
and the $z$-component of the angular momentum,
respectively.
The allowed ranges of these three indices 
and the counting of the coefficients
are given in Table \ref{kaf_ranges}.
The superscripts $E$ and $B$ refer
to the operator parity,
while the superscript numerals 0 or 1 
preceding either $E$ or $B$ 
specify the spin weight of the operator.
The total number of coefficients for given odd $d$ 
is $\half (d+1)(d-1)(d-2)$,
matching the group-theoretic result 
from Sec.\ \ref{sec_theory}.

\subsection{General CPT-even coefficients}
\label{sec_even_coeffs}

The spherical decomposition 
of the CPT-even Lorentz-violating operators 
$(\kf)^{\ka\la\mu\nu}$
follows a procedure similar to that for the CPT-odd case
discussed in the previous subsection.
However,
instead of working directly with $(\kf)^{\ka\la\mu\nu}$,
it is more convenient to decompose 
the three matrices $\kde$, $\khb$, and $\kdb$
given in Eq.\ \rf{kappas},
which have equivalent content.

\begin{table*}
\tabcolsep 3pt
\renewcommand{\arraystretch}{0.5}
\begin{tabular}{c||*{8}{c}l|*{7}{c}l|*{6}{c}l|*{6}{c}l}
\multicolumn{1}{c||}{}&
\multicolumn{9}{c|}{$\EzE,\DzE,$}&
\multicolumn{8}{c|}{$\EoE,\DoE,$}&
\multicolumn{7}{c|}{$\EtE,\DtE,$}&
\multicolumn{7}{c}{}\\[4pt]
\multicolumn{1}{c||}{}&
\multicolumn{9}{c|}{$\BzB$}&
\multicolumn{8}{c|}{$\BoB$}&
\multicolumn{7}{c|}{$\BtB$}&
\multicolumn{7}{c}{$\ezE,\dzE$}\\[4pt]
\hline
\hline
\multicolumn{1}{c||}{$n$}&
\multicolumn{9}{c|}{$j$}&
\multicolumn{8}{c|}{$j$}&
\multicolumn{7}{c|}{$j$}&
\multicolumn{7}{c}{$j$}\\[4pt]
\hline
0  &0& &2& & & & & &  & &2& & & & & &  &2& & & & & &  &0& & & & & &\\
1  & &1& &3& & & & &  &1& &3& & & & &  & &3& & & & &  & &1& & & & &\\
2  &0& &2& &4& & & &  & &2& &4& & & &  &2& &4& & & &  &0& &2& & & &\\
3  & &1& &3& &5& & &  &1& &3& &5& & &  & &3& &5& & &  & &1& &3& & &\\
4  &0& &2& &4& &6& &  & &2& &4& &6& &  &2& &4& &6& &  &0& &2& &4& &\\
$\vdots$&
$\vdots$& & & & & & &$\ddots$& &
$\vdots$& & & & & &$\ddots$& &
$\vdots$& & & & &$\ddots$& &
$\vdots$& & & & &$\ddots$& \\
$d-4$
&0& &2& & &$\cdots$& & &$d-2$
& &2& &4& &$\cdots$ & &$d-2$ 
&2& &4& &$\cdots$& &$d-2$
&0& &2& &$\cdots$& &$d-4$\\[4pt]
\end{tabular}
\end{table*}
\begin{table*}
\tabcolsep 3pt
\renewcommand{\arraystretch}{0.5}
\begin{tabular}{c||*{6}{c}l|*{5}{c}l|*{7}{c}l|*{5}{c}l}
\multicolumn{1}{c||}{}&
\multicolumn{7}{c|}{$\EoB,\DoB$}&
\multicolumn{6}{c|}{$\EtB,\DtB$}&
\multicolumn{8}{c|}{$$}&
\multicolumn{6}{c}{}\\[4pt]
\multicolumn{1}{c||}{}&
\multicolumn{7}{c|}{$\BoE,\HoE$}&
\multicolumn{6}{c|}{$\BtE$}&
\multicolumn{8}{c|}{$\HzE$}&
\multicolumn{6}{c}{$\HoB$}\\[4pt]
\hline
\hline
\multicolumn{1}{c||}{$n$}&
\multicolumn{7}{c|}{$j$}&
\multicolumn{6}{c|}{$j$}&
\multicolumn{8}{c|}{$j$}&
\multicolumn{6}{c}{$j$}\\[4pt]
\hline
0  &1& & & & & &  & & & & & &  & &1& & & & & &  & & & & & &\\
1  & &2& & & & &  &2& & & & &  &0& &2& & & & &  &1& & & & &\\
2  &1& &3& & & &  & &3& & & &  & &1& &3& & & &  & &2& & & &\\
3  & &2& &4& & &  &2& &4& & &  &0& &2& &4& & &  &1& &3& & &\\
4  &1& &3& &5& &  & &3& &5& &  & &1& &3& &5& &  & &2& &4& &\\
$\vdots$&
$\vdots$& & & & &$\ddots$& &
$\vdots$& & & &$\ddots$& &
$\vdots$& & & & & &$\ddots$& &
$\vdots$& & & &$\ddots$& \\
$d-4$
&1& &3& &$\cdots$& &$d-3$
& &3& &$\cdots$& &$d-3$
& &1& &3& &$\cdots$& &$d-3$
& &2& &$\cdots$& &$d-4$\\[4pt]
\end{tabular}
\caption{\label{codepranges}
Summary of the allowed ranges of indices $n$ and $j$
for the 20 sets of codependent spherical coefficients
associated with general CPT-even operators.
The dimension $d$ is even with $d\geq4$,
while $n\leq d-4$.
The index $m$ satisfies the usual restrictions
$-j\leq m\leq j$,
so there are $2j+1$ coefficients for each $j$. }
\end{table*}

We begin by separating 
the three matrices $\kde$, $\khb$, $\kdb$
into their SO(3)-irreducible trace, 
symmetric traceless, 
and antisymmetric parts.
These irreducible parts can be  
expressed in the helicity basis via 
$\ka_{ab} = \mbf{\hat e}_a\cdot \ka \cdot \mbf{\hat e}_b$,
where $a,b = +,r,-$.
Each component can then be expanded
in the appropriate spin-weighted spherical harmonics,
recalling that $\kde$, $\khb$ are tensors
while $\kdb$ is a pseudotensor. 
At this stage,
we find the results can be expressed as 13 equations
involving sums over twelve sets of $E$-type 
and eight sets of $B$-type coefficients,
for a total of 20 sets of codependent coefficients.
The challenge is to use these equations
and the symmetries of $(\kf)^{\ka\la\mu\nu}$
to identify the independent sets of coefficients
among the 20 codependent ones.
The 13 equations are given explicitly below.
In them,
all sums are restricted to 
even dimensions $d\geq 4$.
The maximum range of the indices
spans $0\leq n \leq d-4$ for $n$
and $0\leq j \leq n+2$, $-j\leq m \leq j$
for the eigenvalue indices of the angular momentum.
Details of the allowed index ranges
for each of the 20 sets of codependent coefficients 
are given in Table \ref{codepranges}.

For the matrix $\kde$,
six sets of coefficients are needed,
and the result is
\begin{subequations}
\begin{align}
(\kde)_{rr} &= 
\sum_{dnjm} \om^{d-4-n} p^n\, \syjm{0}{jm}(\mbf{\hat p})\
\notag \\
&\qquad 
\times \big( \EzE + \ezE \big) , 
\label{kde1} 
\db\\[10pt]
(\kde)_{+-} &= 
\sum_{dnjm} \om^{d-4-n} p^n\, \syjm{0}{jm}(\mbf{\hat p})\
\notag \\
&\qquad 
\times \big( -\half\EzE +\ezE \big) , 
\label{kde2}
\db\\[10pt]
(\kde)_{\pm r} &= 
\sum_{dnjm} \om^{d-4-n} p^n\, \syjm{\pm1}{jm}(\mbf{\hat p})\
\notag \\
&\quad 
\times \Frac{1}{\sqrt{2j(j+1)}} \big( \pm\EoE + i\EoB \big) , 
\label{kde3}
\db\\[10pt]
(\kde)_{\pm\pm} &= 
\sum_{dnjm} \om^{d-4-n} p^n\, \syjm{\pm2}{jm}(\mbf{\hat p})\
\notag \\
&\qquad 
\times \sqrt\Frac{(j-2)!}{(j+2)!} \big( \EtE \pm i \EtB \big) . 
\label{kde4}  
\end{align}
\label{kf_exp}%
\end{subequations}
In this decomposition,
the coefficients $\ezE$ emerge 
from the trace component of $\kde$.

Another six sets of coefficients are needed
for the matrix $\khb$.
The corresponding equations are
\begin{subequations}
\begin{align}
(\khb)_{rr} &= 
\sum_{dnjm} \om^{d-4-n} p^n\, \syjm{0}{jm}(\mbf{\hat p})\
\notag \\
&\qquad 
\times \big( \DzE + \dzE \big) , 
\label{khb1}
\db\\[10pt]
(\khb)_{+-} &= 
\sum_{dnjm} \om^{d-4-n} p^n\, \syjm{0}{jm}(\mbf{\hat p})\
\notag \\
&\qquad 
\times \big( -\half\DzE + \dzE \big) , 
\label{khb2}
\db\\[10pt]
(\khb)_{\pm r} &= 
\sum_{dnjm} \om^{d-4-n} p^n\, \syjm{\pm1}{jm}(\mbf{\hat p})\
\notag \\
&\quad 
\times \Frac{1}{\sqrt{2j(j+1)}} \big( \pm\DoE + i\DoB \big) , 
\label{khb3}  
\db\\[10pt]
(\khb)_{\pm\pm} &= 
\sum_{dnjm} \om^{d-4-n} p^n\, \syjm{\pm2}{jm}(\mbf{\hat p})\
\notag \\
&\qquad 
\times \sqrt\Frac{(j-2)!}{(j+2)!} \big( \DtE \pm i\DtB \big) .
\label{khb4}  
\end{align}
\label{kf_exp2}%
\end{subequations}
The coefficients $\dzE$
correspond to the trace component of $\khb$.

Finally, 
eight sets of coefficients are required
in the expansion of $\kdb$.
The results are
\begin{subequations}
\begin{align}
(\kdb)_{rr} &= 
\sum_{dnjm} \om^{d-4-n} p^n\, \syjm{0}{jm}(\mbf{\hat p})\
\BzB , 
\label{kdb1}
\db\\[10pt]
(\kdb)_{\pm\mp} &= 
\sum_{dnjm} \om^{d-4-n} p^n\, \syjm{0}{jm}(\mbf{\hat p})\
\notag \\
&\qquad 
\times \big( -\half\BzB \mp i\HzE \big) ,
\label{kdb2}
\db\\[10pt]
(\kdb)_{r\pm} &= 
\sum_{dnjm} \om^{d-4-n} p^n\, \syjm{\pm1}{jm}(\mbf{\hat p})\
\notag \\
&\quad 
\times \Frac{1}{\sqrt{2j(j+1)}} \big( \pm\BoB + i\BoE 
\notag \\
&\quad \qquad 
\qquad 
\pm\HoB - i\HoE \big) , 
\label{kdb3}
\db\\[10pt]
(\kdb)_{\pm r} &= 
\sum_{dnjm} \om^{d-4-n} p^n\, \syjm{\pm1}{jm}(\mbf{\hat p})\
\notag \\
&\quad 
\times \Frac{1}{\sqrt{2j(j+1)}} \big( \pm\BoB + i\BoE 
\notag \\
&\quad \qquad 
\qquad 
\mp\HoB + i\HoE \big) , 
\label{kdb4}
\db\\[10pt]
(\kdb)_{\pm\pm} &= 
\sum_{dnjm} \om^{d-4-n} p^n\, \syjm{\pm2}{jm}(\mbf{\hat p})\
\notag \\
&\qquad 
\times \sqrt\Frac{(j-2)!}{(j+2)!} \big( \BtB \pm i\BtE \big) . 
\label{kdb5} 
\end{align}
\label{kf_exp3}%
\end{subequations}
The sets of coefficients $\HoB$, $\HoE$, and $\HzE$
are associated with the antisymmetric part of $\kde$,
which corresponds to $\kop$ in Eq.\ \rf{kappastwo}.

The next step is to use the symmetries of the tensors 
$(\kfd{d})^{\ka\la\mu\nu\al_1\ldots\al_{(d-4)}}$
to determine the independent sets of coefficients.
These symmetries are described in Sec.\ \ref{sec_theory},
following Eq.\ \rf{kfd}.
As in the CPT-odd case,
the independent coefficients
can be identified by expressing the symmetries 
as differential equations in momentum space.
It turns out that all the symmetries of $\kfd{d}$ 
are implicit in the above 13 equations,
except for the condition of vanishing 
antisymmetrization on any three indices of 
$(\kfd{d})^{\ka\la\mu\nu\al_1\ldots\al_{(d-4)}}$.
In momentum space, 
this symmetry can be imposed by requiring
\beq
0=\prt^{[\rh}(\kf)^{\mu\nu]\ka\la} ,
\label{kf_sym1}
\eeq
where $\prt^\ka = \prt/\prt p_\ka$.
In terms of $\hat\ka$ matrices, 
this requirement becomes the four conditions
\begin{align}
\nabla_b (\khb)^{ab} &= 0 , 
\notag \\
\nabla_b (\kdb)^{ab} &= 0 , 
\notag \\
\frac{\prt\phantom{\om}}{\prt\om} (\kdb)^{ab} &=
\ve^{bcd}\nabla_c {(\kde)_d}^a , 
\notag \\
\frac{\prt\phantom{\om}}{\prt\om} (\khb)^{ab} &=
-\ve^{bcd}\nabla_c{(\kdb)_d}^a 
\label{kf_sym2}
\end{align}
in the notation of the Appendix.
Note the similarity to the Maxwell equations.

The symmetry constraints \rf{kf_sym2}
lead to relations among the 20 codependent coefficients 
in the expansions \rf{kf_exp}, \rf{kf_exp2}, and \rf{kf_exp3}.
The procedure for determining these relations
involves expressing the differential equations \rf{kf_sym2}
in terms of the operators
$\prt/\prt\om$, $\prt/\prt p$, and $J_\pm$,
and then inserting the expansions.
Following this procedure,
we find that Eq.\ \rf{kf_sym2} 
provides nine constraints on the eight $B$-type coefficients
and eleven constraints on the twelve $E$-type coefficients.
However,
these constraints are not all independent.

\begin{table}
\tabcolsep 3pt
\renewcommand{\arraystretch}{0.5}
\begin{tabular}{c||*{8}{c}|*{6}{c}}
\multicolumn{1}{c||}{}&
\multicolumn{8}{c|}{$\kfoB$}&
\multicolumn{6}{c}{$\kftB$}\\[4pt]
\hline
\hline
\multicolumn{1}{c||}{$n$}&
\multicolumn{8}{c|}{$j$}&
\multicolumn{6}{c}{$j$}\\[4pt]
\hline
0  &&2&&&&&&    &&&&&&\\
1  &1&&3&&&&&   &2&&&&&\\
2  &&2&&4&&&&   &&3&&&&\\
3  &1&&3&&5&&&  &2&&4&&&\\
4  &&2&&4&&6&&  &&3&&5&&\\
$\vdots$ &$\vdots$&&& &&        
  &$\ddots$&     &$\vdots$&&&        &$\ddots$&\\
$d-4$    &&2       &&4&&$\cdots$&        
  &$d-2$&&3       &&$\cdots$&        &$d-3$\\[4pt]
\hline
\hline
\multicolumn{1}{c||}{total}&
\multicolumn{8}{c|}{$\Frac16 (d^3-4d-18)$}&
\multicolumn{6}{c}{$\Frac16 (d+3)(d-2)(d-4)$}
\end{tabular}
\caption{\label{kf_B_ranges}
Summary of the allowed ranges of indices $n$ and $j$
for the independent $B$-type spherical coefficients
associated with CPT-even birefringent operators.
The dimension $d$ is even with $d\geq4$,
while $n\leq d-4$.
The index $m$ satisfies the usual restrictions
$-j\leq m\leq j$,
so there are $2j+1$ coefficients for each $j$.
For a given dimension $d$,
the number of coefficients of each type
is given in the last row.}
\end{table}  

Some calculation with the nine constraints 
on the $B$-type coefficients
shows that there exist only two linearly independent sets 
of $B$-type coefficients for Lorentz violation,
which we denote by $\kfoB$ and $\kftB$.
The index ranges for these two sets
are summarized in Table \ref{kf_B_ranges}.
All the corresponding Lorentz-violating operators
produce birefringent effects. 
In terms of these two sets,
the eight codependent $B$-type coefficients appearing 
in the expansions \rf{kf_exp}, \rf{kf_exp2}, and \rf{kf_exp3}
are given by
\begin{subequations}
\begin{align}
\EoB &= (d-3-n)n \kfoBdnjm{d}{(n-1)jm} ,  
\db\\[10pt]
\EtB &= -(d-3-n)(j+2)(j-1)\kfoBdnjm{d}{(n-1)jm} 
\notag \\
&\hspace{-12pt} 
- (d-3-n)(d-2-n)\kftBdnjm{d}{(n-2)jm} ,
\db\\[10pt]
\DoB &= (n+2)\kftB ,  
\db\\[10pt]
\DtB &= -(n+2)(n+3)\kftB ,  
\db\\[10pt]
\BzB &= (n+1)\kfoB ,  
\db\\[10pt]
\BoB &= -\half(n(n+3)+j(j+1))\kfoB 
\notag \\
&\hspace{-12pt} 
-\half(d-3-n) \kftBdnjm{d}{(n-1)jm} ,  
\db\\[10pt]
\HoB &= -\half((n+2)(n+3)-j(j+1))\kfoB 
\notag \\
&\hspace{-12pt} 
+\half(d-3-n)\kftBdnjm{d}{(n-1)jm} ,  
\db\\[10pt]
\BtB &= \half(n+3)(j+2)(j-1)\kfoB 
\notag \\
&\hspace{-12pt} 
+ (d-3-n)(n+2)\kftBdnjm{d}{(n-1)jm} .
\end{align}
\label{kf_B_coeffs}%
\end{subequations}

The same procedure applied 
to the eleven constraints on the $E$-type coefficients 
reveals that there are four linearly independent
sets of $E$-type coefficients for Lorentz violation.
In this case,
however,
some components of the corresponding operators
are birefringent while others are nonbirefringent.
For ease of application in searches for Lorentz violation,
it would be useful to have a decomposition 
that separates birefringent and nonbirefringent components.
At leading order,
this separation is achieved by the Weyl decomposition 
\rf{decompconstit}.
Unfortunately,
the representations from the Weyl decomposition 
are incompatible with 
the $E$-type coefficient representations 
contained in the symmetry constraints \rf{kf_sym1}.
Although the latter indeed include some operators 
that are purely birefringent
and others that are purely nonbirefringent,
there are typically also birefringent operators
with coefficients that contribute 
to the nonbirefringent matrices 
$\kem$, $\kop$, $\ktp$, and $\ktm$.
Among the four independent sets of $E$-type coefficients 
for Lorentz violation,
there are thus ones that contribute to $(\cf)^{\mu\nu}$,
ones that contribute to $\hat C^{\ka\la\mu\nu}$,
and ones that contribute to both.
However,
we do find that only a small fraction of the coefficients 
contribute solely to $\hat C^{\ka\la\mu\nu}$.
Consequently,
it is reasonable to seek a decomposition 
separating the nonbirefringent terms 
appearing only in $(\cf)^{\mu\nu}$ 
from the birefringent terms 
contributing to $\hat C^{\ka\la\mu\nu}$
and possibly also to $(\cf)^{\mu\nu}$.

\begin{table}
\tabcolsep 3pt
\renewcommand{\arraystretch}{0.5}
\begin{tabular}{c||*{7}{c}}
\multicolumn{1}{c||}{}&
\multicolumn{7}{c}{$\cfzE$}\\[4pt]
\hline
\hline
\multicolumn{1}{c||}{$n$}&
\multicolumn{7}{c}{$j$}\\[4pt]
\hline
0  &0&&&&&&\\
1  &&1&&&&&\\
2  &0&&2&&&&\\
3  &&1&&3&&&\\
4  &0&&2&&4&&\\
$\vdots$  &$\vdots$&&&&&$\ddots$&\\
$d-2$  &0&&2&&4&$\cdots$&$d-2$\\[4pt]
\hline
\hline
\multicolumn{1}{c||}{total}&
\multicolumn{7}{c}{$\Frac16 (d+1)d(d-1)$}
\end{tabular}
\caption{\label{cf_ranges}
Summary of the allowed ranges of indices $n$ and $j$
for the independent spherical coefficients
associated with CPT-even nonbirefringent operators.
The dimension $d$ is even with $d\geq4$,
while $n\leq d-4$.
The index $m$ satisfies the usual restrictions
$-j\leq m\leq j$,
so there are $2j+1$ coefficients for each $j$.
For a given dimension $d$,
the number of coefficients is given in the last row.}
\end{table}

\begin{table*}
\tabcolsep 3pt
\renewcommand{\arraystretch}{0.5}
\begin{tabular}{c||*{9}{c}|*{7}{c}|*{5}{c}}
\multicolumn{1}{c||}{}&
\multicolumn{9}{c|}{$\kfzE$}&
\multicolumn{7}{c|}{$\kfoE$}&
\multicolumn{5}{c}{$\kftE$}\\[4pt]
\hline
\hline
\multicolumn{1}{c||}{$n$}&
\multicolumn{9}{c|}{$j$}&
\multicolumn{7}{c|}{$j$}&
\multicolumn{5}{c}{$j$}\\[4pt]
\hline
0  &&&2&&&&&&    &&&&&&&    &&&&&   \\
1  &&1&&3&&&&&   &&2&&&&&   &&&&&   \\
2  &0&&2&&4&&&&  &1&&3&&&&  &2&&&&  \\
3  &&1&&3&&5&&&  &&2&&4&&&  &&3&&&  \\
4  &0&&2&&4&&6&& &1&&3&&5&& &2&&4&& \\
$\vdots$&$\vdots$&&&&&&&$\ddots$&&$\vdots$
  &&&&&$\ddots$&&$\vdots$&&&$\ddots$&\\
$d-4$&0&&2&&&$\cdots$&&&$d-2$&1&&3&&$\cdots$&&$d-3$  
  &2&&4&$\cdots$&$d-4$\\[4pt]
\hline
\hline
\multicolumn{1}{c||}{total}&
\multicolumn{9}{c|}{$\Frac16 (d^3-d-30)$}&
\multicolumn{7}{c|}{$\Frac16 (d-4)(d^2+d+3)$}&
\multicolumn{5}{c}{$\Frac16 (d-4)(d^2-2d-9)$}
\end{tabular}
\caption{\label{kf_E_ranges}
Summary of the allowed ranges of indices $n$ and $j$
for the independent $E$-type spherical coefficients
associated with CPT-even birefringent operators.
The dimension $d$ is even with $d\geq4$,
while $n\leq d-4$.
The index $m$ satisfies the usual restrictions
$-j\leq m\leq j$,
so there are $2j+1$ coefficients for each $j$.
For a given dimension $d$,
the number of coefficients of each type
is given in the last row.}
\end{table*}

We first consider the nonbirefringent case.
With zero leading-order birefringence,
Lorentz violation associated 
with the quadratic action for electrodynamics 
is completely characterized by $(\cf)^{\mu\nu}$ 
via
\begin{align}
(\kf)^{\mu\nu\rh\si} 
& \to\half\big(\et^{\mu\rh}(\cf)^{\nu\si}-\et^{\nu\rh}(\cf)^{\mu\si}
\notag\\
&\qquad\qquad
+\et^{\nu\si}(\cf)^{\mu\rh}-\et^{\mu\si}(\cf)^{\nu\rh}\big).
\label{nonbirefcoeff}
\end{align}
The symmetry constraints \rf{kf_sym1}
applied to this expression yield the equation
$0=\prt^{[\rh}(\cf)^{\mu]\nu}$
in the nonbirefringent case.
The form of this constraint suggests
solving via a scalar potential $\hat\Ph_F$,
where
\beq
(\cf)^{\mu\nu} \to \prt^\mu\prt^\nu \hat\Ph_F,
\label{cnonbiref}
\eeq
where the derivatives act in momentum space.
This result holds in the nonbirefringent case
but is false in general.
Expanding the potential $\hat\Ph_F$ in spherical harmonics,
\beq
\hat\Ph_F 
= \sum_{dnjm} \om^{d-2-n}p^n\, \syjm{0}{jm}(\mbf{\hat p})\, 
\cfzE ,
\label{phi}
\eeq
we arrive at the minimal set $\cfzE$
of nonbirefringent spherical coefficients
for Lorentz violation.
This expansion reveals that
no $B$-type spherical coefficients appear
in the absence of leading-order birefringence.
The index ranges and component counting 
for the coefficients $\cfzE$ 
are summarized in Table \ref{cf_ranges}.

Using the expansion \rf{phi},
the coefficients $(\cf)^{\mu\nu}$
can be expressed in terms of the spherical coefficients $\cfzE$.
We can also find the contributions
to the matrix components 
in the expansions \rf{kf_exp}, \rf{kf_exp2}, and \rf{kf_exp3}
in the absence of birefringence.
Some of these components vanish and others become equal,
yielding
\begin{align}
\BoE & \to 0, 
\notag\\
\BtE & \to 0,
\notag\\
\EzE & \leftrightarrow -\DzE, 
\notag\\
\EoE & \leftrightarrow -\DoE, 
\notag\\
\EtE & \leftrightarrow -\DtE.
\end{align}
These conditions lead to vanishing $\kep$ and $\kom$,
as required by nonbirefringence.
The nonzero contributions are given by
\begin{subequations}
\begin{align}
\ezE &\supseteq \Frac{-(n+2)(n+3)+j(j+1)}3 
\cfzEdnjm{d}{(n+2)jm} \notag \\
&\quad 
+{\scriptstyle (d-3-n)(d-2-n)} \cfzE , 
\db\\[10pt]
\EzE &\supseteq \Frac{-2n(n+2)-j(j+1)}3 
\cfzEdnjm{d}{(n+2)jm} ,  
\db\\[10pt]
\EoE &\supseteq 
{\scriptstyle (n+1)j(j+1) }\cfzEdnjm{d}{(n+2)jm} ,  
\db\\[10pt]
\EtE &\supseteq \Frac{-(j+2)(j+1)j(j-1)}2 
\cfzEdnjm{d}{(n+2)jm} ,  
\db\\[10pt]
\dzE &\supseteq \Frac{-2(n+2)(n+3)+2j(j+1)}3 
\cfzEdnjm{d}{(n+2)jm} ,
\db\\[10pt]
\DzE &\supseteq \Frac{2n(n+2)+j(j+1)}3 
\cfzEdnjm{d}{(n+2)jm} ,  
\db\\[10pt]
\DoE &\supseteq 
-{\scriptstyle (n+1)j(j+1) } \cfzEdnjm{d}{(n+2)jm} ,  
\db\\[10pt]
\DtE &\supseteq \Frac{(j+2)(j+1)j(j-1)}2 
\cfzEdnjm{d}{(n+2)jm} ,  
\db\\[10pt]
\HzE &\supseteq 
{\scriptstyle (d-3-n)(n+1)} \cfzEdnjm{d}{(n+1)jm} ,  
\db\\[10pt]
\HoE &\supseteq 
-{\scriptstyle (d-3-n)j(j+1)} \cfzEdnjm{d}{(n+1)jm} .
\end{align}
\label{cf_coeffs}%
\end{subequations}
The above analysis completely characterizes
the nonbirefringent Lorentz-violating operators 
associated with the quadratic action for electrodynamics.

In the presence of birefringence,
three additional independent sets of $E$-type coefficients appear.
We can therefore seek four mutually independent sets 
of spherical coefficients,
with one given by $\cfzE$ and the remaining three
covering the birefringent portion of coefficient space.
It turns out that the three new sets of coefficients
for birefringence have spin weights zero, one, and two,
and we denote them by $\kfzE$, $\kfoE$, and $\kftE$.
The index ranges and component counting 
for these three sets are summarized in Table \ref{kf_E_ranges}.

Imposing the symmetry conditions \rf{kf_sym2},
a lengthy calculation yields explicit expressions 
in terms of $\kfzE$, $\kfoE$, and $\kftE$
for the twelve sets of codependent $E$-type coefficients
appearing in the expansions 
\rf{kf_exp}, \rf{kf_exp2}, and \rf{kf_exp3}.
We find contributions to the codependent coefficients 
arising from the trace components
of $\kde$ and $\khb$ given by
\begin{subequations}
\begin{align}
\ezE &\supseteq 
\Frac{(n+2)(n+3)-j(j+1)}3 \kfzE
\notag \\
&\quad 
+{\scriptstyle (d-3-n)(d-2-n)} 
\kfzEdnjm{d}{(n-2)jm} 
\notag \\
&\quad 
\hskip -30pt
-\Frac{2(d-3-n)((n+1)(n+2)+(j+2)(j-1))}3 \kfoEdnjm{d}{(n-1)jm}
\notag \\
&\quad 
-\Frac{2(d-3-n)(d-2-n)}3 \kftEdnjm{d}{(n-2)jm} ,  
\db\\[10pt]
\dzE &\supseteq 
\Frac{-2((n+2)(n+3)-j(j+1))}3 \kfzE .  
\end{align}%
\end{subequations}
The contributions to the remaining codependent coefficients 
in $\kde$ are 
\begin{subequations}
\begin{align}
\EzE &\supseteq 
\Frac{2n(n+2)+j(j+1)}3 \kfzE
\notag \\
&\quad 
\hskip -30pt
+\Frac{2(d-3-n)(4(n+1)(n+2)+(j+2)(j-1))}3 
\kfoEdnjm{d}{(n-1)jm} 
\notag \\
&\quad 
+\Frac{2(d-3-n)(d-2-n)}3 \kftEdnjm{d}{(n-2)jm} ,  
\db\\[10pt]
\EoE &\supseteq 
-{\scriptstyle (n+1)j(j+1) } \kfzE
\notag \\
&\quad 
\hskip -33pt
-{\scriptstyle 
2(d-3-n)((n+1)(n+2)+(j+2)(j-1))(n+1)}
\kfoEdnjm{d}{(n-1)jm} 
\notag \\
&\quad 
\hskip -33pt
-{\scriptstyle 2(d-3-n)(d-2-n)(n+1)}
\kftEdnjm{d}{(n-2)jm} ,  
\db\\[10pt]
\EtE &\supseteq 
\Frac{(j+2)(j+1)j(j-1)}2 \kfzE
\notag \\
&\quad 
\hskip -33pt
+{\scriptstyle (d-3-n)(j+2)(j-1)(2(n+2)^2+(j+2)(j-1)) }
\kfoEdnjm{d}{(n-1)jm} \notag \\
&\quad 
\hskip -33pt
+{\scriptstyle (d-3-n)(d-2-n)(2(n+2)^2+(j+2)(j-1)) }
\kftEdnjm{d}{(n-2)jm} ,
\end{align}%
\end{subequations}
while the contributions to the remaining codependent
coefficients in $\khb$ are
\begin{subequations}
\begin{align}
\DzE &\supseteq 
\Frac{2n(n+2)+j(j+1)}3 \kfzE 
\notag \\
&\quad 
+{\scriptstyle 2(n+3)(n+4) } \kftE , 
\db\\[10pt]
\DoE &\supseteq 
-{\scriptstyle (n+1)j(j+1)} \kfzE 
\notag \\
&\quad 
-{\scriptstyle 
2(n+3)^2(n+4) }
\kftE ,  
\db\\[10pt]
\DtE &\supseteq
\Frac{(j+2)(j+1)j(j-1)}2 \kfzE 
\notag \\
&\quad 
\hskip -30pt
+{\scriptstyle (2(n+3)^2-j(j+1))(n+3)(n+4) } \kftE .  
\end{align}%
\end{subequations}
Finally,
the contributions to $\kdb$ are 
\begin{subequations}
\begin{align}
\HzE &\supseteq 
{\scriptstyle (d-3-n)(n+1) } \kfzEdnjm{d}{(n-1)jm} 
\notag \\
&\quad 
-{\scriptstyle (n+2)(n+3)(n+4) } \kfoE ,  
\db\\[10pt]
\HoE &\supseteq 
-{\scriptstyle (d-3-n)j(j+1) } \kfzEdnjm{d}{(n-1)jm} 
\notag \\
&\quad 
+{\scriptstyle (n+2)^2(n+3)(n+4) } \kfoE ,  
\db\\[10pt]
\BoE &\supseteq 
-{\scriptstyle n(n+2)(n+3)(n+4)} \kfoE  
\notag \\
&\quad 
\hskip -30pt
-{\scriptstyle 2(d-3-n)(n+2)(n+3) } \kftEdnjm{d}{(n-1)jm} ,  
\db\\[10pt]
\BtE &\supseteq 
{\scriptstyle (j+2)(j-1)(n+2)(n+3)(n+4) } \kfoE 
\notag \\
&\quad 
\hskip -30pt
+{\scriptstyle 
2(d-3-n)(n+2)(n+3)^2 }
\kftEdnjm{d}{(n-1)jm} .
\end{align}
\label{kf_E_coeffs}%
\end{subequations}
The above twelve equations completely characterize 
the birefringent CPT-even Lorentz-violating operators
associated with the quadratic action for electrodynamics.

To summarize this subsection,
we find that the coefficients 
controlling the CPT-even Lorentz-violating operators
separate into
one set of nonbirefringent $E$-type spherical coefficients,
three sets of birefringent $E$-type spherical coefficients,
and two sets of birefringent $B$-type spherical coefficients.
The component counting for each of these sets
can be found in Tables 
\ref{kf_B_ranges}, \ref{cf_ranges}, and \ref{kf_E_ranges}.
Adding the totals from each table,
we find a net total of $(d+1)d(d-3)$
independent spherical coefficients at each even dimension $d$.
As expected,
this matches the group-theoretic result 
obtained in Sec.\ \ref{sec_theory}.
Some properties of all these coefficients
are summarized in Table \ref{summary_table}
of Sec.\ \ref{Summary and discussion}.

\section{Special models}
\label{sec_models}

For a given observational or experimental procedure, 
particular sensitivity may be achieved
to specific combinations of spherical coefficients.
Also,
the nine basic sets of spherical coefficients
contain a large number of independent components,
which suggests that some types of analyses 
may be challenging to perform in the general context.
It is therefore useful to identify special models 
that are limiting cases
and are relevant to specific measurements.

In this section,
we first present several convenient limiting cases.
Five basic types of model are considered.
One is the minimal SME, 
which restricts attention to renormalizable terms.
Another is the isotropic limit,
in which rotational invariance is preserved
in a preferred frame.
A third is the vacuum model,
which encompasses the subset of effects 
detectable in searches using dispersion or birefringence
from astrophysical sources.
The fourth is the nonbirefringent nondispersive model,
for which the Lorentz-violating photon propagation 
exhibits several parallels to the conventional case.
The fifth is the vacuum-orthogonal model,
which is the complement of the vacuum model
and contains coefficients that can be detected
only in laboratory experiments.
In a final subsection,
we identify particular subsets of the spherical SME coefficients
that correspond to some special cases 
appearing in the literature.

\subsection{Minimal SME}
\label{sec_renorm}

The first limiting case we consider
is the pure-photon sector of the minimal SME,
obtained by restricting attention 
to Lorentz-violating operators 
of renormalizable dimension $d\leq 4$.
Numerous measurements of the corresponding coefficients 
have been performed to date 
\cite{datatables},
with most analyses using cartesian components 
rather than spherical ones. 
Discussions of the minimal SME 
can be found in the literature
\cite{ck,km},
so we restrict the focus of this subsection
to establishing the relationship
between cartesian and spherical coefficients.
Working in a fixed inertial frame
with cartesian coordinates $(t,x,y,z)$,
we determine the linear combinations of spherical coefficients 
that correspond to the cartesian coefficients.

For the Lorentz-violating operators with $d=3$,
which are CPT odd, 
there are four cartesian components of $(\kafd{3})^\ka$.
They are linear combinations 
of the four spherical coefficients
$\kafzBdnjm{3}{000}$ and $\kafoBdnjm{3}{01m}$, 
where $m=0,\pm1$.
In terms of a $4\times 4$ matrix $S$,
we can write the relationship generically as
\beq
{\cal K}_{\mbox{\scriptsize cartesian}} = \frac{1}{\sqrt{4\pi}}
\, S\cdot{\cal K}_{\mbox{\scriptsize spherical}} ,
\label{renorm}
\eeq
where 
${\cal K}_{\mbox{\scriptsize cartesian}}$
and
${\cal K}_{\mbox{\scriptsize spherical}}$  
are four-dimensional column matrices
containing the cartesian and spherical components,
respectively.
Table \ref{renorm-odd} gives the elements of the matrix $S$.
This table also displays the correspondence 
between the four spherical coefficients
$\kafzBdnjm{3}{000}$, $\kafoBdnjm{3}{01m}$
and the four vacuum coefficients $\kVdjm{3}{jm}$,
which are defined in Eq.\ \rf{kV} below
in the context of the vacuum model
discussed in Sec.\ \ref{sec_vac}.

\begin{table}
\renewcommand{\arraystretch}{1.8}
\setlength{\tabcolsep}{0pt}
\def\p#1{\parbox{0.218\columnwidth}{\centering #1}}
\begin{tabular*}{\columnwidth}{c|*{4}{|c}} 
& 
\p{$\kafzBdnjm{3}{000}$}&
\p{$\kafoBdnjm{3}{011}$}&
\p{$\kafoBdnjm{3}{010}$}&
\p{$\kafoBdnjm{3}{01(-1)}$}\\ 
&
\p{$-\kVdjm{3}{00}$}& 
\p{$2\kVdjm{3}{11}$}&
\p{$2\kVdjm{3}{10}$}&
\p{$2\kVdjm{3}{1(-1)}$}\\
\hline\hline
$(\kafd{3})^t\ $ & $1$ & $0$ & $0$ & $0$ \\
$(\kafd{3})^x\ $ & $0$ & $\sqrt{\Frac38}$ 
  & $0$ & $-\sqrt{\Frac38}$ \\
$(\kafd{3})^y\ $ & $0$ & $i\sqrt{\Frac38}$ 
  & $0$ & $i\sqrt{\Frac38}$ \\
$(\kafd{3})^z\ $ & $0$ & $0$ & $-\sqrt{\Frac34}$ & $0$
\end{tabular*}
\caption{\label{renorm-odd}
Matrix elements relating cartesian to spherical coefficients
in the CPT-odd part of the photon sector in the minimal SME. 
Note that an overall factor of $\sqrt{1/4\pi}$
is omitted.}
\end{table}

\begin{table*}
\def\f{\sqrt{\Frac{15}{2}}}
\renewcommand{\arraystretch}{1.8}
\setlength{\tabcolsep}{0pt}
\def\p#1{\parbox{0.0845\textwidth}{\centering #1}}
\begin{tabular*}{\textwidth}{l|*{11}{|c}}
&
\p{$\scriptstyle\cfzEdnjm{4}{200}$}&
\p{$\scriptstyle\cfzEdnjm{4}{222}$}&
\p{$\scriptstyle\cfzEdnjm{4}{221}$}&
\p{$\scriptstyle\cfzEdnjm{4}{220}$}&
\p{$\scriptstyle\cfzEdnjm{4}{22-1}$}&
\p{$\scriptstyle\cfzEdnjm{4}{22-2}$}&
\p{$\scriptstyle\kfzEdnjm{4}{022}$}&
\p{$\scriptstyle\kfzEdnjm{4}{021}$}&
\p{$\scriptstyle\kfzEdnjm{4}{020}$}&
\p{$\scriptstyle\kfzEdnjm{4}{02-1}$}&
\p{$\scriptstyle\kfzEdnjm{4}{02-2}$}\\
&
\p{$\scriptstyle\fr14\kIdjm{4}{00}$}&
\p{$\scriptstyle\kIdjm{4}{22}$}&
\p{$\scriptstyle\kIdjm{4}{21}$}&
\p{$\scriptstyle\kIdjm{4}{20}$}&
\p{$\scriptstyle\kIdjm{4}{2-1}$}&
\p{$\scriptstyle\kIdjm{4}{2-2}$}&
\p{$$\scriptstyle\fr{-\kEdjm{4}{22}}{\sqrt{6}}$$}&
\p{$$\scriptstyle\fr{-\kEdjm{4}{21}}{\sqrt{6}}$$}&
\p{$$\scriptstyle\fr{-\kEdjm{4}{20}}{\sqrt{6}}$$}&
\p{$$\scriptstyle\fr{-\kEdjm{4}{2-1}}{\sqrt{6}}$$}&
\p{$$\scriptstyle\fr{-\kEdjm{4}{2-2}}{\sqrt{6}}$$}\\
\hline\hline
$(\kdbd{4})^{xx}\,$&$4$&$-\f$&$0$&$\sqrt5$
  &$0$&$-\f$&$\f$&$0$&$-\sqrt5$&$0$&$\f$\\
$(\kded{4})^{xy}\,$&$0$&$-i\f$&$0$&$0$&$0$
  &$i\f$&$i\f$&$0$&$0$&$0$&$-i\f$\\
$(\kded{4})^{xz}\,$&$0$&$0$&$\f$&$0$&$-\f$
  &$0$&$0$&$-\f$&$0$&$\f$&$0$\\
$(\kded{4})^{yy}\,$&$4$&$\f$&$0$&$\sqrt5$
  &$0$&$\f$&$-\f$&$0$&$-\sqrt5$&$0$&$-\f$\\
$(\kded{4})^{yz}\,$&$0$&$0$&$i\f$&$0$&$i\f$
  &$0$&$0$&$-i\f$&$0$&$-i\f$&$0$\\
$(\kded{4})^{zz}\,$&$4$&$0$&$0$&$-2\sqrt5$
  &$0$&$0$&$0$&$0$&$2\sqrt5$&$0$&$0$\\
$(\khbd{4})^{xx}\,$&$-4$&$\f$&$0$&$-\sqrt5$
  &$0$&$\f$&$\f$&$0$&$-\sqrt5$&$0$&$\f$\\
$(\khbd{4})^{xy}\,$&$0$&$i\f$&$0$&$0$&$0$
  &$-i\f$&$i\f$&$0$&$0$&$0$&$-i\f$\\
$(\khbd{4})^{xz}\,$&$0$&$0$&$-\f$&$0$&$\f$
  &$0$&$0$&$-\f$&$0$&$\f$&$0$\\
$(\khbd{4})^{yy}\,$&$-4$&$-\f$&$0$&$-\sqrt5$
  &$0$&$-\f$&$-\f$&$0$&$-\sqrt5$&$0$&$-\f$\\
$(\khbd{4})^{yz}\,$&$0$&$0$&$-i\f$&$0$&$-i\f$
  &$0$&$0$&$-i\f$&$0$&$-i\f$&$0$\\
$(\khbd{4})^{zz}\,$&$-4$&$0$&$0$&$2\sqrt5$
  &$0$&$0$&$0$&$0$&$2\sqrt5$&$0$&$0$ 
\end{tabular*}
\caption{\label{renorm-even-even}
Matrix elements relating cartesian to spherical coefficients
in the CPT-even and parity-even part 
of the photon sector in the minimal SME. 
We have assumed $\cfzEdnjm{4}{000} = 3\cfzEdnjm{4}{200}$.
Note that an overall factor of $\sqrt{1/4\pi}$
is omitted.}
\end{table*}

\begin{table*}
\renewcommand{\arraystretch}{1.8}
\def\f{\sqrt{\Frac{15}{8}}}
\def\F{\sqrt{\Frac32}}
\setlength{\tabcolsep}{0pt}
\def\p#1{\parbox{0.115\textwidth}{\centering #1}}
\begin{tabular*}{\textwidth}{c|*{8}{|c}}
&
\p{$\scriptstyle\cfzEdnjm{4}{111}$}&
\p{$\scriptstyle\cfzEdnjm{4}{110}$}&
\p{$\scriptstyle\cfzEdnjm{4}{11-1}$}&
\p{$\scriptstyle\kfoBdnjm{4}{222}$}&
\p{$\scriptstyle\kfoBdnjm{4}{221}$}&
\p{$\scriptstyle\kfoBdnjm{4}{220}$}&
\p{$\scriptstyle\kfoBdnjm{4}{22-1}$}&
\p{$\scriptstyle\kfoBdnjm{4}{22-2}$}\\[2pt]
&
\p{$\scriptstyle-2\kIdjm{4}{11}$}&
\p{$\scriptstyle-2\kIdjm{4}{10}$}&
\p{$\scriptstyle-2\kIdjm{4}{1-1}$}&
\p{$\scriptstyle-\sqrt{\fr23}\, \kBdjm{4}{22}$}&
\p{$\scriptstyle-\sqrt{\fr23}\, \kBdjm{4}{21}$}&
\p{$\scriptstyle-\sqrt{\fr23}\, \kBdjm{4}{20}$}&
\p{$\scriptstyle-\sqrt{\fr23}\, \kBdjm{4}{2-1}$}&
\p{$\scriptstyle-\sqrt{\fr23}\, \kBdjm{4}{2-2}$}\\[4pt]
\hline
\hline
$(\kdbd{4})^{xx}\ $ & $0$ & $0$ & $0$ 
  & $\f$ & $0$ & $-\Frac{\sqrt5}{2}$ & $0$ & $\f$ \\
$(\kdbd{4})^{xy}\ $ & $0$ & $\sqrt3$ 
  & $0$ & $i\f$ & $0$ & $0$ & $0$ & $-i\f$ \\
$(\kdbd{4})^{xz}\ $ & $i\F$ & $0$ 
  & $i\F$ & $0$ & $-\f$ & $0$ & $\f$ & $0$ \\
$(\kdbd{4})^{yx}\ $ & $0$ & $-\sqrt3$ 
  & $0$ & $i\f$ & $0$ & $0$ & $0$ & $-i\f$ \\
$(\kdbd{4})^{yy}\ $ & $0$ & $0$ & $0$ 
  & $-\f$ & $0$ & $-\Frac{\sqrt5}{2}$ & $0$ & $-\f$ \\
$(\kdbd{4})^{yz}\ $ & $-\F$ & $0$ & $\F$ 
  & $0$ & $-i\f$ & $0$ & $-i\f$ & $0$ \\
$(\kdbd{4})^{zx}\ $ & $-i\F$ & $0$ 
  & $-i\F$ & $0$ & $-\f$ & $0$ & $\f$ & $0$ \\
$(\kdbd{4})^{zy}\ $ & $\F$ & $0$ & $-\F$ 
  & $0$ & $-i\f$ & $0$ & $-i\f$ & $0$ 
\end{tabular*}
\caption{\label{renorm-even-odd}
Matrix elements relating cartesian to spherical coefficients
in the CPT-even and parity-odd part 
of the photon sector in the minimal SME. 
Note that an overall factor of $\sqrt{1/4\pi}$
is omitted.}
\end{table*}

As an example from Table \ref{renorm-odd},
consider the cartesian coefficient
$(\kaf)^x$.
In terms of spherical coefficients,
the table gives
\begin{align}
(\kaf)^x & = 
\frac{1}{\sqrt{4\pi}}\big(
\sqrt{\Frac38}\, \kafoBdnjm{3}{011} -
\sqrt{\Frac38}\, \kafoBdnjm{3}{01(-1)}\big)
\notag\\
& =
\frac{1}{\sqrt{4\pi}}\big(
\sqrt{\Frac32}\, \kVdjm{3}{11}
- \sqrt{\Frac32}\, \kVdjm{3}{1(-1)} \big).
\end{align}
Note that the spherical
coefficients are complex in general,
but the property 
$\kafzBdnjm{3}{011}= -[\kafzBdnjm{3}{01(-1)}]^*$
contained in the phase condition \rf{ccrel}
ensures a total of four real degrees of freedom.
Indeed,
the components $(\kaf)^x$ and $(\kaf)^y$ correspond,
respectively,
to the real and imaginary parts 
of the spherical coefficient $\kafzBdnjm{3}{011}$. 

For the renormalizable CPT-even Lorentz-violating operators,
all of which have $d=4$,
the operator $\kf$ contains 20 real independent constants.
From Tables 
\ref{kf_B_ranges}, \ref{cf_ranges}, and \ref{kf_E_ranges}, 
we see that these include two scalar singlets
$\cfzEdnjm{4}{000}$ and $\cfzEdnjm{4}{200}$,
one triplet $\cfzEdnjm{4}{11m}$,
and three quintuplets
$\cfzEdnjm{4}{22m}$, $\kfzEdnjm{4}{22m}$, $\kfoBdnjm{4}{22m}$.
In terms of the $\hat\kappa$ matrices of Eq.\ \rf{kappas},
the two singlets can be viewed as corresponding
to the two trace components,
while the triplet corresponds to the
antisymmetric part of $\kdb$.
The three quintuplets match to the traceless parts 
of $\kde$ and $\khb$ and the symmetric part of $\kdb$.

We can again write the relationship 
between cartesian and spherical coefficients 
using the generic form \rf{renorm}.
The elements of the matrix $S$
connecting the parity-even coefficients $\kde$ and $\khb$
and the spherical coefficients are given in 
Table \ref{renorm-even-even}.
The elements of $S$ for the parity-odd coefficients $\kdb$
are given in Table \ref{renorm-even-odd}.
These tables also relate the 20 spherical coefficients with $d=4$ 
to the relevant 20 vacuum coefficients 
$\kIdjm{4}{jm}$, $\kEdjm{4}{jm}$, $\kBdjm{4}{jm}$, 
which are defined 
in Eqs.\ \rf{kI}, \rf{kE}, and \rf{kB}
in the context of the vacuum model
presented in Sec.\ \ref{sec_vac}.

In the photon sector of the minimal SME,
the nonbirefringent operators 
are controlled by the ten spherical coefficients 
$\cfzEdnjm{4}{njm}$,
as shown in Sec.\ \ref{sec_even_coeffs}.
Nine combinations of these have been measured 
in laboratory-based experiments.
The remaining coefficient combination,
\beq
\ktp = \Frac{1}{\sqrt{4\pi}}
\big[\cfzEdnjm{4}{000} - 3\cfzEdnjm{4}{200}\big],
\eeq
is Lorentz invariant
and leads to a rescaling of the electric and magnetic fields.
This combination is therefore typically 
taken to vanish, 
which implies the condition
\beq
\cfzEdnjm{4}{000} = 3\cfzEdnjm{4}{200}.
\eeq
This property has been assumed in constructing
the entries for Table \ref{renorm-even-even}.

Using the matrix elements 
in Tables \ref{renorm-odd}, \ref{renorm-even-even},
and \ref{renorm-even-odd},
it is comparatively straightforward
to convert between cartesian and spherical representations
of the photon-sector coefficients in the minimal SME.
Note,
however,
that the spherical coefficients 
represent angular-momentum eigenstates
and therefore have simpler rotational properties.
The behavior of the spherical coefficients 
under rotations is discussed in Sec.\ \ref{sec_rotations}.

\subsection{Isotropic models}
\label{sec_fc}

In any given inertial frame,
a small subset of Lorentz-violating operators
preserve rotational invariance.
Restricting attention to these operators
defines an interesting limiting case
of the general theory.
In these models,
sometimes called `fried-chicken' models
due to their popularity and simplicity,
the isotropic inertial frame must be specified
because observer boosts to other frames 
destroy the rotational invariance.
One natural choice for the preferred frame 
is the frame of the cosmic microwave background (CMB),
but other choices are possible.
Note,
however,
that isotropy in the CMB frame
implies anisotropy in a Sun-centered frame
and in laboratory experiments.

Within our analysis,
imposing rotational invariance is immediate.
The general isotropic model
is obtained by imposing the condition 
that all spherical coefficients 
vanish in the preferred frame, 
except those with $j=0$.
This condition drastically reduces 
the number of available coefficients.
The only nonzero coefficients are:
\begin{align}
\cffc &= \cfzEdnjm{d}{n00} , 
\notag \\
\kffc &= \kfzEdnjm{d}{n00} , 
\notag \\
\kaffc &= \kafzBdnjm{d}{n00} .
\label{fccoeffdef}
\end{align}
Following standard convention
\cite{nuexpt},
these isotropic coefficients are identified
by a ring diacritic.
Note that the coefficient $\cffc$ controls 
isotropic nonbirefringent Lorentz-violating operators
in the preferred frame,
while the others control leading-order birefringent effects.

In the general isotropic model,
the nonzero components of $\kaf$ are
\begin{subequations}
\begin{align}
(\kaf)_0 &= \sum_{dn} \frac{\om^{d-3-n}p^n}{\sqrt{4\pi}}\
\kaffc , 
\label{kaf0fc} \db\\[10pt]
(\kaf)_r &= 
- \sum_{dn} \frac{\om^{d-3-n}p^n}{\sqrt{4\pi}}\
\frac{(d-2-n)}{n+2} \kaffcdn{d}{(n-1)} .
\label{kafrfc} 
\end{align}%
\end{subequations}
For the $\hat \ka$ matrices,
we find the expressions 
\begin{subequations}
\begin{align}
(\kde)_{rr} &= 
\sum_{dn} \frac{\om^{d-4-n}p^n}{\sqrt{4\pi}}\ \big[ -(n+1)(n+2) 
\notag \\    
&\qquad\quad \times
\big( \cffcdn{d}{(n+2)} - \kffc \big)
\notag \\
&\quad\quad
+(d-3-n)(d-2-n)
\notag \\    
&\qquad\quad \times
\big( \cffc + \kffcdn{d}{(n-2)} \big) \big]
\ , \label{kde1fc} \db\\[10pt]
(\kde)_{+-} &= 
\sum_{dn} \frac{\om^{d-4-n}p^n}{\sqrt{4\pi}}\ \big[ (-1)(n+2)
\notag \\    
&\qquad\quad \times
\big( \cffcdn{d}{(n+2)} - \kffc \big)
\notag \\    
&\quad\quad
+(d-3-n)(d-2-n)
\notag \\    
&\qquad\quad \times
\big( \cffc + \kffcdn{d}{(n-2)} \big) \big] , 
\label{kde2fc} \db\\[10pt]
(\khb)_{rr} &= 
\sum_{dn} \frac{\om^{d-4-n}p^n}{\sqrt{4\pi}}\ (-2)(n+2) 
\notag \\
&\qquad\quad \times
\big( \cffcdn{d}{(n+2)} + \kffc \big) , 
\label{khb1fc} \db\\[10pt]
(\khb)_{+-} &= 
\sum_{dn} \frac{\om^{d-4-n}p^n}{\sqrt{4\pi}}\ (-1)(n+2)^2 
\notag \\
&\qquad\quad \times
\big( \cffcdn{d}{(n+2)} + \kffc \big) , 
\label{khb2fc} \db\\[10pt]
(\kdb)_{\pm\mp} &= 
\sum_{dn} \frac{\om^{d-4-n}p^n}{\sqrt{4\pi}}\ (\mp i)(d-3-n)(n+1)  
\notag \\
&\qquad\quad \times
\big( \cffcdn{d}{(n+1)} +\kffcdn{d}{(n-1)} \big) . 
\label{kdb2fc}
\end{align}
\label{fc_coeffs}%
\end{subequations}

The index ranges and component counting 
for the isotropic coefficients
can be determined from the $j=0$ columns 
of Tables \ref{kaf_ranges}, \ref{kf_B_ranges},
\ref{cf_ranges}, and \ref{kf_E_ranges}.
For a given even dimension $d$,
there are $d/2$ coefficients $\cffc$
and $(d-4)/2$ coefficients $\kffc$,
for a total of $(d-2)$ independent components.
For odd $d$,
there are $(d-1)/2$ coefficients $\kaffc$.
A summary of the properties of these coefficients
is provided as part of 
Table \ref{limiting_table}
in Sec.\ \ref{Summary and discussion}.

\subsection{Vacuum models}
\label{sec_vac}

For observations of light from distant sources,
far-field solutions apply.
The electromagnetic fields are well approximated 
by vacuum plane waves with $\om=p$.
Although Lorentz-violating operators 
typically generate a nontrivial dispersion relation
with $\om\neq p$,
we can impose $\om=p$ in the operators $\kf$ and $\kaf$
to obtain leading-order results.
The resulting vacuum coefficients for Lorentz violation
comprise the set of linear combinations
of spherical coefficients
relevant for photon propagation in the vacuum.
They are of direct interest
for studies of light from astrophysical sources,
and they can also be important in Earth-based tests.  

Imposing $\om=p$ in the expansions \rf{kafs} and \rf{kfs}
for $\kaf$ and $\kf$
dramatically simplifies the spherical-harmonic structure.
It turns out that the relevant combinations of coefficients 
are those associated with the Stokes parameters
of the eigenmodes for vacuum photon propagation,
because these eigenmodes characterize
the effects of Lorentz violation
on the overall photon polarization
\cite{km}.
To find the eigenmodes,
we first need plane-wave solutions
of the modified Maxwell equations \rf{eqnmot}.

At leading order,
the covariant dispersion relation Eq.\ \rf{dr}
becomes
\begin{align}
&( p^\mu p_\mu - (\cf)^{\mu\nu}p_\mu p_\nu)^2
-2(\hat \ch_w)^{\al\be\ga\de}(\hat \ch_w)_{\al\mu\ga\nu} 
p_\be p_\de p^\mu p^\nu
\notag\\
&\quad
-4 (p^\mu (\kaf)_\mu )^2 \simeq 0 .
\label{dr2}
\end{align}
The two solutions at leading order can be written as
\beq
p^0 \simeq \big(1-\vs^0 
\pm \sqrt{(\vs^1)^2+(\vs^2)^2+(\vs^3)^2}\, \big) p ,
\label{dr3}
\eeq
where the three rotationally invariant dimensionless combinations
\begin{align}
\vs^0   &= 
\half (\cf)^{\mu\nu}p_\mu p_\nu / \om^2 ,
\notag \\
(\vs^1)^2 + (\vs^2)^2 &= 
\half (\hat \ch_w)^{\al\be\ga\de}(\hat \ch_w)_{\al\mu\ga\nu}
p_\be p_\de p^\mu p^\nu / \om^4 , 
\notag \\
\vs^3 &= - p^\mu (\kaf)_\mu / \om^2 
\label{stokes_params}
\end{align}
contain coefficients for 
nonbirefringent CPT-even,
birefringent CPT-even, 
and birefringent CPT-odd effects,
respectively.
The three combinations
$\vs^1$, $\vs^2$, $\vs^3$
turn out to determine the Stokes parameters of the two eigenmodes.
The momentum structure of Eq.\ \rf{stokes_params}
reveals that the vacuum coefficients for odd $d$
are simply obtained as the totally symmetric and traceless part 
of $(\kafd{d})^{\ka\al_1\ldots\al_{(d-3)}}$.
For even $d$,
a similar result holds for the vacuum coefficients 
$(\cf)^{\mu\nu}$
controlling nonbirefringent effects.

The procedure for finding the Stokes parameters
$(s^1,s^2,s^3) = (Q,U,V)$
associated with the two solutions in Eq.\ \rf{dr3}
follows the same basic steps as in Ref.\ \cite{km}.
Using the plane-wave equation \rf{meq},
we find a solution with unique polarization
for each of the signs in Eq.\ \rf{dr3}.
It is advantageous to work in the temporal gauge,
for which $A_0 =0$ and $\mbf E \propto \mbf A$.
We can then construct the specific Stokes parameters 
for each eigenmode.
Adopting the orthonormal spherical coordinate system 
described in Appendix \ref{sec_am},
we define a general Stokes vector $\mbf s$ 
and its corresponding three Stokes parameters by
\beq
\mbf s = 
\left(\begin{array}{c} s^1 \\ s^2\\ s^3
\end{array}\right)
=
\left(\begin{array}{c}
|E^\th|^2-|E^\ph|^2  \\ 
2\mbox{Re} E^{\th *}E^\ph \\ 
2\mbox{Im} E^{\th *}E^\ph
\end{array}
\right) .
\label{stokes}
\eeq
This Stokes vector completely characterizes
the polarization of an electromagnetic wave.
It is also useful to define a fourth Stokes parameter:
\beq 
s^0 = |E^\th|^2+|E^\ph|^2 ,
\eeq
which corresponds to the intensity $I$.

Due to their orthogonality,
the specific Stokes vectors for the two plane-wave solutions
differ by a sign.
It therefore suffices to construct 
a single specific Stokes vector $\mbf\vs$,
which can be associated 
with the faster of the two birefringent modes.
In spherical coordinates,
this vector is given by
\beq
\mbf\vs = \left(\begin{array}{c} 
\vs^1\\ \vs^2\\ \vs^3
\end{array}\right)
\simeq \left(\begin{array}{c} 
-\half\big((\kep)^{\th\th}-(\kep)^{\ph\ph}\big)-(\kom)^{\th\ph} 
\\ 
-(\kep)^{\th\ph} + \half\big((\kom)^{\th\th}-(\kom)^{\ph\ph}\big)
\\
-\big((\kaf)^0-(\kaf)^r\big)/\om
\end{array}\right) ,
\label{stokes_rot_axis}
\eeq
where we have adopted a convenient normalization.
At leading order,
the components of $\mbf \vs$ are consistent 
with the combinations \rf{stokes_params}.
In spherical coordinates,
the fourth combination is
\begin{align}
\vs^0 &\simeq -\half(\kem)^{rr} + (\ktm) + (\kop)^{\th\ph} 
\notag \\
&= \half(\kde - \khb + i\kdb + i\khe)_{+-} 
\label{rho}
\end{align}
at leading order.
This combination is analogous to the
scalar Stokes parameter $s^0$.

The combinations $\vs^0$ and $\vs^3$ 
are invariant under rotations about 
the direction $\mbf{\hat p}$ of the photon 3-momentum,
like the general Stokes parameters.
They therefore have zero spin weight.
The remaining combinations $\vs^1$ and $\vs^2$
transform under rotations as a rank-2 traceless tensor
in the tangent space of the sphere 
and so are combinations of components
with spin weight $\pm2$.
As a result,
the expansion in spin-weighted spherical harmonics
of the Stokes parameters
for Lorentz-violating photon propagation 
takes the form
\cite{km_cmb,km_apjl}
\begin{align}
\vs^0 &= 
\sum_{djm} \om^{d-4} (-1)^j\, \syjm{0}{jm}(\mbf{\hat p})\, \kI ,
\notag \\
\vs^1\pm i \vs^2 &= 
\sum_{djm} \om^{d-4} (-1)^j\, \syjm{\pm2}{jm}(\mbf{\hat p})
\big(\kE\mp i\kB\big) ,
\notag \\
\vs^3 &= 
\sum_{djm} \om^{d-4} (-1)^j\, \syjm{0}{jm}(\mbf{\hat p})\, \kV .
\label{vac_exp}
\end{align}
This result is independent 
of the index $n$ because 
at leading order we can take $\om = p$
in Eq.\ \rf{stokes_params}.
We remark that the notation $\kI$ 
adopted here for one of the sets of vacuum coefficients 
differs from that of Ref.\ \cite{km_apjl},
where the quantity 
\beq
\koldI \equiv \kI
\eeq
is used instead.
The improved present notation $\kI$ reflects
the absence of birefringence 
from the corresponding Lorentz-violating operators.
Also,
the factors of $(-1)^j$ 
in Eq.\ \rf{vac_exp}
have been introduced
for convenience and to match the definitions
in Ref.\ \cite{km_apjl}.
The point is that vacuum models
are well suited for studies involving 
radiation from astrophysical sources,
for which the source direction $\mbf{\hat n}$
is normally specified rather than the 
propagation direction $\mbf{\hat p} = -\mbf{\hat n}$.
For these studies,
it is therefore more natural to work with spin weight 
defined with respect to $\mbf{\hat n}$ 
instead of $\mbf{\hat p}$.
The correspondence
\beq
(-1)^j \syjm{s}{jm}(\mbf{\hat p}) 
= \syjm{-s}{jm}(\mbf{\hat n}),
\eeq
which follows from Eq.\ \rf{parity},
can be used to transform between the two pictures.

The net result of the above discussion
is that the vacuum behavior is controlled
by four sets of vacuum coefficients
$\kI$, $\kE$, $\kB$, and $\kV$.
These are related to 
the general coefficients for Lorentz violation
by taking the limit $\om=p$ in the various expansions 
provided in Sec.\ \ref{sec_gen_coeffs}.
Via this limit, 
we find the results 
\begin{subequations}
\begin{align}
\kI&=\sum_n \half (-1)^j \big(
-\half\EzE + \half\DzE 
\notag \\
&\hspace{-10pt}
+ \ezE - \dzE + 2\HzE \big) ,
\label{kI}
\db\\[10pt]
\kE&=\sum_n \half (-1)^{j+1} \sqrt\Frac{(j-2)!}{(j+2)!}
\notag \\
&\hspace{-10pt}
\times \big( \EtE + \DtE + 2\BtE  \big) ,
\label{kE}
\db\\[10pt]
\kB&=\sum_n \half (-1)^j \sqrt\Frac{(j-2)!}{(j+2)!}
\notag \\
&\hspace{-10pt}
\times \big( \EtB + \DtB -2\BtB  \big) ,
\label{kB}
\db\\[10pt]      
\kV&=\sum_n (-1)^{j+1} \big( 
\Frac{d}{n+3} \kafzB
+\Frac{1}{n+2} \kafoB \big) .
\label{kV}
\end{align}
\label{vac_coeffs}%
\end{subequations}
As expected,
no frequency or wavelength dependence
appears in these expressions.
These equations reveal that the vacuum coefficients 
are linear combinations 
of the general coefficients
involving identical $d$, $j$, $m$ indices
but different $n$ values.

The index ranges for the vacuum coefficients
$\kI$, $\kE$, $\kB$, and $\kV$
can be found using general symmetry arguments
or the relations \rf{vac_coeffs}.
For even $d$,
there are 
$(d-1)^2$ coefficients $\kI$,
$(d-1)^2-4$ coefficients $\kE$,
and $(d-1)^2-4$ coefficients $\kB$,
giving a total of 
$3(d-1)^2 - 8$
independent components
associated with CPT-even Lorentz violation
in vacuum propagation. 
For odd $d$,
the coefficients $\kV$ for CPT-odd Lorentz violation
have $(d-1)^2$ independent components.
Some properties of these coefficients
are summarized as part of 
Table \ref{limiting_table}
in Sec.\ \ref{Summary and discussion}.

The phenomenological effects controlled
by each set of vacuum coefficients are different.
The coefficients $\kE$ and $\kB$ 
are associated with CPT-even operators
that lead to birefringence,
with the propagating modes being linearly polarized.
The coefficients $\kV$ control CPT-odd birefringence,
and the corresponding eigenmodes are circularly polarized.
In contrast,
the CPT-even operators associated with the coefficients $\kI$
are nonbirefringent.
These physical differences suggest 
it may be of value to introduce special vacuum models 
as particular limiting cases. 
For example,
a nonbirefringent vacuum model
involving only nonzero coefficients $\kI$
can be countenanced.
This model has Lorentz-violating operators
only in even dimensions $d = 4,6,8, \ldots$,
with the number of independent coefficients
being 9, 25, 49 $\ldots$, 
respectively.
Another model of interest 
can be obtained as a hybrid of
the general vacuum model and the isotropic limit,
by introducing a preferred frame
in which attention is restricted only to $j=0$ coefficients.
This vacuum isotropic model 
involves only the coefficients
$\kIdjm{d}{00}$ and $\kVdjm{d}{00}$.
It therefore contains exactly one Lorentz-violating operator
for each value of $d$.

\subsection{Camouflage models}
\label{sec_nonbire}

The vacuum models considered in the previous subsection
are well suited for astrophysical observations.
In this and the following subsections,
we consider instead the subset of spherical coefficients 
that are challenging to detect via studies
of astrophysical birefringence or dispersion 
but that produce observable effects in laboratory experiments.
We begin here
by focusing on Lorentz-violating models 
without leading-order birefringent or vacuum-dispersive effects,
which we call camouflage models. 
The next subsection generalizes this treatment 
to identify explicitly all vacuum-orthogonal coefficients,
including ones that control non-vacuum birefringence.
The key idea motivating these constructions
is that studies of astrophysical birefringence yield 
among the best sensitivities to Lorentz violation 
in any sector,
due to the accumulation of polarization changes 
that occur over the cosmological propagation times.
This also applies to studies of astrophysical dispersion,
albeit typically at lesser sensitivity.
As a result,
searches for Lorentz violation can naturally be split
into ones sensitive to operators
generating vacuum birefringence or vacuum dispersion
and ones with sensitivity to other operators.

We begin by restricting attention to operators
without leading-order birefringence,
which are associated with the spherical coefficients $\cfzE$.
Some key results for this case
are discussed in Sec.\ \ref{sec_even_coeffs} above.
Following Eq.\ \rf{nonbirefcoeff},
the Lagrange density reduces to
\beq
\cl  =  -\Frac 1 4 F_{\mu\nu}F^{\mu\nu}
-\Frac 1 2  F_{\mu\rh}(\cf)^{\mu\nu}{F_\nu}^\rh .
\eeq
Note that CPT invariance holds because $\kaf = 0$,
and that only even-dimensional operators 
with $E$-type parity contribute.

Using Eqs.\ \rf{kf_exp} and \rf{cf_coeffs},
we can write the $\hat\ka$ matrices
in terms of the coefficients $\cfzE$ 
by taking the equality in \rf{cf_coeffs}.
However,
it is often more convenient to use
the explicit expansion \rf{cnonbiref} and \rf{phi}
via the generating function $\hat\Ph$.
The results are summarized as
\begin{subequations}
\begin{align}
& \half \big(3\ktm+\ktp\big) = 
\sum_{dnjm} \om^{d-4-n} p^n\, \syjm{0}{jm}(\mbf{\hat p}) 
\notag \\
&\qquad\qquad\quad   
\times(d-2-n)(d-3-n) \ \cfzE , 
\db\\[10pt]
& \Frac32 \big(\ktm-\ktp\big) = 
\sum_{dnjm} \om^{d-4-n} p^n\, \syjm{0}{jm}(\mbf{\hat p}) 
\notag \\
&\qquad 
\times\big((n+2)(n+3)-j(j+1)\big) \ \cfzEdnjm{d}{(n+2)jm} , 
\db\\[10pt]
&(\kem)_{rr} = -2(\kep)_{+-} = 
\sum_{dnjm} \om^{d-4-n} p^n\, \syjm{0}{jm}(\mbf{\hat p}) 
\notag \\
&\qquad\qquad\qquad   
\times\Frac{-(2n(n+2)+j(j+1))}3 \ \cfzEdnjm{d}{(n+2)jm} , 
\db\\[10pt]
&(\kem)_{\pm r} = (\kem)_{r\pm} = 
\sum_{dnjm} \om^{d-4-n} p^n\, \syjm{\pm 1}{jm}(\mbf{\hat p}) 
\notag \\
&\qquad\qquad  
\times(\pm 1)(n+1)\sqrt{\Frac{j(j+1)}2} \ \cfzEdnjm{d}{(n+2)jm} , 
\db\\[10pt]
&(\kem)_{\pm\pm} = 
\sum_{dnjm} \om^{d-4-n} p^n\, \syjm{\pm 2}{jm}(\mbf{\hat p}) 
\notag \\
&\qquad\qquad\qquad\qquad   
\times\Frac{-1}2\sqrt{\Frac{(j+2)!}{(j-1)!}}\ \cfzEdnjm{d}{(n+2)jm} ,
\db\\[10pt]
&(\kop)_{+-} =  -(\kop)_{-+} =
\sum_{dnjm} \om^{d-4-n} p^n\, \syjm{0}{jm}(\mbf{\hat p}) 
\notag \\
&\qquad 
\times (-i)(d-3-n)(n+1) \ \cfzEdnjm{d}{(n+1)jm} , 
\db\\[10pt]
&(\kop)_{\pm r} = -(\kop)_{r \pm} = 
\sum_{dnjm} \om^{d-4-n} p^n\, \syjm{\pm 1}{jm}(\mbf{\hat p}) 
\notag \\
&\qquad 
\times (-i)(d-3-n)\sqrt{\Frac{j(j+1)}2} \ \cfzEdnjm{d}{(n+1)jm} .
\end{align}%
\end{subequations}
Recall that the matrix $\kem$ is traceless and symmetric,
while $\kop$ is antisymmetric.
For no leading-order birefringence,
the relations
\begin{align}
\kde &= \kem + \ktm +\ktp , 
\notag \\
\khb &= -\kem -\ktm +\ktp , 
\notag \\
\kdb &= \kop 
\end{align}
also hold.

The next step is to consider dispersive effects.
For dimensions $d>4$, 
only a subset of independent combinations 
of the coefficients $\cfzE$ are associated 
with leading-order vacuum dispersion.
The nondispersive operators
are precisely the ones of interest
that define the camouflage models. 
At leading order,
the condition that ensures no vacuum dispersion is
\beq
0=p_\mu p_\nu(\cf)^{\mu\nu}|_{p^2=0}.
\eeq
This is satisfied if the generating
function $\hat \Phi_F$ in Eq.\ \rf{phi} is of the form
$\hat \Phi_F=p^2 \tilde\Phi_F$.
We can therefore define the camouflage coefficients $\cftzE$
through the expansion
\beq
\tilde\Phi_F=\sum_{dnjm} \om^{d-4-n}p^n\,
\syjm{0}{jm}(\mbf{\hat p})\, \cftzE.
\eeq
This result leads to the comparatively simple relation
\beq
\cfzE =  \cftzE - \cftzEdnjm{d}{(n-2)jm} 
\label{ct}
\eeq
in the limit of no leading-order birefringence 
or vacuum dispersion.

The index ranges for the camouflage coefficients ${\cftzE}$
are given in Table \ref{cft_ranges}.
For each even dimension $d$,
there are $(d-1)(d-2)(d-3)/6$ independent components.
They represent combinations of the  
$(d+1)d(d-1)/6$ coefficients $\cfzE$ 
that are complementary to the $(d-1)^2$ vacuum coefficients $\kI$
introduced in Sec.\ \ref{sec_vac}.
Note that a subset of the camouflage operators 
are rotation invariant,
so a hybrid camouflage isotropic model exists
that has $(d-2)/2$ independent coefficients for each even $d$.
 
The camouflage coefficients $\cftzE$
are challenging to detect via astrophysical observations
of birefringence or dispersion
because their effects arise only at higher order.
Along with the minimal-SME coefficients $\cfzEdnjm{4}{njm}$,
they are best sought via alternative methods 
such as laboratory experiments.
A class of experiments sensitive
to the effects of these coefficients
is discussed in Sec.\ \ref{sec_cavities}.
 
\begin{table}
\tabcolsep 3pt
\renewcommand{\arraystretch}{0.5}
\begin{tabular}{c||*{7}{c}}
\multicolumn{1}{c||}{}&
\multicolumn{7}{c}{$\cftzE$}\\[4pt]
\hline
\hline
\multicolumn{1}{c||}{$n$}&
\multicolumn{7}{c}{$j$}\\[4pt]
\hline
0  &0&&&&&&\\
1  &&1&&&&&\\
2  &0&&2&&&&\\
3  &&1&&3&&&\\
4  &0&&2&&4&&\\
$\vdots$  &$\vdots$&&&&&$\ddots$&\\
$d-4$  &0&&2&&4&$\cdots$&$d-4$\\[4pt]
\hline
\hline
\multicolumn{1}{c||}{total}&
\multicolumn{7}{c}{$\Frac16 (d-1)(d-2)(d-3)$}
\end{tabular}
\caption{\label{cft_ranges}
Summary of the allowed ranges of indices $n$ and $j$
for the independent camouflage coefficients.
The dimension $d$ is even with $d\geq4$,
while $n\leq d-4$.
The index $m$ satisfies the usual restrictions
$-j\leq m\leq j$,
so there are $2j+1$ coefficients for each $j$.
For a given dimension $d$,
the total number of coefficients 
is given in the last row.}
\end{table}

\subsection{Vacuum-orthogonal models}
\label{vacorthog}

In Sec.\ \ref{sec_vac},
we identified the subset of spherical coefficients
relevant for photon propagation in the vacuum.
These are the vacuum coefficients 
$\kI$, $\kE$, $\kB$, and $\kV$.
The number of independent vacuum coefficients
grows as $d^2$ for large $d$,
whereas the total number of coefficients grows as $d^3$.
Consequently,
the vacuum coefficients comprise a comparatively small portion 
of the total coefficient space.
In this subsection,
we construct a complete set of independent coefficients 
spanning the complementary part of the coefficient space.
We refer to these as the vacuum-orthogonal coefficients.
At leading order,
the corresponding vacuum-orthogonal operators 
induce neither dispersion nor birefringence
for photon propagation in the vacuum.
Instead,
these Lorentz-violating effects
can become relevant whenever the boundary conditions
or the properties of macroscopic media 
differ from those for vacuum propagation.

To extract the vacuum-orthogonal coefficients,
we begin by considering 
the $E$-type vacuum coefficients $\kI$ and $\kE$.
Expanding Eqs.\ \rf{kI} and \rf{kE} 
in terms of the general spherical coefficients,
we find that each is a combination of 
$\cfzE$, $\kfzE$, $\kfoE$, and $\kftE$.
Careful consideration shows that
the restriction $\kE=0$ can be achieved 
by writing $\kfzE$ as combinations of $\kfoE$ and $\kftE$
and a new set of vacuum-orthogonal coefficients $\kftzE$.
The net result is the replacement of
$\kfzE$ with $\kftzE$ 
in the nine sets of general coefficients,
implemented via the substitution
\begin{align}
\kfzE &\to \kftzE - \kftzEdnjm{d}{(n+2)jm}
\notag \\
&\quad
- a_1 \kfoEdnjm{d}{(n+1)jm} - a_2 \kftE ,
\label{nonvac_kE}
\end{align}
where the numerical factors $a_1$ and $a_2$ are 
\begin{align}
& a_1 = 
\Frac{(d-5-n)(2(n+4)^2+(j+2)(j-1)) + 2(n+3)(n+4)(n+5)}{j(j+1)},
\notag\\
& a_2 = 
\Frac{1}{(j+2)(j+1)j(j-1)} 
\scriptstyle{\big[ (d-5-n)(d-4-n)(2(n+4)^2+(j+2)(j-1))} 
\notag\\
&\qquad
\scriptstyle{
+ 4(d-4-n)(n+3)(n+4)^2 + (2(n+3)^2-j(j+1))(n+3)(n+4)\big]} .
\end{align}
The result \rf{nonvac_kE} represents combinations
of $E$-type coefficients 
that are complementary to the vacuum coefficients $\kE$.
The index ranges and counting for the 
vacuum-orthogonal coefficients ${\kftzE}$ 
are given in Table \ref{kftE_ranges}.

The restriction $\kI=0$ yields a
similar result for the coefficients $\cfzE$,
\begin{align}
\cfzE &\to \cftzE - \cftzEdnjm{d}{(n-2)jm} 
- b_1 \kfzEdnjm{d}{(n-2)jm}
\notag \\
& + b_2 \kfoEdnjm{d}{(n-1)jm} + b_3 \kftE .
\label{nonvaccE}
\end{align}
where the numerical factors $b_1$, $b_2$, and $b_3$ are
\begin{align}
& b_1 = 
\Frac{(d-3-n)(d-2-n) + 2(d-2-n)n + n(n+1) -j(j+1)}{(d-3)(d-2)} ,
\notag\\
& b_2 = 
\Frac{(d-3-n)(2(n+1)(n+2)+(j+2)(j-1)) +2(n+1)(n+2)(n+3)}{(d-3)(d-2)} ,
\notag\\
& b_3 = \Frac{(d-5-n)(d-4-n) -(n+3)(n+4)}{(d-3)(d-2)} .
\end{align}
The coefficients $\cftzE$ appearing in Eq.\ \rf{nonvaccE}
are the camouflage coefficients 
introduced in the previous subsection.
Their index ranges and counting 
are given in Table \ref{cft_ranges}.
Note that $\kfzE$ appears in Eq.\ \rf{nonvaccE},
so the expression \rf{nonvac_kE} is needed to
fully reduce $\cfzE$.
Note also that Eq.\ \rf{nonvaccE} generalizes Eq.\ \rf{ct} 
to include coefficients for birefringent operators
that leave unaffected vacuum propagation at leading order.
This shows explicitly that
these operators contribute both to birefringent 
and to nonbirefringent effects,
an effect discussed in Sec.\ \ref{sec_even_coeffs}.

\begin{table}
\tabcolsep 3pt
\renewcommand{\arraystretch}{0.5}
\begin{tabular}{c||*{7}{c}}
\multicolumn{1}{c||}{}&
\multicolumn{7}{c}{$\kftzE$}\\[4pt]
\hline
\hline
\multicolumn{1}{c||}{$n$}&
\multicolumn{7}{c}{$j$}\\[4pt]
\hline
0 \\
1  &&1&&&&&\\
2  &0&&2&&&&\\
3  &&1&&3&&&\\
4  &0&&2&&4&&\\
$\vdots$  &$\vdots$&&&&&$\ddots$&\\
$d-4$  &0&&2&&4&$\cdots$&$d-4$\\[4pt]
\hline
\hline
\multicolumn{1}{c||}{total}&
\multicolumn{7}{c}{$\Frac16 (d-1)(d-2)(d-3)-1$}
\end{tabular}
\caption{\label{kftE_ranges}
Summary of the allowed ranges of indices $n$ and $j$
for the $E$-type vacuum-orthogonal coefficients $\kftzE$
associated with CPT-even nonbirefringent operators.
The dimension $d$ is even with $d\geq4$,
while $n\leq d-4$.
The index $m$ satisfies the usual restrictions
$-j\leq m\leq j$,
so there are $2j+1$ coefficients for each $j$.
For a given dimension $d$,
the number of coefficients is given in the last row.}
\end{table}

Determining the combinations of spherical coefficients
that are complementary to the vacuum coefficient $\kB$ 
is more straightforward.
We can write 
\begin{align}
\kB &= 
\sum_n \Frac{(-1)^{j+1}(d-1)}{2}\sqrt{\Frac{(j-2)!}{(j+2)!}}
\big((d-2)\kftB 
\notag \\
&\qquad\qquad\qquad
+(j+2)(j-1)\kfoB \big) .
\end{align}
Taking into account the index ranges of
$\kfoB$ and $\kftB$,
we find that a suitable set of coefficients
compatible with the restriction $\kB=0$ 
is obtained by replacing $\kfoB$ 
with a new set $\kftoB$ of vacuum-orthogonal coefficients. 
The combination leading to vanishing $\kB$ is given by 
\begin{align}
\kfoB &\to 
\kftoB - \kftoBdnjm{d}{(n+2)jm}
\notag \\
&\qquad\qquad
-\Frac{(d-2)}{(j+2)(j-1)}\kftBdnjm{d}{(n+1)jm} .
\end{align}
The index ranges and counting 
for the coefficients ${\kftoB}$ 
are given in Table \ref{kftB_ranges}.
Note that the coefficients $\kftoB$ and $\kftB$ 
are absent from Eq.\ \rf{nonvaccE},
so the corresponding operators have no nonbirefringent effects. 

\begin{table}
\tabcolsep 3pt
\renewcommand{\arraystretch}{0.5}
\begin{tabular}{c||*{6}{c}}
\multicolumn{1}{c||}{}&
\multicolumn{6}{c}{$\kftoB$}\\[4pt]
\hline
\hline
\multicolumn{1}{c||}{$n$}&
\multicolumn{6}{c}{$j$}\\[4pt]
\hline
0 \\
1  &1&&&&&\\
2  &&2&&&&\\
3  &1&&3&&&\\
4  &&2&&4&&\\
$\vdots$  &$\vdots$&&&&$\ddots$&\\
$d-4$  &&2&&4&$\cdots$&$d-4$\\[4pt]
\hline
\hline
\multicolumn{1}{c||}{total}&
\multicolumn{6}{c}{$\Frac16 d(d-2)(d-4)$}
\end{tabular}
\caption{\label{kftB_ranges}
Summary of the allowed ranges of indices $n$ and $j$
for the $B$-type vacuum-orthogonal coefficients $\kftoB$
associated with CPT-even nonbirefringent operators.
The dimension $d$ is even with $d\geq4$,
while $n\leq d-4$.
The index $m$ satisfies the usual restrictions
$-j\leq m\leq j$,
so there are $2j+1$ coefficients for each $j$.
For a given dimension $d$,
the number of coefficients is given in the last row.}
\end{table}

Finally,
we construct the combinations of spherical coefficients
that are complementary to $\kV$.
Although the vacuum coefficients $\kV$
have the comparatively simple form \rf{kV},
finding combinations that cover the coefficient space 
under the restriction $\kV=0$ involves some calculation. 
It turns out to involve two new sets 
$\kaftzB$, $\kaftoB$ 
of vacuum-orthogonal coefficients,
which appear via the substitutions
\begin{align}
\kafzB &\to
\Frac{(d-2-n)(n+3)}{d(d-2-n+j)}
\big( \kaftzB - \kaftzBdnjm{d}{(n-2)jm} \big)
\notag \\
&\qquad 
- \Frac{1}{n+1} \kaftoBdnjm{d}{(n-1)jm} ,
\\
\kafoB &\to 
\Frac{j(n+2)}{d-3-n+j}
\big( \kaftzBdnjm{d}{(n+1)jm} - \kaftzBdnjm{d}{(n-1)jm} \big)
\notag \\
&\qquad 
+ \Frac{d}{n+4} \kaftoB .
\end{align}
The index ranges and the numbers of independent components
for the vacuum-orthogonal coefficients
${\kaftzB}$ and ${\kaftoB}$ are shown in Table \ref{kaft_ranges}.

\begin{table}
\tabcolsep 3pt
\renewcommand{\arraystretch}{0.5}
\begin{tabular}{c||*{7}{c}|*{7}{c}}
\multicolumn{1}{c||}{}&
\multicolumn{7}{c|}{$\kaftzB$}&
\multicolumn{7}{c}{$\kaftoB$}\\[4pt]
\hline
\hline
\multicolumn{1}{c||}{$n$}&
\multicolumn{7}{c|}{$j$}&
\multicolumn{7}{c}{$j$}\\[4pt]
\hline
0  &0&&&&&&    &1&&&&&&\\
1  &&1&&&&&    &&2&&&&&\\
2  &0&&2&&&&   &1&&3&&&&\\
3  &&1&&3&&&   &&2&&4&&&\\
4  &0&&2&&4&&  &1&&3&&5&&\\
$\vdots$  &$\vdots$&&&&&$\ddots$&  &$\vdots$&&&&&$\ddots$&\\
$d-4$  &&1&&3&&$\cdots$&$d-4$      &&2&&4&&$\cdots$&$d-3$\\[4pt]
\hline
\hline
\multicolumn{1}{c||}{total}&
\multicolumn{7}{c|}{$\Frac16 (d-1)(d-2)(d-3)$}&
\multicolumn{7}{c}{$\Frac16 (d+1)(d-1)(d-3)$}
\end{tabular}
\caption{\label{kaft_ranges}
Summary of the allowed ranges of indices $n$ and $j$
for the vacuum-orthogonal coefficients 
$\kaftzB$ and $\kaftoB$
associated with CPT-odd nonbirefringent operators.
The dimension $d$ is even with $d\geq4$,
while $n\leq d-4$.
The index $m$ satisfies the usual restrictions
$-j\leq m\leq j$,
so there are $2j+1$ coefficients for each $j$.
For a given dimension $d$,
the number of coefficients is given in the last row.}
\end{table}

To summarize,
in this subsection we have completed the decomposition
of the original nine sets of general spherical coefficients
into those that generate leading-order effects
in the vacuum propagation of light
and those that comprise the complementary subset.
The four sets of vacuum coefficients are 
$\kI$, $\kE$, $\kB$, and $\kV$.
They are discussed in Sec.\ \ref{sec_vac}.
The complement consists of
the nine reduced sets of vacuum-orthogonal coefficients
$\cftzE$, $\kftzE$, $\kfoE$, $\kftE$, $\kftoB$, $\kftB$,
$\kaftzB$, $\kaftoB$, and $\kafoE$.
Note that these are nonzero only for $d>4$.
This decomposition is summarized as part of 
Table \ref{limiting_table}
in Sec.\ \ref{Summary and discussion}.

Except for the camouflage coefficients $\cftzE$
discussed in the previous subsection,
all the vacuum-orthogonal coefficients
control birefringence effects
that cannot be detected at leading order via vacuum propagation.
To measure these coefficients,
alternative methods such as laboratory experiments 
are therefore desirable. 
In contrast,
all four sets of vacuum coefficients 
are detectable in astrophysical tests
involving birefringence or dispersion,
with the exception 
of the special $d=4$ coefficients $\kIdjm{4}{jm}$.
The latter have been extensively studied
through a variety of different methods
\cite{cav1,cav2,cav3,cav4,cav5,cav6}.

Note also that various hybrid models 
involving the vacuum-orthogonal coefficients
can be countenanced.
For example,
a general vacuum-orthogonal isotropic model
is obtained upon further restricting attention
to the isotropic coefficients
$\cftzEdnjm{d}{n00}$, 
$\kftzEdnjm{d}{n00}$, 
and ${\kaftzBdnjm{d}{n00}}$.
This model has $(d-2)/2$ nonbirefringent 
and $(d-4)/2$ 
birefringent operators for even $d$,
along with $(d-3)/2$ 
birefringent operators for odd $d$.

\subsection{Connections to other formalisms}
\label{sec_formalisms}

Several specialized models
involving particular Lorentz-violating photon operators with $d>4$ 
have been considered in the literature.
The generality of the SME implies
that any realistic model for the photon propagator
compatible with standard field theory
is encompassed via special values
of the coefficients $\kf$ and $\kaf$.
In this subsection,
we outline some of these connections.
For a selection of specific models defined via field theory,
we provide explicit limiting values 
of the SME coefficients for Lorentz violation
that reproduce the models. 
We also offer here some remarks 
about the relationship between the SME 
and the kinematical approach to Lorentz violation,
which is based on altering the transformation laws.
Comments on the links between 
the photon sector of the SME 
and Lorentz-violating modifications 
of the photon dispersion relation
outside the context of standard field theory
can be found in Sec.\ \ref{sec_disp}.

\subsubsection{Field-theoretic models}

We begin by discussing models
defined via a Lagrange density in field theory.
Several specialized models exist that involve photon fields 
with a small number of specific Lorentz-violating operators
of mass dimensions $d=5$ and in some cases also $d=6$.
We provide here brief comments
identifying the match between these models
and the SME coefficients for Lorentz violation.

One such model is presented by Gambini and Pullin
\cite{gp}.
Lorentz violation in this model 
is controlled by the parameter $\ch l_P$.
To see the connection to the SME,
we can make the field redefinition
\beq
\mbf E + 2\chi l_P\mbf\nabla\times\mbf E \rightarrow \mbf E.
\eeq
This model then is equivalent
to taking the special limit of the SME
with the nonzero coefficients being
$(\kafd{5})^{0jk}$ and $(\kfd{6})^{jklmpq}$,
given by
\begin{align}
(\kafd{5})^{0jk} & = 2\chi l_P \de^{jk},
\notag\\
(\kfd{6})^{jklmpq} & = 
-4\chi^2 l_P^2( \ep^{jkn}\ep^{nlm}\de^{pq}
-\fr14\ep^{jk(p}\ep^{q)lm}).
\end{align}
This model involves only isotropic Lorentz violations,
so it is a special limit of the isotropic models
discussed in Sec.\ \ref{sec_fc}.
In terms of the coefficients for Lorentz violation
defined in Eq.\ \rf{fccoeffdef},
the match
\begin{align}
\kaffcdn{5}{2} & = -2\sqrt{4\pi}~\chi l_P,
\notag\\
\cffcdn{6}{2} & = \kffcdn{6}{2}=-\sqrt{4\pi}~\chi^2 l_P^2/5
\end{align}
provides a complete specification of this model
within the SME.
It involves three nonzero isotropic coefficients
containing a single degree of freedom.
A nonlinear generalization of this model is obtained
in Ref.\ \cite{amu},
in which the propagator component  
is a special limit of the isotropic models
discussed in Sec.\ \ref{sec_fc}
with three degrees of freedom.

Another specialized model incorporating
photon-sector Lorentz violation
is introduced by Myers and Pospelov
\cite{rmmp}.
This model has a unit timelike background vector $n^\mu$
that defines a preferred frame
and a parameter $\xi/M_P$ setting the scale
of the Lorentz violation.
Only one $d=5$ Lorentz-violating operator affects
the photon propagator.
In terms of SME coefficients,
the model is obtained by taking 
the nonzero coefficients for Lorentz violation 
to be 
\beq
(\kafd{5})^{\ka\mu\nu}=
-\fr{\xi}{M_P}
( n^\ka n^\mu n^\nu - \fr15 n^2 n^{(\mu}\et^{\nu)\ka}).
\eeq
In the preferred frame,
the vector takes the form $n^\mu = (1,0,0,0)$.
This model can therefore also be defined uniquely
in terms of isotropic Lorentz violation
and understood as a special limit
of the isotropic models
discussed in Sec.\ \ref{sec_fc}.
It is equivalent to taking as nonzero 
only the coefficient $\kaffcdn{5}{0}$,
with the specific choice
\beq
\kaffcdn{5}{0} = \frac {3\xi \sqrt{4\pi}~} {5M_P}.
\eeq 

A model focusing on a Lagrange density
that is explicitly gauge invariant
and involves Lorentz-violating operators
with $d=5$ is considered by Bolokhov and Pospelov
\cite{pbmp}.
By construction,
the component of the model
relevant to the photon propagator
is restricted to Lorentz violation 
affecting leading-order vacuum propagation.
It consists of $d=5$ operators governed by 
a totally symmetric and traceless parameter
$C^{\mu\nu\rh}$.
This parameter has 16 independent components,
corresponding to the 16 vacuum coefficients $\kVdjm{5}{jm}$
among the 36 independent coefficients
$(\kafd{5})^{\mu\nu\rh}$
for Lorentz violation at $d=5$ in the SME.
In terms of the latter coefficients,
the model is fixed by 
\begin{align}
\kVdjm{5}{jm}\sim
(\kafd{5})^{\mu\nu\rh}
\Big\vert_{{\rm symmetric}\atop{\rm traceless~}}
&= -2C^{\mu\nu\rh},
\end{align}
where the correspondence on the left-hand side
is determined by matching 
Eqs.\ \rf{stokes_params} and \rf{vac_exp},
and the restriction to the 16 relevant coefficients
in $(\kafd{5})^{\mu\nu\rh}$
is obtained by imposing total symmetry and tracelessness.
The 20 SME coefficients at $d=5$ that are absent from this model
are the vacuum-orthogonal coefficients
$\kaftzBdnjm{5}{njm}$, $\kaftoBdnjm{5}{njm}$, 
and $\kafoEdnjm{5}{njm}$,
which leave unaffected leading-order vacuum propagation
but nonetheless produce leading-order effects
in suitable laboratory experiments. 

The series structure of the SME Lagrange density \rf{lagrangian}
implies that models defined 
via nonpolynomial but analytic functionals of field operators 
can be matched to the SME by Taylor expansion.
This includes models with apparent singularities,
provided the expansion is taken about a nonsingular point.
In the latter case,
distinct analytic continuations may correspond 
to different values of SME coefficients,
which may lead to unusual dynamical effects.

As an illustration,
consider the class of Lorentz-violating models 
called very special relativity (VSR)
\cite{vsr},
in which the Lorentz group is broken explicitly
to the four-parameter subgroup SIM(2).
It is useful to introduce the operator 
$N^\mu = n^\mu/(n\cdot\prt)$,
where $n^\mu$ is a unit null vector
that establishes a preferred lightlike frame.
It follows that the combination $N\cdot T$ 
involving a tensor field operator $T$
is SIM(2) covariant but violates Lorentz covariance.
We can construct a generic on-shell linear VSR electrodynamics 
preserving gauge invariance by specifying field equations 
of the form 
\beq
\prt_\mu F^\mn + K_\mu (N) F^\mn = 0,
\label{vsrem}
\eeq 
where $K^\mu (N)$ is a nontrival 4-vector function 
of $N^\mu$ that can also depend on the metric, 
the Levi-Civita tensor, and derivatives.
Note that it is problematic to obtain 
a gauge-invariant action for these equations.

To investigate the correspondence to the SME,
it is convenient to work in momentum space,
where the operator $iN^\mu = n^\mu/(n\cdot p)$
is singular whenever $n\cdot p = 0$.
Matching the momentum-space version of 
the VSR electrodynamics \rf{vsrem}
to the momentum-space version of 
the SME equations of motion \rf{eqnmot}
therefore involves expanding $K_\mu (iN) F^\mn(p)$ 
about a nonsingular point $Q$ in momentum space.
Assigning coordinates $p_Q^\mu$ to $Q$,
we find the expansion converges 
either for 
$0 < n\cdot p < 2n\cdot p_Q$ 
or for
$2n\cdot p_Q < n\cdot p < 0$,
corresponding to two different analytic continuations
of $K_\mu$.
Each expansion can be matched 
to a set of SME coefficients,
but the two sets of coefficients differ.
This implies the VSR electrodynamics \rf{vsrem}
is represented by two different limits of the SME,
according to the value of the photon momentum 
relative to the preferred lightlike frame.

\subsubsection{Robertson-Mansouri-Sexl model}

In a different vein,
some authors adopt a kinematical approach to Lorentz violation
that is based on modifications of the transformation laws.
An older test model of this type
that is encompassed by the SME approach
is the kinematical formalism of
Robertson, Mansouri, and Sexl (RMS)
\cite{rms,rms2}.
This approach assumes that
there is a preferred universal inertial frame U
in which light propagates conventionally
as measured using a definite set of rods and clocks.
In other frames,
which include any inertial frame E relevant for experiment,
light can behave anisotropically
with respect to the boosted rods and clocks.
The RMS approach assumes that the lengths of rods 
and the ticking rates of clocks
are invariant in inertial frames related to U
by RMS coordinate transformations ${T^\mu}_\nu$.
These are deformations of the Lorentz transformations
involving three functions of the boost $\mbf v$,
conventionally denoted as 
$a(\mbf v)$, $b(\mbf v)$, $d(\mbf v)$.

The RMS formalism can be translated into the SME framework.
Consider first the preferred universal frame U.
Since light is conventional in U by definition,
the Maxwell action must be valid in this frame,
so that
\beq
\cl_{\rm RMS,~U}^{\rm photon} =  
-\Frac 1 4 \et^{\mu\rh}\et^{\nu\si} F_{\mu\nu}F_{\rh\si}.
\eeq
This Lagrange density is Lorentz invariant,
so all Lorentz violation in this frame resides
in the physics describing the chosen rods and clocks.
For the photons,
we can match the RMS formalism to the SME 
by imposing the condition
\beq
\cl_{\rm SME,~U}^{\rm photon} = \cl_{\rm RMS,~U}^{\rm photon},
\eeq
which eliminates all Lorentz-violating effects 
in the photon sector of the SME as seen in U.

For the RMS rods and clocks,
any realistic set is made 
of constituent particles and fields.
Since the SME describes general Lorentz violation
for all particles and fields,
it follows that the properties 
of any definite set of rods and clocks
can be derived from the full SME action,
at least in principle.
This action contains 
infinitely many coefficients for Lorentz violation 
outside the photon sector,
so there is plenty of room for anomalous behavior.
However,
to preserve the observed isotropy of light in U
required by the RMS formalism
for any possible choice of rods and clocks,
only violations of Lorentz invariance in U
that produce isotropic effects on the rods and clocks 
should be countenanced.
We conclude that the match 
between the RMS formalism and the SME in the frame U 
requires restricting the SME 
to a subset of coefficients outside the photon sector.
For definiteness in what follows,
we denote these SME coefficients 
collectively by $\{k\}$.
Note that $\{k\}$ can include anisotropic coefficients,
provided they have no anisotropic effects 
on the chosen rods and clocks.

Next,
consider an experimentally relevant frame E 
that is moving with velocity $\mbf v$
relative to U.
In the RMS formalism,
the properties of light in E
are obtained by performing 
an RMS coordinate transformation using ${T^\mu}_\nu$.
This leads to a modified Maxwell action with
\beq
\cl_{\rm RMS, ~E}^{\rm photon} =  
-\Frac 1 4 \sqrt{|g_{\rm RMS}|}~ 
(g_{\rm RMS}^{-1})^{\mu\rh} (g_{\rm RMS}^{-1})^{\nu\si}
F_{\mu\nu}F_{\rh\si},
\eeq
where
\beq
(g_{\rm RMS})_{\mu\nu} = 
\et_{\rh\si} {(T^{-1})^\rh}_\mu {(T^{-1})^\si}_\nu 
\label{rmsmetric}
\eeq
is an effective metric that depends 
on the three functions $a$, $b$, $d$.
Note that physically different choices for the rods and clocks
in the frame U 
imply different invariant RMS transformations 
and hence different $a$, $b$, $d$.

In contrast,
the SME properties of light and the boosted rods and clocks
in the frame E are obtained 
by performing a particle Lorentz transformation
with the velocity $\mbf v$.
Since $\cl_{\rm SME,~U}^{\rm photon}$
is invariant under particle Lorentz transformations,
light in the SME must obey conventional electrodynamics 
in the frame E too.
However,
the rods and clocks involve Lorentz-violating operators
that change under the particle Lorentz transformation.
This produces a deformation 
of the standard rods and clocks in E
that depends on the SME coefficients $\{k\}$.

We thus find that the RMS formalism and the SME
naturally generate two different coordinate systems
for describing physics in the frame E.
To relate the two,
we can redefine the length and time intervals 
specified by the boosted SME rods and clocks
to match numerically 
those of the boosted RMS rods and clocks,
and we can choose the SME synchronization 
to match the RMS one.
The redefinition can be implemented
by scaling the spacetime coordinates.
In the photon sector of the SME,
the scaling produces an effective metric 
$(g_{\rm SME,~eff})_{\mu\nu}$ 
that depends on the coefficients $\{k\}$.
Since measurements made with rods and clocks
using either coordinate system now agree,
this metric must match the RMS metric 
\rf{rmsmetric} in the frame E.
This yields the result
\beq
(g_{\rm SME,~eff})_{\mu\nu} (\{k\}) 
= (g_{\rm RMS})_{\mu\nu} (a,b,d),
\label{smerms}
\eeq
which provides a direct correspondence 
between the SME coefficients and the RMS functions.

The above match shows that the RMS formalism 
can be understood as a special limit of the SME
in which normal light behavior 
together with 
Lorentz violation affecting rods and clocks isotropically
are assumed in the frame U.
This limit excludes infinitely many Lorentz-violating effects.
Since the RMS formalism is a special-relativistic test model,
the gravitational sector of the SME must also be disregarded. 
Note that the three functions $a$, $b$, $d$
can be expanded in powers of the velocity
to yield a triple infinity of constant parameters,
which can be absorbed into the multiple infinity 
of coefficients $\{k\}$.
In the idealized case
of rods and clocks formed from a scalar particle 
with only one isotropic dimension-zero coefficient $k$ 
for Lorentz violation in the frame U,
explicit expressions for the three RMS functions
in terms of $k$ are given in Sec.\ III C
of Ref.\ \cite{km}.

Another point of interest is that 
physically different rods and clocks 
are associated with different functions $a$, $b$, $d$,
and so involve different combinations of 
the coefficients $\{k\}$.
Within the RMS formalism,
measurements of $a$, $b$, $d$ in a given experiment 
cannot meaningfully be compared to those in another experiment
unless physically identical rods and clocks are used in both.
Note that the rods and clocks 
must also be in the same physical state,
since state changes in the presence of Lorentz violation 
can deform physical properties.
This major disadvantage of the RMS formalism
is circumvented by the SME.
The SME coefficients for Lorentz violation
are specific to particles and interactions 
and can therefore be reported in an experiment-independent way
in a conveniently chosen frame,
which conventionally
is taken as the Sun-centered frame
described in Sec.\ \ref{sec_rotations}.

To illustrate the above reasoning
with an explicit example,
we can consider a scenario
in which the photons, the chosen rods, and the chosen clocks
each have properties governed 
by different effective metrics.
Following the RMS assumptions,
the effective metric 
$(g_{\rm photon,~U})_{\mu\nu}$
for photons in the frame U 
must be taken as the Minkowski metric,
while the effective metric  
$(g_{\rm rod,~U})_{\mu\nu}$
for the rods
and the effective metric 
$(g_{\rm clock,~U})_{\mu\nu}$
for the clocks
characterize the Lorentz-violating physics in U.
We examine here a simple model for which 
\begin{align}
(g_{\rm photon,~U})_{\mu\nu}
&=\et_\mn , 
\notag\\
(g_{\rm rod,~U})_{\mu\nu}
&=\et_\mn - (c_{\rm rod,~U})_\mn ,
\notag\\
(g_{\rm clock,~U})_{\mu\nu}
&=\et_\mn - (c_{\rm clock,~U})_\mn .
\end{align}
The quantities 
$(c_{\rm rod,~U})_\mn$ and $(c_{\rm clock,~U})_\mn$
can be viewed as effective coefficients for Lorentz violation
defined in the frame U.
Depending on the nature of the chosen rods and clocks,
these coefficients can be identified either 
directly with specific SME coefficients
or indirectly as suitable combinations of SME coefficients.
We also assume the conditions 
\begin{align}
(c_{\rm rod,~U})_{j0} &= (c_{\rm rod,~U})_{0k} = 0, 
\notag\\
(c_{\rm rod,~U})_{jk} 
&= \frac13 (c_{\rm rod,~U})_{00} \de_{jk},
\notag\\
(c_{\rm clock,~U})_{j0} &= (c_{\rm clock,~U})_{0k} = 0, 
\notag\\
(c_{\rm clock,~U})_{jk} 
&= \frac13 (c_{\rm clock,~U})_{00} \de_{jk},
\end{align}
which ensure isotropic properties
of the rods and clocks in U,
as required by the RMS formalism.
For simplicity,
the coefficients are taken 
to be independent of spacetime position
or, 
equivalently, 
independent of frequency and momentum.
This simple model therefore 
has only two degrees of freedom
controlling Lorentz violation.

Suppose for definiteness
that the experimentally relevant frame E
is moving with velocity $\mbf v = (v,0,0)$
relative to U.
Following the general reasoning above,
we can obtain the SME description of the model
in E by performing 
a standard particle Lorentz transformation $\La$ 
from U to E.
The resulting effective metrics in E 
are given by expressions of the form 
\beq
g_{\rm E} = \La^{-1\, T} g_{\rm U} \La^{-1} .
\eeq
In the frame E,
light remains conventional 
but the rods and clocks are distorted.
To match to the RMS formalism,
we must therefore seek
alternative coordinates in E
in which the rods are isotropic 
and of the same length as in U,
and in which the clocks tick at the original rate in U.
The appropriate coordinate transformation $C$
leaves the origin of E in place
but acts on the effective metrics as
\begin{align}
g_{\rm E} &\to C^{-1\, T} g_{\rm E} C^{-1} 
\equiv 
C^{-1\, T} \La^{-1\, T} g_{\rm U} \La^{-1} C^{-1} .
\end{align}
Since the components $(g_{\rm rod})_{jk}$
determine the rod length
and the component $(g_{\rm clock})_{00}$
determines the clock ticking rate,
and since both these quantities
are assumed invariant in the RMS formalism,
the required transformation $C$
is fixed by demanding that 
\begin{align}
[C^{-1\, T} \La^{-1\, T} (g_{\rm rod, U}) \La^{-1} C^{-1}]_{jk}
&= (g_{\rm rod, U})_{jk},
\notag\\
[C^{-1\, T} \La^{-1\, T} (g_{\rm clock, U}) \La^{-1} C^{-1}]_{00}
&= (g_{\rm clock, U})_{00}.
\end{align}
With the usual definition
$\ga \equiv 1/\sqrt{1-v^2}$,
we find that the nonzero elements of $C$ take the form
\begin{align}
{C^0}_0&=
\sqrt\frac{1-(c_{\rm clock,~U})_{00}\ga^2(1+\frac13 v^2)}
{1-(c_{\rm clock,~U})_{00}},
\notag\\
{C^1}_1&=
\sqrt\frac{1+(c_{\rm rod,~U})_{00}\ga^2(\frac13+v^2)}
{1+\frac13 (c_{\rm rod,~U})_{00}},
\notag\\
{C^2}_2&= {C^3}_3= 1,
\end{align}
which amounts to performing different dilations of time 
and of space in the direction of the boost.

With this information in hand,
we can construct the RMS transformation $T\equiv C\La$
from U to E.
This provides the direct correspondence
between the SME coefficients 
and the RMS functions $a$, $b$, $d$
for this simple model.
We obtain 
\begin{align}
a&=\frac 1 \ga
\sqrt\frac{1-(c_{\rm clock,~U})_{00}\ga^2(1+\frac13 v^2)}
{1-(c_{\rm clock,~U})_{00}},
\notag\\
b&=\ga
\sqrt\frac{1+(c_{\rm rod,~U})_{00}\ga^2(\frac13+v^2)}
{1+\frac13 (c_{\rm rod,~U})_{00}}, 
\notag\\
d&=1,
\notag\\
\ep&=\frac{-a\ga^2v}{b},
\end{align}
where $\ep$ is the RMS synchronization function
in Einstein synchronization.
Unlike the elementary single-coefficient example 
given in Ref.\ \cite{km},
the present model has $a\neq 1/b$. 
It also has $d=1$,
but allowing frequency dependence
in $(c_{\rm rod,~U})_{00}$
can generate $d\neq 1$
via the mixing of frequency with momentum 
resulting from the $\La$ boost
and its consequent effects on $C$.
A frequency or momentum dependence may arise 
directly from the inclusion of matter-sector operators 
of nonrenormalizable dimension 
\cite{kmfermion},
or indirectly 
from combinations of SME coefficients of renormalizable dimension
when the motions of the component particles in the rod
are incorporated.

Expanding the above results for $a$, $b$, $d$
to leading order in $v^2$ 
and to leading order in coefficients for Lorentz violation 
yields 
\begin{align}
a \approx 1 + \al v^2,
\quad
\al&=-\half-\Frac{5}{12}(c_{\rm clock,~U})_{00} ,
\notag\\
b \approx 1 + \be v^2,
\quad
\be&=\half+\Frac{7}{12}(c_{\rm rod,~U})_{00} ,
\notag\\
d \approx 1 + \de v^2,
\quad
\de&=0 .
\end{align}
The combination of the RMS parameters
$\al$, $\be$, $\de$ that can be tested 
in Michelson-Morley experiments is known to be 
\cite{rms2} 
\beq
\be+\de-\half=\Frac{7}{12}(c_{\rm rod,~U})_{00} ,
\eeq
and the combination tested in Kennedy-Thorndike experiments is
\beq
\al-\be+1 = 
- \Frac{7}{12}(c_{\rm rod,~U})_{00} 
-\Frac{5}{12}(c_{\rm clock,~U})_{00} ,
\eeq
while Ives-Stilwell experiments are sensitive to $\al$.
Evidently,
two of the three types of experiments
are required to disentangle all effects,
even for this simple model.
Various special cases can be considered.
For example,
if either 
$(c_{\rm rod,~U})_{00}$ or $(c_{\rm clock,~U})_{00}$ 
vanishes,
then either Michelson-Morley or Ives-Stilwell experiments
have no signal. 
If $(c_{\rm rod,~U})_{00}=(c_{\rm clock,~U})_{00}$,
all experiments have interdependent signals.

The example verifies that the RMS formalism
concerns Lorentz-violation residing in the matter sector
rather than in the photon sector.
Note that different choices of rods and clocks
generically involve different values of
$(c_{\rm rod,~U})_{00}$
and $(c_{\rm clock,~U})_{00}$
and therefore are associated with distinct 
Lorentz-violating effects,
confirming that results from RMS experiments performed 
using different rods and clocks 
cannot meaningfully be compared.
Note also that 
$(c_{\rm rod,~U})_{00}$ and $(c_{\rm clock,~U})_{00}$
are defined in the frame U,
which typically differs from the
canonical Sun-centered frame S 
in which SME coefficients are reported.
A conventional particle Lorentz boost 
from U to S can be implemented 
to identify the relevant coefficient combinations in S.
Interpreting experiments in the RMS formalism requires
a choice of U,
which is often taken as the frame
of the cosmic microwave background,
in which case the relevant boost to S is of order $10^{-3}$
and so to a good approximation
$(c_{\rm rod,~U})_{00}=(c_{\rm rod,~S})_{TT}$ 
and 
$(c_{\rm clock,~U})_{00}=(c_{\rm clock,~S})_{TT}$.
Alternatively,
the frame U can simply be chosen to be the frame S,
since to date no compelling evidence 
for anisotropic Lorentz violation in S has been identified.
Other choices are also possible.
The requirement that U be specified to fix the RMS formalism
is a disadvantage that is avoided in the SME,
where any universal frame U is acceptable
and moreover the existence of U is unnecessary 
for interpreting experimental data.

\subsubsection{Deformed special relativities}

Another kinematical approach
involves requiring the invariance of all physics
under some specified modification 
of the Lorentz transformations.
Recent efforts along these lines
are generically called deformed special relativity (DSR),
or in some cases doubly special relativity
or kappa-deformed relativity
\cite{dsr}.
They posit that all physics is invariant 
under a set of deformed nonlinear Lorentz transformations,
usually one that introduces a maximum energy scale.
A generic DSR model is defined by replacing 
the 4-momentum $p_\mu$ with a modified 4-momentum $\pi_\mu$,
typically with a Planck-scale suppression factor 
for the deformation
\cite{jv}.
The momentum-space Lorentz transformations 
act conventionally on $\pi_\mu$.
This induces unconventional DSR transformations on $p_\mu$,
which are required to leave invariant
the physics of the model.

Since the SME contains
arbitrary polynomial Lorentz-violating operators 
at all mass dimensions,
it must be possible to express any nonsingular DSR model
involving realistic fields in the SME framework.
The deformations normally are assumed
to preserve rotation invariance, 
in which case the match to the SME
involves the isotropic coefficients
discussed in Sec.\ \ref{sec_fc}.
Within the context of the present work,
we can investigate this correspondence explicitly 
in the photon sector.

Consider a generic DSR model
in which the replacement of the 4-momentum $p_\mu$
is specified by the nonlinear transformation 
\beq
p_\mu \to \pi_\mu = {M_\mu}^\nu (p) p_\nu
\label{dsrM}
\eeq
acting in momentum space,
where $M_\mu^{\pt{\mu}\nu}(p)$
is a nonsingular matrix
that is a deformation of the identity.
By definition,
the Lorentz transformations ${\La_\al}^\be$ in momentum space
act on $\pi_\mu$ as usual,
$\pi_\mu \to \pi'_\mu = {\La_\mu}^\nu \pi_\nu$.
It follows that the expressions
\beq
p_\mu \to p'_\mu = {S_\mu}^\nu p_\nu,
\qquad {S_\mu}^\nu = {(M^{-1}\La M)_\mu}^\nu 
\label{dsrtrans}
\eeq
specify the nonlinear DSR transformations ${S_\mu}^\nu$ of $p_\mu$.

In the photon sector,
we can determine the DSR-covariant dispersion relation 
for the photon
by applying the replacement \rf{dsrM}
to the standard Maxwell dispersion relation 
$p_\mu \et^{\mu\nu} p_\nu = 0$.
This gives 
\beq
p_\mu (g_{\rm DSR}^{-1})^{\mu\nu} p_\nu = 0 ,
\label{dsrdisp}
\eeq
where the effective metric
\beq
(g_{\rm DSR})_{\mu\nu} = 
\et_{\rh\si} {(M^{-1})_\mu}^\rh {(M^{-1})_\nu}^\si 
\label{dsrmetric}
\eeq
is defined in momentum space.
By construction,
the dispersion relation \rf{dsrdisp} is invariant 
under the DSR transformations ${S_\mu}^\nu$ 
in Eq.\ \rf{dsrtrans}.

In the SME context,
the DSR dispersion relation \rf{dsrdisp}
is recovered as a limiting case 
of the scalar covariant dispersion relation \rf{dr}.
The match arises 
in the special limit with $(\kaf)_\mu =0$
and with the momentum-space operator $(\kf)^{\mu\nu\rh\si}$ 
given by
\begin{align}
(\kf)^{\mu\nu\rh\si} &=
\sqrt{g_{\rm DSR}}~ 
(g_{\rm DSR}^{-1})^{\mu\rh} (g_{\rm DSR}^{-1})^{\nu\si} 
\notag\\
&
-\half(\et^{\mu\rh}\et^{\nu\si}-\et^{\nu\rh}\et^{\mu\si}).
\label{dsrcoeff}
\end{align}
We see that the nonlinear transformation 
$M_\mu^{\pt{\mu}\nu}(p)$
generates a subset of 
the Lorentz-violating operators in the SME,
typically in the form of an infinite series.

An alternative derivation yielding the same result
can be performed at the level of field theory.
We can construct the DSR-invariant modified action
for this generic model 
by taking the Maxwell action in momentum space
and implementing the replacement \rf{dsrM}.
The transformation of the photon field
under ${S_\mu}^\nu$ is taken as 
\beq
A_\mu (p) \to A'_\mu (p') = {S_\mu}^\nu (p) A_\nu (p),
\label{dsrAtrans}
\eeq
corresponding to the replacement 
$A_\mu \to {M_\mu}^\nu A_\nu$.
This gives the momentum-space Lagrange density
\beq
\cl_{\rm DSR}^{\rm photon}(p)=
-\Frac 1 4 \sqrt{|g_{\rm DSR}|}~ 
(g_{\rm DSR}^{-1})^{\mu\rh} (g_{\rm DSR}^{-1})^{\nu\si} 
F_{\mu\nu} F_{\rh\si},
\label{dsrlag}
\eeq
where the effective metric 
$(g_{\rm DSR})_{\mu\nu}$
in momentum space
is given by Eq.\ \rf{dsrmetric}.
By construction,
the Lagrange density \rf{dsrlag} is invariant 
under the DSR transformations ${S_\mu}^\nu$ 
in Eq.\ \rf{dsrtrans}.
In the SME context,
this same Lagrange density 
is obtained as the special limit of the SME photon sector 
of the form
\begin{align}
\cl_{\rm SME}^{\rm photon}\Big|_{\rm DSR} 
&\equiv \cl_{\rm DSR}^{\rm photon}
\notag\\
&
= -\Frac 1 4 F^{\mu\nu} F_{\mu\nu} 
-\Frac 1 4  
F_{\mu\nu} (\kf)^{\mu\nu\rh\si} F_{\rh\si} ,
\label{smedsrgen}
\end{align}
where the momentum-space operator $(\kf)^{\mu\nu\rh\si}$ is 
given by Eq.\ \rf{dsrcoeff}.
This SME construction applies 
for a generic DSR model,
and it confirms that the nonlinear transformation 
$M_\mu^{\pt{\mu}\nu}(p)$
produces a series of SME coefficients for Lorentz violation. 

Although the model \rf{smedsrgen} transforms nontrivially 
under conventional particle Lorentz transformations,
observable physical effects cannot arise.
This is because in any frame
the inverse momentum replacement can be applied 
to all particles and fields
to recover an action of the usual Lorentz-invariant form
in terms of the momentum $\pi^\mu$,
and any experiment naturally identifies this momentum 
as the physical one. 
The situation in this respect differs 
from that of the RMS formalism,
where only the rods and clocks are assumed
invariant under modified Lorentz transformations
while light behaves differently,
so physical effects can arise.
It also differs from the SME,
where distinct particles and fields
can break Lorentz symmetry in different ways 
and only a subset of coefficients are unobservable.
The absence of observable effects
when the conventional momentum is adopted 
is a known characteristic of DSR models
\cite{lnjs},
which follows from their definition
as nonlinear momentum-space representations 
of the usual Lorentz transformations.

More generally,
the above considerations reveal that
modifications of the dynamics are unphysical
whenever they arise from 
a universal and reversible momentum substitution.
Observable effects 
might in principle be possible in special models 
if singularities in the physics obstruct the recovery 
of the usual 4-momentum via the inverse momentum replacement,
although the requirement 
of physical singularities seems unappealing.
However,
an alternative approach does exist.
Observable effects in models with deformed Lorentz transformations 
can be obtained by imposing the deformed invariance
only on a subset of particles or fields,
while either conventional Lorentz invariance 
or a different deformed invariance holds for others.
This idea has not been investigated in the literature,
perhaps because the idea 
of two or more mutually incompatible invariances 
in nature runs counter to the DSR philosophy.
Note that any such models are subsets of the SME,
so constraints from SME coefficients apply.

As one simple exotic example of a model with multiple invariances,
one can consider `spinning' special relativity
in which the deformed symmetry differs for fermions and bosons.
The quadratic action for a fermion
would then be invariant under one transformation,
while that for a boson would have a different invariance.
The deformed transformations
can also be chosen to depend explicitly 
on the representation,
so that multiple invariances would be involved
in the various pieces of the quadratic action.
Different versions of this idea could be considered.
For instance,
`flavorful' and `colorful' special relativities 
could be constructed by choosing the deformations 
to vary with the particle species,
or more specifically 
with the representation of the internal symmetry group.
In any case,
the full action generically breaks the individual invariances,
thereby leading to detectable signals.
As before,
any subgroup of the deformed transformations 
that leaves invariant the full action
is associated with unobservable effects.

\section{Reference frames and rotations}
\label{sec_rotations}

For comparative purposes,
it is useful to adopt a standard inertial frame
in reporting measurements 
of coefficients for Lorentz violation.
The canonical frame used in the literature
is a Sun-centered celestial equatorial frame
\cite{km,spaceexpt}.
Cartesian coordinates in this frame
are denoted $(T,X,Y,Z)$.
The $Z$ axis lies along the rotation axis of the Earth,
while the $X$-$Y$ plane coincides with the Earth's equatorial plane. 
The $X$ axis is directed from the Earth
to the Sun at the vernal equinox.
One advantage of this conventional choice 
is that transforming between the Sun-centered frame 
and a laboratory frame is comparatively simple.

In a typical application,
a measurement of coefficients for Lorentz violation
is made in a laboratory frame of reference.
However,
the rotation and revolution of the Earth 
imply this frame is noninertial,
so the coefficients for Lorentz violation
in the laboratory vary with time.
These varying coefficients are related
to the constant ones in the Sun-centered frame
by an observer Lorentz transformation,
which predominantly involves rotations.
In this section,
we construct the relevant rotation transformations
to an arbitrary laboratory reference frame, 
including one based on a rotating turntable.
We apply the results to obtain the transformation 
of spherical coefficients for Lorentz violation
between the laboratory and Sun-centered frames.
Since the spherical coefficients 
represent a decomposition based on angular momentum,
their transformation is comparatively straightforward.
Moreover,
the spin weight is unaffected by rotations
because helicity commutes with the angular momentum $\mbf J$.
The only index that changes under rotations
is therefore the $J_z$ eigenvalue $m$.

\subsection{Rotation matrices}

To construct the rotation transformations,
we adopt Euler angles $\al$, $\be$, and $\ga$ 
that relate two arbitrary cartesian frames 
with coordinates
$(x,y,z)$ and $(x',y',z')$
through the rotation
\beq
R=e^{-i\al J_z}e^{-i\be J_y}e^{-i\ga J_z} .
\eeq
This rotation can be visualized
by starting with the two frames coinciding,
rotating the second frame by $\ga$ about the $z$ axis,
then by $\be$ about the $y$ axis,
and finally by $\al$ about the $z$ axis again.
Acting on coordinates,
this combination of rotations
can be shown to be implemented by the matrix equation
\begin{align}
\left(\begin{array}{c} x'\\y'\\z' \end{array}\right) &=
\left(\begin{array}{ccc} 
\cos\ga&\sin\ga&0\\-\sin\ga&\cos\ga&0\\0&0&1 \end{array}\right)
\left(\begin{array}{ccc} 
\cos\be&0&-\sin\be\\0&1&0\\\sin\be&0&\cos\be \end{array}\right)
\notag \\ 
& \quad\quad\quad\times
\left(\begin{array}{ccc} 
\cos\al&\sin\al&0\\-\sin\al&\cos\al&0\\0&0&1 \end{array}\right)
\left(\begin{array}{c} x\\y\\z \end{array}\right) .
\label{R}
\end{align}
Note that this result implies that the net rotation 
can alternatively be viewed as a rotation 
by $\al$ about $z$,
followed by a rotation by $\be$
about the rotated $y$ axis, 
and then by a rotation
by $\ga$ about the rotated $z$ axis. 

To determine the rotation rules for spherical harmonics,
consider a spin-weighted function ${}_s f$
expanded in both coordinate systems:
\beq
{}_s f=\sum_{jm} f_{jm}\, \syjm{s}{jm}(\Om)
= \sum_{jm} f'_{jm}\, \syjm{s}{jm}(\Om') .
\eeq
The spherical harmonics are related by 
\beq
\syjm{s}{jm}(\Om') = R\ \syjm{s}{jm}(\Om) ,
\eeq
which implies the transformation
\begin{align}
f_{jm} &=  
\sum_{m'} D^{(j)}_{mm'}(\al,\be,\ga) f'_{jm'} , 
\notag \\
f'_{jm} &=  
\sum_{m'} D^{(j)}_{mm'}(-\ga,-\be,-\al) f_{jm'} .
\label{Rcoeffs}
\end{align}
In these expressions, 
the quantities
\beq
D^{(j)}_{mm'}(\al,\be,\ga) = 
\int \syjm{s}{jm}^* 
e^{-i\al J_z}e^{-i\be J_y}e^{-i\ga J_z}\,
\syjm{s}{jm'}\, d\Om 
\eeq
are the Wigner rotation matrices
\cite{wigner}.
It can be shown that these matrices are independent of $s$,
which is to be expected because 
$\mbf J$ commutes with the helicity operator.
Consequently, 
all spherical coefficients are rotated
using the same set of matrices,
regardless of their spin weight.

The Wigner matrices are often written in the form
\beq
D^{(j)}_{mm'}(\al,\be,\ga) = 
e^{-i\al m} e^{-i\ga m'}\, d^{(j)}_{mm'}(\be) , 
\eeq
where $d^{(j)}_{mm'}(\be)=D^{(j)}_{mm'}(0,\be,0)$
are called the little Wigner matrices.
The phases in this equation correspond
to the two rotations about the $z$ axis,
while $d^{(j)}_{mm'}$ accounts for the rotation about $y$.
Explicitly, 
the little-matrix elements are given by
\begin{align}
d^{(j)}_{mm'}(\be) &= 
\sum_k 
\scriptstyle{(-1)^{k+m+m'}}
\Frac{\sqrt{(j+m)!(j-m)!(j+m')!(j-m')!}}
{(j-m-k)!(m-m'+k)!(j+m'-k)!k!}
\notag \\ 
&\qquad\qquad\quad
\times
\big(\cos\Frac\be2\big)^{2j}
\big(\tan\Frac\be2\big)^{2k+m-m'} ,  
\end{align}
where the sum is restricted to all $k$ 
for which the arguments of the factorials are nonnegative.

\subsection{Laboratory frame}

We next apply the Wigner matrices to 
relate the coefficients for Lorentz violation
in the laboratory frame to those in the Sun-centered frame.
For this purpose,
it is useful to work with a canonical laboratory frame
\cite{km}.
Cartesian coordinates in this frame are denoted $(x,y,z)$.
The $z$ direction is directed towards the zenith,
and the $x$ axis lies 
at an angle $\ph$ measured east of south.
The colatitude of the laboratory is denoted $\ch$.
The orientation of the laboratory
with respect to the Sun-centered coordinates $(X,Y,Z)$
is determined by the local sidereal time $T_\oplus$.
Since the $Z$ axis points towards the celestial north pole
while the $X$ and $Y$ axes lie in the equatorial plane
with right ascension $0^\circ$ and $90^\circ$,
respectively,
it follows that the laboratory $z$ axis 
points toward right ascension 
$\om_\oplus T_\oplus$,
where $\om_\oplus\simeq 2\pi/(\rm 23\ hr\ 56\ min)$
is the Earth's sidereal frequency.

With these conventions,
the angles $\ph$, $\ch$, and $\om_\oplus T_\oplus$
represent three Euler angles giving the relevant net rotation
between frames.
From the perspective of the Sun-centered frame,
the laboratory frame is obtained
by rotating by $\ph$ about $Z$, 
then by $\ch$ about $Y$,
and lastly by $\om_\oplus T_\oplus$ about $Z$.
The Euler angles relating the Sun-centered frame
to the laboratory frame are therefore
\beq
\al=\om_\oplus T_\oplus,
\quad
\be=\ch,
\quad
\ga=\ph.
\eeq
It follows that the explicit rotation 
relating the two sets of coordinates is 
\begin{align}
\left(\begin{array}{c} x\\y\\z \end{array}\right) &=
\left(\begin{array}{ccc} 
\cos\ph&\sin\ph&0\\-\sin\ph&\cos\ph&0\\0&0&1 \end{array}\right)
\left(\begin{array}{ccc} 
\cos\ch&0&-\sin\ch\\0&1&0\\\sin\ch&0&\cos\ch \end{array}\right)
\notag \\ & \times
\left(\begin{array}{ccc} 
\cos\om_\oplus T_\oplus&\sin\om_\oplus T_\oplus&0\\
-\sin\om_\oplus T_\oplus&\cos\om_\oplus T_\oplus&0\\ 0&0&1 
\end{array}\right)
\left(\begin{array}{c} X\\Y\\Z \end{array}\right) .
\label{Rlab}
\end{align}
The spherical coefficients in the laboratory frame 
can now be expressed as Sun-frame coefficients
through the relation
\begin{align}
{\cal K}_{jm}^{\rm lab}
&= \sum_{m'} D^{(j)}_{mm'}(-\ph,-\ch,-\om_\oplus T_\oplus)\,
{\cal K}_{jm'}^{\rm Sun} 
\notag \\
&= \sum_{m'} e^{im\ph}e^{im'\om_\oplus T_\oplus} 
d^{(j)}_{mm'}(-\ch)\,
{\cal K}_{jm'}^{\rm Sun} ,
\label{k_rot1}
\end{align}
where ${\cal K}_{jm}^{\rm lab}$
and ${\cal K}_{jm}^{\rm Sun}$
represent arbitrary spherical coefficients for Lorentz violation 
in the laboratory and Sun-centered frames,
respectively.

The result \rf{k_rot1} is auspicious.
A key signal in many experiments is the sidereal variation 
introduced by the rotation of the Earth.
Here, 
this rotation is expressed in a simple form 
involving time-dependent phases 
$e^{im\om_\oplus T_\oplus}$.
Only the colatitude $\ch$ appears in the little Wigner matrices
$d^{(j)}_{m'm}(-\ch)$,
which are time independent.
Consequently,
for typical applications these time-independent factors
need be calculated only once for a given experiment 
at fixed $\ch$.

Some experiments involve turntables
rotating about the vertical axis.
This situation can be incorporated 
into the above rotation 
by fixing the laboratory frame 
with respect to the turntable.
This implies the azimuthal angle acquires a time dependence
of the form $\ph=\om_{\rm tt} T_{\rm tt}$,
where $\om_{\rm tt}$ is the turntable rotation frequency 
and $T_{\rm tt}$ 
is measured from a time when the $x$ axis points south.
Again, 
the time dependence enters through simple phases.

As a simple example,
consider the rotations of vector coefficients
in an experiment involving a turntable.
Calculating the Wigner matrices for $j=1$,
we find that the rotation between
spherical coefficients in the laboratory and Sun-centered frames
is given by
\begin{align}
\left(\begin{array}{c}
{\cal K}_{11}^{\rm lab}\\
{\cal K}_{10}^{\rm lab}\\
{\cal K}_{1(-1)}^{\rm lab}
\end{array}\right) &=
\left(\begin{array}{ccc}
e^{i\om_{tt} T_{tt}} & 0 & 0 \\
0 & 1 & 0 \\
0 & 0 & e^{-i\om_{tt} T_{tt}}
\end{array}\right)
\notag \\
& \hspace*{-15pt} \times
\left(\begin{array}{ccc}
\cos^2\Frac\ch2 & -\Frac1{\sqrt2}\sin\ch & \sin^2\Frac\ch2 \\
\Frac1{\sqrt2}\sin\ch & \cos\ch & -\Frac1{\sqrt2}\sin\ch \\
\sin^2\Frac\ch2 &  \Frac1{\sqrt2}\sin\ch & \cos^2\Frac\ch2 \\
\end{array}\right)
\notag \\
& \hspace*{-15pt} \times
\left(\begin{array}{ccc}
e^{i\om_\oplus T_\oplus} & 0 & 0 \\
0 & 1 & 0 \\
0 & 0 & e^{-i\om_\oplus T_\oplus}
\end{array}\right)
\left(\begin{array}{c}
{\cal K}_{11}^{\rm Sun}\\
{\cal K}_{10}^{\rm Sun}\\
{\cal K}_{1(-1)}^{\rm Sun}
\end{array}\right) 
\end{align}
in matrix form.

In some situations,
it may be convenient to define a third frame 
that is fixed with respect to the apparatus.
The advantage of this third frame 
is that it may be chosen to simplify calculations.
For example,
a laboratory apparatus often has a symmetry axis,
so adopting a third apparatus frame 
with one coordinate axis 
along the symmetry direction may be convenient.
To incorporate this in the above formalism, 
it suffices to identify a suitable apparatus frame 
and to determine the corresponding Euler angles 
relating it to the laboratory frame.
The laboratory-frame and apparatus-frame spherical coefficients 
are then related through Eq.\ \rf{Rcoeffs}.
Thus,
if the apparatus-frame coordinates $(x',y',z')$ 
are related to laboratory-frame coordinates by 
\begin{align}
\left(\begin{array}{c} x'\\y'\\z' \end{array}\right) &=
\left(\begin{array}{ccc} 
\cos\ga&\sin\ga&0\\-\sin\ga&\cos\ga&0\\0&0&1 \end{array}\right)
\left(\begin{array}{ccc} 
\cos\be&0&-\sin\be\\0&1&0\\\sin\be&0&\cos\be \end{array}\right)
\notag \\ 
& \quad\quad\quad\times
\left(\begin{array}{ccc} 
\cos\al&\sin\al&0\\-\sin\al&\cos\al&0\\0&0&1 \end{array}\right)
\left(\begin{array}{c} x\\y\\z \end{array}\right) ,
\end{align}
where $\al$, $\be$, $\ga$ are appropriate Euler angles,
then the apparatus-frame spherical coefficients
can be written in terms of the
laboratory-frame spherical coefficients through
\beq
{\cal K}^{\rm app}_{jm} =  \sum_{m'}
D^{(j)}_{mm'}(-\ga,-\be,-\al) {\cal K}^{\rm lab}_{jm'} .
\label{k_rot2}
\eeq
Assuming the orientation
of the system is fixed in the
laboratory frame,
the Wigner matrices for this
rotation are constant.

As a simple illustration,
consider a system with symmetry axis 
oriented along the $x$ axis of the laboratory frame.
In this case,
it may be beneficial to choose an apparatus frame 
having angular-momentum projection axis $z'$ 
aligned with the symmetry axis.
This can be achieved by taking $x'=-z$, $y'=y$, and $z'=x$.
A suitable choice of Euler angles
is then given by $\al=0$, $\be=90^\circ$, and $\ga=0$.

\section{Astrophysical tests}
\label{sec_astro}

In this section,
we discuss searches for Lorentz violation
involving observation of radiation 
from sources at cosmological distances.
Due to the large baselines involved,
searches for birefringent and dispersive effects
can in principle achieve high sensitivities
to almost all the vacuum coefficients 
$\kI$, $\kE$, $\kB$, and $\kV$
introduced in Sec.\ \ref{sec_vac}.

Where relevant,
vacuum birefringence provides considerably greater
sensitivity than vacuum dispersion,
as is shown below.
It is therefore natural to separate 
the relevant Lorentz-violating operators into two classes.
The first class is controlled by
the vacuum coefficients $\kE$, $\kB$, $\kV$
and produces leading-order birefringence.
The second consists of vacuum operators 
causing dispersion without leading-order birefringence,
and is associated with the coefficients $\kI$ for $d>4$.
Observations of dispersion
are therefore well suited to measurements of $\kI$,
while studies of birefringence 
are appropriate for measurements of the remaining coefficients.

We begin in Sec.\ \ref{sec_disp}
with a discussion of dispersion tests.
The basic theory and results are summarized,
and new constraints are obtained 
using recent results.
In Sec.\ \ref{sec_bire},
we consider birefringence tests.
A general treatment is first outlined,
and then applications to point sources
and to the cosmic microwave background (CMB)
are presented.
We obtain new constraints from gamma-ray bursts
on coefficients of mass dimensions five, seven, and nine,
and we discuss some general features 
of the effects of Lorentz violation on the CMB.

\subsection{Dispersion tests}
\label{sec_disp}

Frequency-dependent photon velocities
arise from Lorentz-violating operators with $d\neq 4$,
which cause wave dispersion.
Astrophysical searches for vacuum dispersion 
seek these differences in the velocity of light
at different wavelengths.
Typical searches involve explosive or pulsed sources of radiation,
such as gamma-ray bursts, pulsars, or blazars,
that produce light over a wide range of wavelengths 
in a short period of time.
Assuming the temporal structure of the emission 
is sufficiently well understood,
observed but unexpected arrival-time differences
can be interpreted as wavelength dependences in the velocity.
Even if detailed frequency information is unavailable,
limits can still be obtained from the pulse width
because dispersion results in a spreading of wave packets.

A number of astrophysical searches
for a modified photon dispersion relation 
have been performed.
Most of these studies assume isotropic Lorentz violation.
To make the connection between these approaches and the SME,
we momentarily restrict our attention 
to the vacuum isotropic model
discussed in Sec.\ \ref{sec_vac},
which is the relevant limit for astrophysical studies
of isotropic violations.
Recall that this model 
has exactly one nonzero spherical coefficient
for Lorentz violation at each $d$,
consisting of $\kIdjm{d}{00}$ for even $d$ 
and $\kVdjm{d}{00}$ for odd $d$.
In the isotropic limit,
the velocity defect arising from Eq.\ \rf{dr3} is given by 
\beq
\de v \simeq
\frac{1}{\sqrt{4\pi}} \sum_d E^{d-4}
\big(- \kIdjm{d}{00} \pm \kVdjm{d}{00}\big) , 
\eeq
in terms of the photon energy $E$.
The coefficients $\kIdjm{d}{00}$ 
are associated with CPT-even operators
producing dispersion but no leading-order birefringence,
while nonzero coefficients $\kVdjm{d}{00}$ 
imply birefringence and corresponding changes in polarization.
Among the studies involving
modified dispersion relations with isotropic Lorentz violation,
it follows that only those with odd $d$ and birefringence
or even $d$ and no birefringence
are consistent with linear effective field theory 
in flat spacetime
\cite{rl}.

Isotropic Lorentz-violating effects 
in modified dispersion relations
are sometimes described using a velocity deviation
of the form $\de v = \pm \xi_1 E$,
where $\xi_1$ is a constant
\cite{rmmp,jlms}.
This model is phenomenologically equivalent
to the single SME coefficient
$\kVdjm{5}{00} = \sqrt{4\pi}~\xi_1$.
However,
because this particular combination causes birefringence,
its best constraints currently come from polarimetry observations,
which are discussed in the next subsection.
An isotropic higher-order correction of the form 
$\de v = \xi_2 E^2$ has also been considered
\cite{xi2}.
This case corresponds to the $d=6$ coefficient
$\kIdjm{6}{00} = -\sqrt{4\pi}~\xi_2$.
Bounds on this term from the active galaxy Markarian 501 
are currently of order $10^{-21}$ GeV$^{-2}$
\cite{km_apjl},
though some evidence for nonzero
dispersion from this source exists
\cite{magic}.
Other cases that are sometimes considered involve
an isotropic nonbirefringent linear defect 
$\de v = \xi_1 E$
\cite{aemns} 
or an isotropic birefringent defect 
$\de v = \pm\xi_2 E^2$,
where the sign indicates helicity
\cite{im}.
Both these cases are inconsistent 
with the present general analysis 
and may be problematic.

In principle,
searches for Lorentz violation via dispersion
are sensitive to all coefficients with $d\neq 4$.
However,
considerably greater sensitivities 
to coefficients associated with birefringence
are typically accessible via polarimetry.
This can be understood as follows. 
In a vacuum dispersion study involving a source
at baseline distance $L$,
the quantity of interest is the change 
$\de t \simeq \de v L$ in arrival time
of the signal,
which implies a sensitivity to $\de v$
given by $\de t / L$.
In contrast,
for a polarimetric study of the same astrophysical source,
the quantity of interest is the phase difference
$\de\ph \simeq E \de v L$ of the eigenmodes,
which yields sensitivity $\de\ph/LE$.
Comparing these two results,
we see that dispersive sensitivity 
depends on the difference in arrival time,
while polarimetric sensitivity depends 
on the periodicity $\propto 1/E$ of the source radiation.
Consequently,
to achieve similar sensitivity,
an astrophysical dispersion test requires a time resolution 
comparable to the inverse frequency of the photons,
which is infeasible.

We can therefore conclude that astrophysical dispersion studies 
are best suited to searches for the 
nonbirefringent vacuum coefficients $\kI$.
Note,
however,
that birefringence also causes 
a spreading of wave packets,
due to the differences in velocity 
of the two birefringent eigenmodes.
Moreover,
all birefringent operators with $d\neq 4$
are also dispersive.
A definitive interpretation of an observed arrival-time difference
as a dispersive nonbirefringent effect 
associated with the vacuum coefficients $\kI$
therefore requires a polarimetric study of the signal
or elimination of possible contributions
from the other vacuum coefficients
via independent studies. 

Setting all other coefficients to zero,
the velocity defect including anisotropies
is given by
\beq
\de v \simeq -\vs^0 = - \sum_{djm}
E^{d-4} \, \syjm{0}{jm}(\mbf{\hat n})\, \kI .
\eeq
Note that this involves only 
even-dimensional Lorentz-violating operators
and that the $d=4$ case involves no dispersion.
For operators with $d>4$,
the sensitivities increase with energy
and so high-frequency sources 
can be expected to yield the sharpest results.
Note also that limiting attention to isotropic dispersion
disregards a total of $(d^2-2d-2)$ independent types
of vacuum Lorentz violation at each $d$.

\begin{table*}
\renewcommand{\arraystretch}{1}
\begin{tabular}{c|c|c|c}
Model & Coefficients & Result & System \\ 
\hline
\hline
Vacuum
&
$\big|\sum_{jm} \syjm{0}{jm} (99.7^\circ,240^\circ) 
~ \kIdjm{6}{jm}\big|$ 
&
$< 1\times 10^{-16}$
GeV$^{-2}$
&
GRB 021206
\\
&
$\sum_{jm} \syjm{0}{jm} (50.2^\circ,253^\circ) ~\kIdjm{6}{jm}$
&
$3^{+1}_{-2}\times 10^{-22}$ 
GeV$^{-2}$
&
Markarian 501
\\
&
$\sum_{jm} \syjm{0}{jm}(147^\circ,120^\circ) ~\kIdjm{6}{jm}$
&
$< 3.2\times 10^{-20}$
GeV$^{-2}$
&
GRB 080916C
\\
&
$\big|\sum_{jm} \syjm{0}{jm}(330^\circ,-30^\circ) 
~\kIdjm{6}{jm}\big|$
&
$< 7.4\times 10^{-22}$ 
GeV$^{-2}$
&
PKS 2155-304 
\\ &&&
\\
&
$\big|\sum_{jm} \syjm{0}{jm} (99.7^\circ,240^\circ) 
~ \kIdjm{8}{jm}\big|$
&
$< 3\times 10^{-13}$
GeV$^{-4}$
&
GRB 021206
\\
&
$\sum_{jm} \syjm{0}{jm}(147^\circ,120^\circ) ~\kIdjm{8}{jm}$
&
$< 2.6\times 10^{-23}$ 
GeV$^{-4}$
&
GRB 080916C
\\
\hline
Vacuum isotropic 
&
$\big|\kIdjm{6}{00}\big|$
&
$< 4 \times 10^{-16}$
GeV$^{-2}$
&
GRB 021206
\\
&
$\kIdjm{6}{00}$
&
$10^{+4}_{-7}\times 10^{-22}$
GeV$^{-2}$
&
Markarian 501
\\
&
$\kIdjm{6}{00}$ 
&
$< 1.1\times 10^{-19}$ 
GeV$^{-2}$
&
GRB 080916C
\\
&
$\big|\kIdjm{6}{00}\big|$ 
&
$< 2.6\times 10^{-21}$ 
GeV$^{-2}$
&
PKS 2155-304 
\\ &&&
\\
&
$\big| \kIdjm{8}{00} \big|$
&
$< 9 \times 10^{-13}$
GeV$^{-4}$
&
GRB 021206
\\
&
$\kIdjm{8}{00}$ 
&
$< 9.2\times 10^{-23}$ 
GeV$^{-4}$
&
GRB 080916C
\\
\hline
\hline
\end{tabular}
\caption{\label{dispersionbounds}
Constraints on spherical coefficients
from astrophysical dispersion studies.
The first five rows give constraints
on the vacuum coefficients
with $d = 6,8$. 
The next five rows give the constraints
on coefficients in the isotropic limit,
for which there is exactly one nonzero coefficient
for each $d$.
Except for the limits from GRB 080916C
and PKS 2155-304,
which are obtained in the text,
all results in the table are taken from
the analysis of Ref.\ \cite{km_apjl},
which used data from Refs.\ \cite{magic} and \cite{bwhc}.
The bounds shown are at the 95\% confidence level. }
\end{table*}

The difference in velocity
between photons of different energies
leads to an arrival-time difference given by
\cite{km_apjl}
\begin{align}
t_2-t_1 &\approx \int_0^z \frac{v_1-v_2}{H_z}dz 
\notag \\
&
\hskip -10pt
\approx 
(E_2^{d-4}-E_1^{d-4}) 
\int_0^z \frac{(1+z)^{d-4}}{H_z}dz
\sum_{jm}
\syjm{0}{jm}\, \kI ,
\label{disp}
\end{align}
where the source redshift is $z$
and  $t_1$, $t_2$ are the propagation times 
for photons with observed energies $E_1$, $E_2$
and velocities $v_1$, $v_2$.
Also,
\beq
H_z = H_0[\Om_r(1+z)^4 + \Om_m(1+z)^3
+\Om_k(1+z)^2+\Om_\La]^{1/2}
\eeq
is the Hubble expansion rate at $z$,
expressed in terms of
the present-day Hubble constant 
$H_0\simeq 71 \mbox{ km s}^{-1}\mbox{Mpc}^{-1}$,
radiation density $\Om_r\simeq 0.015$,
matter density $\Om_m\simeq 0.27$,
vacuum density $\Om_\La\simeq 0.73$, 
and curvature density 
$\Om_k = 1-\Om_r-\Om_m-\Om_\La$.
In Ref.\ \cite{km_apjl},
these expressions are used to place
direction-dependent bounds
on combinations of the coefficients 
$\kIdjm{6}{jm}$ and $\kIdjm{8}{jm}$,
using observations of GRB 021206
\cite{bwhc}
and of the blazar Markarian 501
\cite{magic}.
These results are summarized in
Table \ref{dispersionbounds}.

As an illustration,
consider the recent measurements 
made by the Fermi Observatory
on the source GRB 080916C 
\cite{fermi}.
This exceptionally energetic source 
produced a burst of photons 
with observed energies ranging to 
$13.22^{+0.70}_{-1.54}$ GeV,
all of which arrived within 16.54 s 
of the initial detection of low-energy photons.
The high photon energies and the large redshift 
of $z=4.35\pm0.15$
make this burst a sensitive probe of Lorentz violation.
A conservative bound on 
the vacuum coefficients for Lorentz violation 
can be obtained using the $2\si$ lower limits
for the energy and redshift.
Performing the integral \rf{disp} for $d=6$
and assuming the lower energy is negligible,
we find the constraint 
\beq
\sum_{jm}
\syjm{0}{jm}(147^\circ,120^\circ)\, \kIdjm{6}{jm}
< 3.2\times 10^{-20} \mbox{ GeV}^{-2} 
\eeq
on a direction-dependent combination of 
operators for Lorentz violation of mass dimension $d=6$.
Taking instead operators of mass dimension $d=8$,
we obtain the constraint
\beq
\sum_{jm}
\syjm{0}{jm}(147^\circ,120^\circ)\,  \kIdjm{8}{jm}
< 2.6\times 10^{-23} \mbox{ GeV}^{-4} .
\eeq
Note that these are one-sided bounds that
suppose the higher-energy photons propagate more slowly
and that disregard possible burst-timing structure 
from the source.
Under the assumption that the observed time difference
is due to Lorentz-violating effects,
a careful study of the leading edge
of the high-energy photons
might also permit the derivation of a lower positive bound
for each of the above coefficient combinations. 

Another example is provided by the recent data 
obtained for the active galaxy PKS 2155-304 
by the High Energy Stereoscopic System (HESS)
\cite{hess}.
This source has redshift $z=0.116$
and a light curve spanning an energy range of a few TeV,
with time delays of a few tens of seconds.
The reported analysis places a constraint
of 41 s TeV$^{-2}$ 
at the 95\% confidence level
on dispersion effects quadratic in the energy.
Performing the integral \rf{disp} as before
and identifying the reported constraint
with $(t_2-t_1)/(E_2^2 - E_1^2)$
implies a conservative bound 
on a direction-dependent combination 
of vacuum coefficients for Lorentz violation with $d=6$.
We obtain 
\beq
\Big| 
\sum_{jm}
\syjm{0}{jm}(330^\circ,-30^\circ)\, \kIdjm{6}{jm}
\Big| 
< 7.4\times 10^{-22} \mbox{ GeV}^{-2} 
\eeq
at the 95\% confidence level.
A more complete study of the existing data from this source
could yield additional constraints
involving operators of mass dimension $d\geq 8$.

The above analyses demonstrate that 
a single point source provides sensitivity 
to only a limited number of direction-dependent combinations
of coefficients for Lorentz violation.
Multiple sources are needed to access all coefficients 
for a given value of $d$.
For example,
there are 25 coefficients $\kIdjm{6}{jm}$ 
and 49 coefficients $\kIdjm{8}{jm}$,
and so even when birefringence is neglected
we see that a corresponding number of sources 
at different locations on the sky 
is required to constrain fully these coefficients
and the corresponding types of Lorentz violation.
Data from gamma-ray bursts and other burst sources
obtanied by existing telescopes,
including 
Fermi,
HESS,
the Major Atmospheric Gamma-ray Imaging Cherenkov Telescope
(MAGIC)
\cite{magic},
and 
the Very Energetic Radiation Imaging Telescope Array System
(VERITAS)
\cite {veritas},
or by future telescopes
such as
the Advanced Gamma-ray Imaging System
(AGIS)
\cite{agis},
the Cherenkov Telescope Array
(CTA)
\cite{cta},
and the High Altitude Water Cherenkov Experiment
(HAWC)
\cite{hawc},
could be combined 
to measure completely the coefficients $\kI$ 
for various fixed values of $d\geq 6$.

In contrast,
in the limit of rotation invariance
we recover the vacuum isotropic model,
which reduces the number of coefficients 
to one for each $d$.
This implies that a single source suffices
to place constraints when only one value of $d$ is considered
at a time.
For example,
in the vacuum isotropic model
a single constraint $<10^{-22}$ GeV$^{-2}$
in any location on the sky
suffices to exclude 
the suggestion of a signal for $d=6$ Lorentz violation 
from Markarian 501,
whereas the general SME treatment
requires at least 25 independent sources
at this constraint level.
For the limiting case of the vacuum isotropic model,
the above bounds from Fermi reduce to the one-sided constraints 
\begin{align}
\kIdjm{6}{00} &< 1.1\times 10^{-19} \mbox{ GeV}^{-2} ,
\end{align}
and
\begin{align}
\kIdjm{8}{00} &< 9.2\times 10^{-23} \mbox{ GeV}^{-4} ,
\end{align}
while the one from HESS reduces to
\begin{align}
\big| \kIdjm{6}{00}\big| &< 2.6\times 10^{-21} \mbox{ GeV}^{-2} .
\end{align}
Both the isotropic and the anisotropic constraints
obtained from GRB 080916C and PKS 2155-304 
are also included in Table \ref{dispersionbounds}.

\subsection{Birefringence tests}
\label{sec_bire}

In birefringent scenarios,
the two eigenmodes propagate at slightly different velocities.
This implies that the superposition of the modes
is altered as light propagates in free space.
Since the two modes differ in polarization, 
the change in superposition causes a change 
in the net polarization of the radiation.
This provides a signature of Lorentz violation.
In the present subsection,
we outline the theory of these polarization changes
and discuss birefringence tests based on polarimetry
using both point sources and the CMB.

\subsubsection{Theory}
\label{bireftheory}

The direction of a Stokes vector 
$\mbf s = (s^1,s^2,s^3)^T$
in the abstract Stokes-parameter space
uniquely characterizes the polarization
of the associated plane wave
\cite{bw}.
A Stokes vector in the $s^1$-$s^2$ plane 
corresponds to linear polarization,
while a Stokes vector along the $s^3$ axis
represents circular polarization.
Other directions represent
general elliptical polarizations.

In this picture,
birefringence can be understood as 
a rotation of the Stokes vector
$\mbf s = (s^1,s^2,s^3)^T$
about the birefringence axis 
$\mbf\vs = (\vs^1,\vs^2,\vs^3)^T$.
The birefringence axis represents 
the polarization of the eigenmodes
and is determined by the properties of the medium.
The total angle of rotation 
is equivalent to the change
in the relative phase between the two eigenmodes.
With the normalization adopted in Eq.\ \rf{vac_exp},
we can write this rotation in differential form:
\beq
d\mbf s/dt = 2E\mbf\vs\times\mbf s = -i\Si\cdot\mbf s ,
\label{stokes_rot}
\eeq
where
$E$ is the photon energy 
and $\Si^{jk} = -2iE\ep^{jkl}\vs^l$
represents the rotation generators.

Some basic features of birefringence
in the context of Lorentz violation
can be extracted from this picture
\cite{km}.
If CPT is conserved,
then $\mbf\vs$ lies in the $s^1$-$s^2$ plane,
and so the birefringent eigenmodes are linearly polarized.
Linearly polarized radiation therefore typically 
rotates out of the $s^1$-$s^2$ plane 
and becomes elliptically polarized.
Similarly, 
circularly polarized radiation rotates away from the $s^3$ axis,
becomes elliptical, 
and may eventually rotate through a linear polarization.
In contrast, 
if CPT is violated,
then the birefringent eigenmodes are circularly polarized
with one being left-handed and the other right-handed.
The rotation axis $\mbf\vs$ is therefore 
aligned with the $s^3$ axis in this case.
As a result,
linear polarizations remain linear,
but a change in the linear-polarization angle occurs.
However,
circular polarizations remain circular
because they are eigenmodes of the propagation.

In a typical application,
one considers a distant source of polarized light 
and integrates the rotation \rf{stokes_rot}
from emission to detection.
This yields changes in polarization
that can depend on both energy and direction of propagation.
To search for birefringence,
we can either model the polarization at the source 
and seek discrepancies in the observed polarization,
or we can test for unexpected energy dependence
in the polarization parameters.

Since we are characterizing Lorentz violation
using spin-weighted spherical coefficients,
it is convenient to reformulate the rotation 
of the Stokes vector
in terms of spin-weighted Stokes parameters.
We can decompose the components of the Stokes vector
according to their spin weight 
with respect to the line of sight $\mbf{\hat n}$.
This yields $s_{(0)} = s^3$
as a Stokes parameter of spin weight $0$ 
and $s_{(\pm2)} = s^1\mp i s^2$
as two Stokes parameters having spin weight $\pm 2$.

In the Stokes basis with 
$\mbf s = (s_{(+2)},s_{(0)},s_{(-2)})^T$,
the rotation generators are given in matrix form as
\beq
\Si = 2E \left(\begin{array}{ccc}
\vs_{(0)} & -\vs_{(+2)} & 0 \\
-\half\vs_{(-2)} & 0 & \half\vs_{(+2)} \\
0 & \vs_{(-2)} & -\vs_{(0)} 
\end{array}\right) .
\label{gen}
\eeq
Here,
$\mbf \vs = (\vs_{(+2)},\vs_{(0)},\vs_{(-2)})^T$
is the birefringence axis in this basis,
with components given by
\begin{align}
\vs_{(\pm2)} &= 
\sum_{djm} E^{d-4}\, \syjm{\pm2}{jm}(\mbf{\hat n})\,
\big(\kE\pm i\kB\big) ,
\notag \\
\vs_{(0)}&= 
\sum_{djm} E^{d-4}\, \syjm{0}{jm}(\mbf{\hat n})\, 
\kV .
\label{bire_stokes}
\end{align}
Since the rotation of the Stokes vector 
is determined by the combination $E\mbf\vs$,
the effects of birefringence 
enter with an energy dependence of $E^{d-3}$.
This implies that an increased sensitivity 
to coefficients with $d>3$ 
can be achieved by using higher-energy photons.
We also see that unconventional energy dependence
is a signal for Lorentz violation.
Only the $d=3$ case leads to
energy-independent birefringence.

For many astrophysical sources,
cosmological expansion is significant
during the time of flight 
and must be incorporated in the analysis.
We implement this by expressing
the differential rotation in terms of redshift:
\beq
d\mbf s = \frac{i\Si_z\cdot\mbf s}{(1+z)H_z}\ dz ,
\label{dstokes}
\eeq
where $\Si_z$ represents the rotation
matrix at the blueshifted energy
$(1+z)E$ and source direction $\mbf{\hat n}$.
To obtain the net polarization change,
we then integrate this expression
from source redshift $z$ to 0.

Some searches for Lorentz violation
investigate Lorentz-violating operators
of a specified mass dimension $d$.
With this assumption,
the calculation of the net rotation is simplified 
because the energy integral 
is independent of the matrix multiplication.
In the CPT-odd case with a single odd value of $d$,
the rotations of the Stokes vector about the $s^3$ axis 
lead to a change in the linear-polarization angle $\psi$
without affecting the degree of linear or circular polarization.
The rotation is diagonal in the spin-weighted basis,
and we obtain the simple result
\begin{align}
s_{(\pm 2)} &= 
e^{\mp i2\de\Psi_z}\, s_{(\pm 2)z} , 
\notag \\
s_{(0)} &= s_{(0)z} ,
\end{align}
relating the present-day polarization
to the original polarization at redshift $z$.
Here,
the change $\de\Psi_z$ in polarization is given by the integral
\beq
\de\Psi_z=
E^{d-3} \int_0^z \frac{(1+z)^{d-4}}{H_z} dz
\sum_{jm} \syjm{0}{jm}(\mbf{\hat n})\ \kV .
\label{delta_psi}
\eeq
The linear-polarization angle $\psi$ at the present epoch
is then related through 
\beq
\psi = \psi_z + \de\Psi_z
\eeq
to the blueshifted angle $\psi_z$.

In the CPT-even case
with a single even value of $d$,
the eigenmodes are linearly polarized
and the rotation is more complicated.
It is convenient in this case
to define the direction-dependent phase
\beq
e^{-i\xi(\mbf{\hat n})} = 
\frac{\vs_{(+2)}(\mbf{\hat n})}{|\vs_{(+2)}(\mbf{\hat n})|} ,
\eeq
which controls the evolution of the polarization.
The phase angle $\xi$ is twice the polarization angle 
of the eigenmode of propagation.
Consequently,
linear polarizations with angle
$\psi=\xi/2$ or $\psi=\xi/2+90^\circ$ 
remain unaffected as the radiation propagates.
Calculation shows that the redshift integral 
can be expressed using this phase and the angle
\begin{align}
\Phi_z &= E^{d-3} \int_0^z \frac{(1+z)^{d-4}}{H_z} dz
\notag \\
&\qquad 
\times \bigg|\sum_{jm} \syjm{2}{jm}(\mbf{\hat n})\,
\big(\kE\pm i\kB\big)\bigg| .
\end{align}
The net rotation is given by 
\beq
\mbf s =m_z\cdot {\mbf s}_z,
\eeq
where the M\"uller matrix $m_z$ takes the form 
\beq
m_z =\left[\begin{array}{ccc}
\cos^2\Phi_z & -i\sin2\Phi_z e^{-i\xi} & \sin^2\Phi_z e^{-2i\xi}\\
-\frac{i}{2}\sin2\Phi_z e^{i\xi} 
  & \cos2\Phi_z & \frac{i}{2}\sin2\Phi_z e^{-i\xi}\\
\sin^2\Phi_z e^{2i\xi} & i\sin2\Phi_z e^{i\xi} & \cos^2\Phi_z
\end{array}\right] 
\eeq
in the spin-weighted Stokes basis.

\subsubsection{Point sources}
\label{sec_point_sources}

\begin{figure}
\begin{center}
\centerline{\psfig{figure=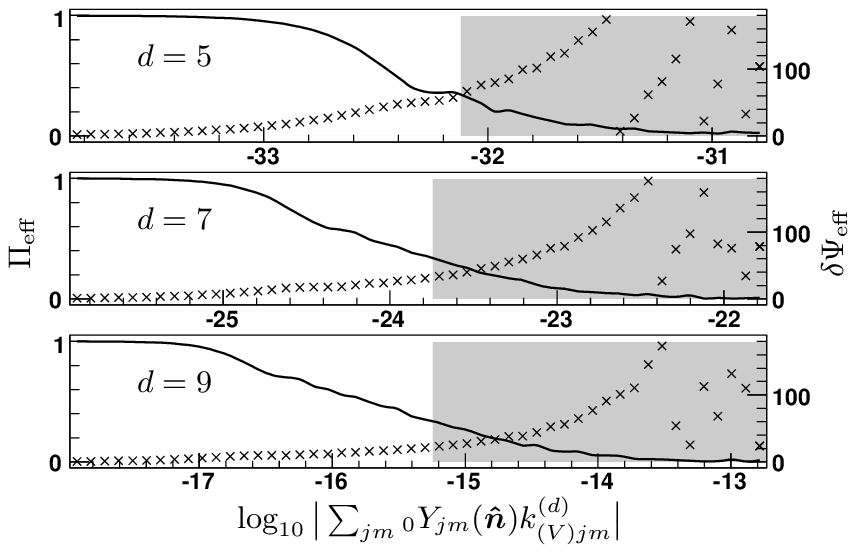,width=\hsize}}
\caption{\label{930131_odd}
Spectropolarimetry bounds from GRB 930131 
on spherical coefficients 
corresponding to Lorentz-violating operators 
of mass dimensions $d=5,7,9$.
The source direction is given by the angles 
$\mbf{\hat n} = (98.2^\circ,182.1^\circ)$.
The solid curve represents the
effective degree of polarization 
$\Pi_{\rm eff}=\sqrt{\vev{s^1}^2+\vev{s^2}^2}$.
The effective polarization angle 
$\psi_{\rm eff}=\tan^{-1}\vev{s^2}/2\vev{s^1}$
in degrees is displayed with crosses.
The shaded area is the disallowed region
with $\Pi_{\rm eff}<35\%$.
All coefficients are in units of GeV$^{4-d}$.}    
\end{center}
\end{figure}

\begin{figure}
\begin{center}
\centerline{\psfig{figure=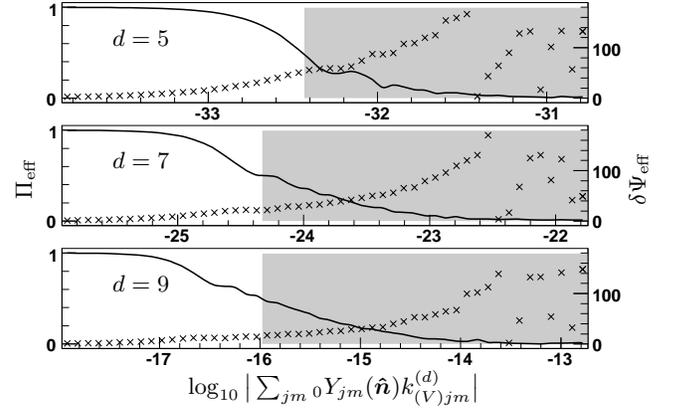,width=\hsize}}
\caption{\label{960924_odd}
Spectropolarimetry bounds from GRB 960924
on spherical coefficients 
corresponding to Lorentz-violating operators 
of mass dimensions $d=5,7,9$.
The source direction is given by the angles 
$\mbf{\hat n} = (87.3^\circ,37.3^\circ)$.
The solid curve represents the
effective degree of polarization 
$\Pi_{\rm eff}=\sqrt{\vev{s^1}^2+\vev{s^2}^2}$.
The effective polarization angle 
$\psi_{\rm eff}=\tan^{-1}\vev{s^2}/2\vev{s^1}$
in degrees is displayed with crosses.
The shaded area is the disallowed region
with $\Pi_{\rm eff}<50\%$.
All coefficients are in units of GeV$^{4-d}$.}    
\end{center}
\end{figure}

We next use the above theoretical results 
to obtain bounds on spherical coefficients 
from polarimetry of astrophysical point sources.
Some of the tightest existing constraints on Lorentz violation
have been achieved in birefringence searches 
involving high-frequency sources such as gamma-ray bursts
\cite{km_grb,im,jlms,grb_bire}.
However,
point sources have the disadvantage 
that a single line of sight $\mbf{\hat n}$ is involved,
which provides sensitivity to only a restricted portion 
of the space of coefficients for Lorentz violation.
As with astrophysical dispersion tests,
multiple sources are therefore required 
to perform a comprehensive search,
even for a fixed value of $d$.
For example,
for the case of $d=3$
a multiple-source search 
involving a large number of radio galaxies
\cite{cfj}
has placed a limit on a quantity $p_\al \equiv -2(\kafd{3})_\al$
corresponding to the constraint 
\beq
\Big| \sum_{jm} \syjm{0}{jm} \kVdjm{3}{jm} \Big| 
< 6\times 10^{-43} {~\rm GeV}
\eeq
at the 95\% confidence level
in terms of spherical coefficients.
In the isotropic limit,
this gives the limit 
$\kVdjm{3}{00} < 2\times 10^{-42}$ GeV,
although sharper bounds have recently emerged 
from CMB polarimetry as discussed below.
Multiple-source searches for the case of $d=4$
have also been performed
\cite{km,km_agn,km_grb,km_apjl}.
For example,
a search using 16 sources 
\cite{km}
places a limit that translates in the present context 
to the rotationally invariant constraint
\beq
\sqrt{
\sum_{m} 
(| \kEdjm{4}{2m} |^2 
+ | \kBdjm{4}{2m} |^2) } 
< 5\times 10^{-32} 
\eeq
at the 95\% confidence level.
For larger values of $d$,
multiple-source birefringence analyses
offer excellent prospects 
for systematic tests of Lorentz violation
at extreme sensitivity. 

In this subsection,
we illustrate the procedure and obtain first constraints
on some spherical coefficients with larger $d$,
by analyzing the recent evidence
for significant polarized components in the radiation 
from the gamma-ray bursts GRB 930131 and GRB 960924
\cite{willis}.
Observations of gamma rays 
associated with these two sources 
suggest that they are polarized at levels of
$\Pi_{930131}>35\%$ and $\Pi_{960924}>50\%$,
respectively.
Ideally, 
if we knew the degree of polarization
and polarization angles at the source,
we could search directly for changes due to birefringence.
Without this information,
however,
we can still place limits on decoherence effects 
caused by birefringence.
The point is that significant birefringence 
would lead to large differences in observed polarizations
at slightly different frequencies,
effectively unpolarizing the radiation.
Evidence for polarization can therefore be used 
to constrain the frequency-dependent birefringence 
caused by violations with dimension $d>3$.

\begin{table*}
\renewcommand{\arraystretch}{1}
\begin{tabular}{c|c|c|c}
Model & Coefficients & \quad GRB 930131 \quad & \quad GRB 960924 \quad \\ 
\hline
\hline
Vacuum
&
$\big| \sum_{jm} \syjm{0}{jm} (\mbf{\hat n}) ~\kVdjm{5}{jm} \big|$
&
$< 7 \times 10^{-33}$
GeV$^{-1}$
&
$< 4 \times 10^{-33}$
GeV$^{-1}$
\\
&
$\big| \sum_{jm} \syjm{0}{jm} (\mbf{\hat n}) ~\kVdjm{7}{jm} \big|$
&
$< 2 \times 10^{-24}$
GeV$^{-3}$
&
$< 5 \times 10^{-25}$
GeV$^{-3}$
\\
&
$\big| \sum_{jm} \syjm{0}{jm} (\mbf{\hat n}) ~\kVdjm{9}{jm} \big|$
&
$< 6 \times 10^{-16}$
GeV$^{-5}$
&
$< 1 \times 10^{-16}$
GeV$^{-5}$
\\
\hline
Vacuum isotropic 
&
$\big| \kVdjm{5}{00} \big|$
&
$< 2 \times 10^{-32}$
GeV$^{-1}$
&
$< 1 \times 10^{-32}$
GeV$^{-1}$
\\
&
$\big| \kVdjm{7}{00} \big|$
&
$< 7 \times 10^{-24}$
GeV$^{-3}$
&
$< 2 \times 10^{-24}$
GeV$^{-3}$
\\
&
$\big| \kVdjm{9}{00} \big|$
&
$< 2 \times 10^{-15}$
GeV$^{-5}$
&
$< 4 \times 10^{-16}$
GeV$^{-5}$
\\
\hline
\hline
Vacuum
&
$\big| \sum_{jm}\,_2Y_{jm} (\mbf{\hat n}) 
~\big(\kEdjm{4}{jm} + i\kBdjm{4}{jm}\big) \big|$
&
$\lsim 10^{-37}$
&
$\lsim 10^{-37}$
\\
&
$\big| \sum_{jm}\,_2Y_{jm} (\mbf{\hat n}) 
~\big(\kEdjm{6}{jm} + i\kBdjm{6}{jm}\big) \big|$
&
$\lsim 10^{-29}$
GeV$^{-2}$
&
$\lsim 10^{-29}$
GeV$^{-2}$
\\
&
$\big| \sum_{jm}\,_2Y_{jm} (\mbf{\hat n}) 
~\big(\kEdjm{8}{jm} + i\kBdjm{8}{jm}\big) \big|$
&
$\lsim 10^{-20}$
GeV$^{-4}$
&
$\lsim 10^{-20}$
GeV$^{-4}$
\\
\hline
\hline
\end{tabular}
\caption{\label{pointbounds}
Constraints on spherical coefficients
from polarization observations 
of the gamma-ray bursts GRB 930131 and GRB 960924.
The first three rows give constraints
on the vacuum coefficients
for the CPT-odd cases with $d = 5,7,9$. 
The arguments of the spherical harmonics
are $\mbf{\hat n} = (98.2^\circ,182.1^\circ)$ 
for GRB 930131
and $\mbf{\hat n} = (87.3^\circ,37.3^\circ)$ 
for GRB 960924.
The next three rows give the constraints
on coefficients in the isotropic limit,
for which there is exactly one nonzero coefficient
for each $d$.
The final three rows give the approximate sensitivities
achieved for vacuum coefficients 
in the CPT-even cases with $d = 4,6,8$.
Unlike the constraints in the first six rows,
the results in the final three rows
cannot be interpreted as definitive bounds
because the amount of birefringence in the CPT-even case 
depends on details of the source polarization. 
All constraints are at the 95\% confidence level.}
\end{table*}

\begin{figure}
\begin{center}
\centerline{\psfig{figure=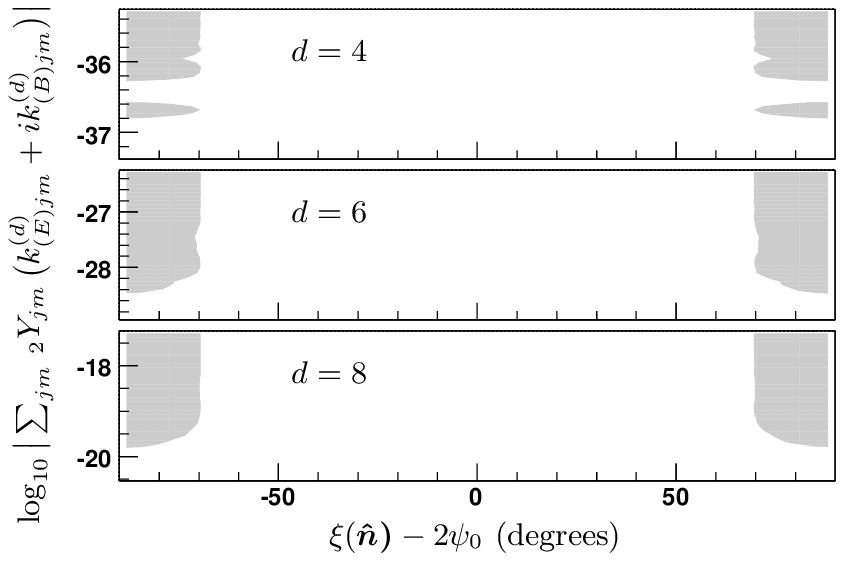,width=\hsize}}
\caption{\label{930131_even}
Spectropolarimetry bounds from GRB 930131 
on spherical coefficients 
corresponding to Lorentz-violating operators 
of mass dimensions $d=4,6,8$.
The source direction is given by the angles 
$\mbf{\hat n} = (98.2^\circ,182.1^\circ)$.
The shaded area is the disallowed region
with $\Pi_{\rm eff}<35\%$.
All coefficients are in units of GeV$^{4-d}$.}    
\end{center}
\end{figure}

\begin{figure}
\begin{center}
\centerline{\psfig{figure=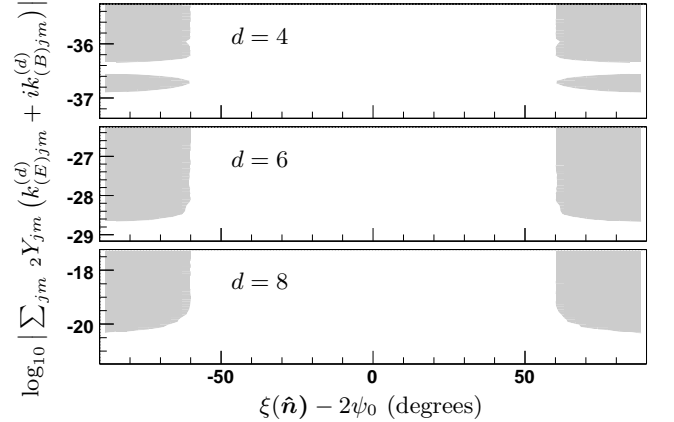,width=\hsize}}
\caption{\label{960924_even}
Spectropolarimetry bounds from GRB 960924 
on spherical coefficients 
corresponding to Lorentz-violating operators 
of mass dimensions $d=4,6,8$.
The source direction is given by the angles 
$\mbf{\hat n} = (87.3^\circ,37.3^\circ)$.
The shaded area is the disallowed region
with $\Pi_{\rm eff}<50\%$.
All coefficients are in units of GeV$^{4-d}$.}    
\end{center}
\end{figure}

Figures \ref{930131_odd} and \ref{960924_odd}
show the calculated effective degrees of polarization 
and the changes in the polarization angles
for the above two gamma-ray bursts
in scenarios with nonzero spherical coefficients
corresponding to CPT-odd Lorentz-violating operators
of mass dimensions $d=5,7,9$.
In constructing these plots,
we assume the radiation is initially 100\% linearly polarized,
which is a maximally conservative assumption
in the present context.
The displayed results are then obtained 
by numerically calculating 
the change in the effective polarization 
$\Pi_{\rm eff}=\sqrt{\vev{s^1}^2+\vev{s^2}^2}$
smeared over observed frequencies,
following the basic procedure 
outlined in Ref.\ \cite{km_grb}.
The shaded regions in
Figs.\ \ref{930131_odd} and \ref{960924_odd}
show the ranges of coefficient space 
that are excluded by the observation 
of polarization in the radiation from these sources.
Coefficients lying in these regions 
would cause depolarization beyond what is observed.
The resulting constraints 
for the vacuum coefficients
and for the limiting case of the vacuum isotropic model 
are summarized in Table \ref{pointbounds}.

In the CPT-even case,
the linear polarization of the source
could in principle coincide 
with one of the eigenmodes of propagation. 
This situation occurs when 
the phase angle $\xi(\mbf{\hat n})$ 
is twice the initial polarization angle $\psi_0$,
as discussed above.
Consequently,
the results for any given point source 
contain unbounded regions of coefficient space,
and so definitive constraints on the spherical coefficients
cannot be obtained.
However,
the analysis does achieve
high sensitivities to Lorentz violation
in certain regions of coefficient space.
Figures \ref{930131_even} and \ref{960924_even}
show the portions of coefficient space 
excluded by the observations of GRB 930131  and GRB 960924 
for CPT-even operators of mass dimensions $d=4,6,8$
\cite{fn1}.
The shaded areas in these figures 
represent disallowed regions.
Their shape and extent demonstrates that 
part of the coefficient space
is excluded at high sensitivity,
while emphasizing the need 
for simultaneous analysis of multiple sources 
to obtain definitive constraints.
The approximate sensitivities to the vacuum coefficients
achieved from the two sources
are summarized in Table \ref{pointbounds}.

The above examples show that
high-frequency polarimetry of gamma-ray bursts 
has the ability to probe Lorentz violation at extreme sensitivity.
The use of multiple sources
offers the potential to constrain
all of the vacuum coefficients
$\kE$, $\kB$, and $\kV$ 
associated with leading-order birefringent Lorentz violation
at large $d$.
In the CPT-odd case,
the limits can completely constrain
the single vacuum isotropic coefficient at each odd $d$ 
because the direction of propagation is irrelevant.

The power of this type of analysis is apparent
in comparing the sensitivities in Table \ref{pointbounds}
to those achieved via the astrophysical dispersion tests 
discussed in Sec.\ \ref{sec_disp}. 
For $d=6$, 
the polarimetry of gamma-ray bursts attains 
a sensitivity roughly a million times beyond 
that of the dispersion tests involving Markarian 501,
despite the million-fold difference in energy.
This confirms the suitability of dispersion tests
to constrain instead only the vacuum coefficients $\kI$,
which are associated with Lorentz-violating operators
that have no leading-order birefringence. 

\subsubsection{CMB tests}
\label{cmb_sec}

In principle,
an extended source can
provide access to all birefringent Lorentz-violating operators
and hence bypass
the major limitation of point sources described above.
A prime example of an extended source is the CMB.
Since the CMB is the oldest observable untainted radiation 
and hence represents the longest available baseline,
an analysis of CMB polarization might be expected 
to yield high sensitivities to Lorentz violation.
However,
this expectation may fail for operators at larger $d$
due to the comparatively low CMB frequencies.
Nonetheless,
the CMB does provide interesting opportunities
for lower-dimensional violations,
and in particular it is the best available source for
studies of coefficients with $d=3$.
In this subsection,
we discuss and illustrate some of the unusual features 
that can arise in the CMB
in the event of significant birefringence
\cite{km_cmb,km_apjl,cmb_bire}.

The accepted description of the CMB
uses a decomposition 
of the temperature $T$ and the Stokes parameters
into spin-weighted spherical harmonics
\cite{zs},
analogous to the discussion in Sec.\ \ref{sec_vac}.
Typically,
power spectra are introduced to characterize
the strength of each mode and the correlations between them,
according to
\beq
C^{X_1X_2}_j = \frac{1}{2j+1}\sum_m
\vev{(a^{(X_1)}_{jm})^*a^{(X_2)}_{jm}},
\eeq
where $X_1$, $X_2$ range over $T,E,B,V$
and where $a^{(X)}_{jm}$ are the coefficients 
in the spherical-harmonic expansion of $X$.
While temperature anisotropies have been firmly established,
the detection of polarization in the CMB
by several experiments
\cite{wmap3yr,wmap5yr,boomerang,cmb_pol_ex}
is of more significance in the present context.

In the conventional Lorentz-invariant picture,
temperature and density fluctuations at 
recombination provide the necessary anisotropies 
to produce a net polarization
\cite{cmb_rev}.
In addition,
these processes lead to a correlation
between temperature and the $E$-parity component of the CMB.
The $E$ component makes up only a tiny fraction
of approximately $10^{-6}$ of the total radiation.
Several measurements 
of this small degree of polarization
have been made.
The $B$-type polarization is expected to be even smaller
and uncorrelated with temperature.
The standard picture predicts no significant $V$ polarization
because Thomson scattering produces only linear polarization.

The presence of Lorentz violation 
may alter many of the above properties.
It can introduce unexpected types of polarization,
and it can induce mixing between initially uncorrelated modes
during the nearly 14 billion years of propagation.
The associated violations of rotational symmetry
can also cause mixing across multipoles in $j$ and $m$,
including for modes of the same polarization type.
Here,
we provide a discussion of general features 
based on a numerical survey
of some Lorentz-violating models exhibiting these effects.
The calculations parallel those
presented in Refs.\ \cite{km_cmb,km_apjl}.

Some qualitative features can be determined directly
using the intuition provided by the Stokes rotations
described in Sec.\ \ref{bireftheory}.
For example,
the vacuum coefficients $\kV$ 
are associated with CPT-violating operators
and produce local rotations of the
Stokes vector about the $s^3$ axis.
This causes a global mixing of
the linearly polarized $E$ and $B$ modes.
The result can be unconventionally large $B$ polarization,
although no circular $V$ modes can appear.
In contrast,
the vacuum coefficients $\kE$ and $\kB$
associated with CPT-even operators
cause both mixing between $E$ and $B$ modes
and also the emergence of $V$ modes,
since in this case
the Stokes vector rotates out of the $s^1$-$s^2$ plane.
Consequently,
mixing between the three types of polarization are possible
in this scenario,
although details of the mixing
depend strongly on the specifics 
of the Lorentz-violating operators involved.

\begin{figure}
\begin{center}
\centerline{\psfig{figure=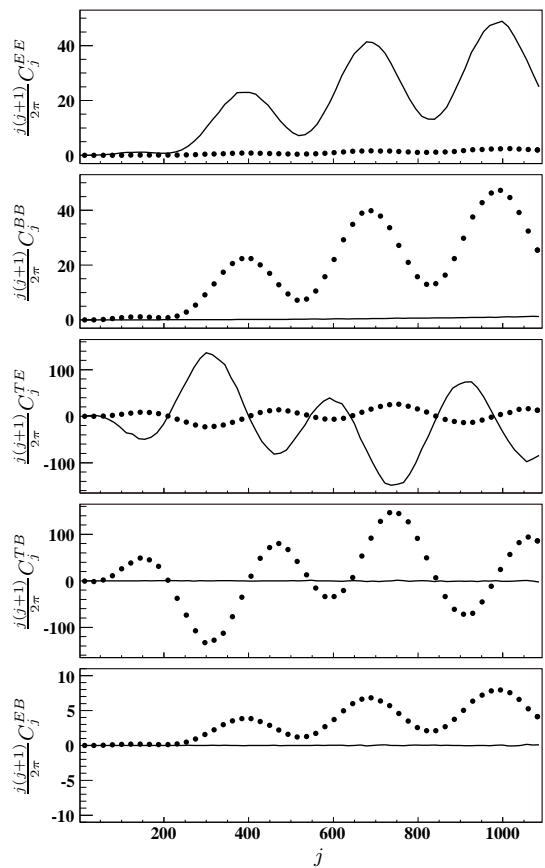,width=0.85\hsize}}
\caption{\label{k300}
Correlation spectra for isotropic CPT-odd Lorentz violation.
Circles show the effects of Lorentz violation
due to the spherical coefficient
$\kVdjm{3}{00}=12\times10^{-42}$ GeV,
while lines show the Lorentz-invariant case.
Power is transferred from $E$ to $B$,
and an $EB$ correlation is generated.
Also,
the initial $TE$ correlation induces a $TB$ correlation.
As expected for CPT-odd Lorentz violation,
no significant $V$ polarization or correlation appears.
The coefficients $C_j$ have units of $\mu\mbox{K}^2$. }    
\end{center}
\end{figure}

Other key features that may be present 
for some types of violations include
dependences on the photon frequency
and birefringence varying with the direction of propagation.
Only the $d=3$ vacuum coefficients 
lead to frequency-independent rotations.
Also,
only the $j=0$ coefficients 
generate direction-independent rotations.
Consequently,
the isotropic vacuum coefficient $\kVdjm{3}{00}$
provides a particularly simple special case.
It causes a frequency-independent mixing
that is uniform across the sky
and that leads to a simple rotation between $E$ and $B$ modes.
Using Eq.\ \rf{delta_psi},
we estimate this rotation for CMB radiation to be 
\beq
\de\Psi\simeq \kVdjm{3}{00}\ 10^{43} \mbox{ degree}/\mbox{GeV} .
\label{cmb_delta_psi}
\eeq
Simple frequency-independent rotations of this type 
have been considered by several groups 
\cite{km_cmb,km_apjl,feng,cabella,komatsu,xia,kahniashvili,wu},
and the existing measurements are listed
as part of Table \ref{cmb3bounds}.
Figure \ref{k300} illustrates the type
of mixing that results.
We see that initial $E$ power 
partially rotates into $B$ power.
Also, 
the initial $TE$ correlation induces a $TB$ correlation.
Furthermore,
since the $B$ polarization is generated 
from the original $E$ polarization,
these two modes become correlated 
and so a significant $EB$ component emerges.

\begin{table}
\renewcommand{\arraystretch}{1}
\begin{tabular}{c|c|c}
Coefficients & Result & Reference \\ 
\hline
\hline
$|\mbf{\kafd{3}}|$
&
$(15\pm 6)\times 10^{-43}$ 
GeV
&
\cite{km_cmb}
\\
&
$(10^{+4}_{-8})\times 10^{-43}$
GeV
&
\cite{km_apjl}
\\
\hline
$\kVdjm{3}{00}$
&
$(6.0 \pm 4.0) \times 10^{-43}$
GeV
&
\cite{feng}
\\
&
$(2.5 \pm 3.0) \times 10^{-43}$
GeV
&
\cite{cabella}
\\
&
$(12\pm 7) \times 10^{-43} $
GeV
&
\cite{km_cmb}
\\
&
$(1.2 \pm 2.2) \times 10^{-43}$
GeV
&
\cite{komatsu}
\\
&
$(2.6 \pm 1.9) \times 10^{-43}$
GeV
&
\cite{xia}
\\
&
$< 2.5 \times 10^{-43}$
GeV
&
\cite{kahniashvili}
\\
&
$(2.3 \pm 5.4) \times 10^{-43}$
GeV
&
\cite{km_apjl}
\\
&
$(-1.4 \pm 0.9 \pm 0.5) \times 10^{-43}$ 
GeV
& 
\cite{wu}	
\\
\hline
\hline
\end{tabular}
\caption{\label{cmb3bounds}
Constraints on spherical coefficients with $d=3$
from CMB studies.
The table lists some existing 1-$\si$ results 
for the scalar magnitude
$|\mbf{\kafd{3}}|$ defined in Eq.\ \rf{scalmag}
and for the isotropic component $\kVdjm{3}{00}$,
all obtained via CMB analyses.
Constraints on these quantities
from other sources are compiled in the data tables
of Ref.\ \cite{datatables}. }
\end{table}

Other isotropic rotations from CPT-odd operators of larger $d$,
such as the one associated with the 
isotropic vacuum coefficient $\kVdjm{5}{00}$,
lead to similar effects.
However, 
the frequency dependence introduced 
by coefficients at larger $d$
implies that the amount of rotation 
depends on the photon frequency.
Spectral signatures of this type should be accessible
to observations having sensitivity
to a wide range of frequencies.

\begin{figure}
\begin{center}
\centerline{\psfig{figure=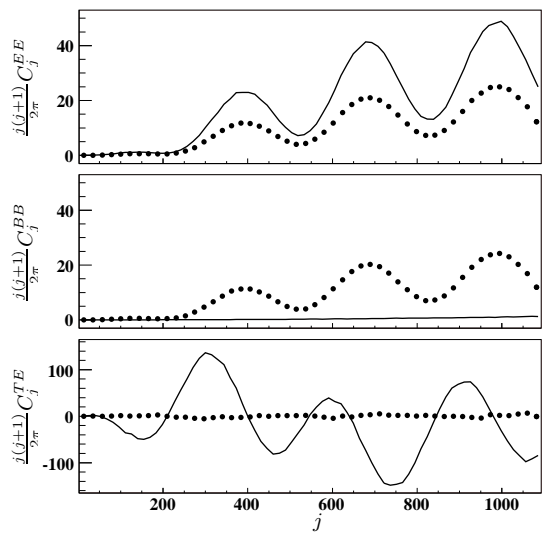,width=0.85\hsize}}
\caption{ \label{k310}
Correlation spectra for anisotropic CPT-odd Lorentz violation.
Circles show the effects of Lorentz violation
due to a large value of the spherical coefficient
$\kVdjm{3}{10}=64\times10^{-42}$ GeV,
while lines show the Lorentz-invariant case.
Power is transferred from $E$ to $B$,
and a loss of $TE$ correlation is evident.
As expected for CPT-odd Lorentz violation,
no significant $V$ polarization or correlation appears.
The coefficients $C_j$ have units of $\mu\mbox{K}^2$. }    
\end{center}
\end{figure}

\begin{figure}
\begin{center}
\centerline{\psfig{figure=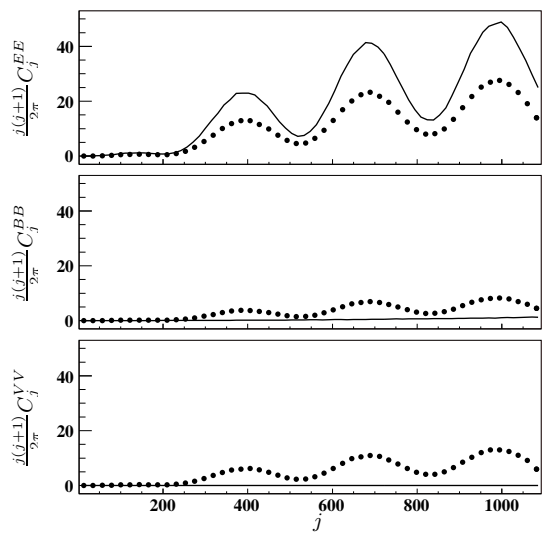,width=0.85\hsize}}
\caption{ \label{ke420}
Correlation spectra for anisotropic CPT-even Lorentz violation
at photon frequency $\om =380$ GHz. 
Circles show the effects of Lorentz violation
due to a large value of the spherical coefficient
$\kEdjm{4}{20}=16\times10^{-29}$,
while lines show the Lorentz-invariant case.
Power is transferred from $E$ to $B$ and $V$,
while no significant correlations appear.
The coefficients $C_j$ have units of $\mu\mbox{K}^2$. }    
\end{center}
\end{figure}

Any CPT-odd Lorentz-violating operator with nonzero $j$
produces direction-dependent rotations.
An example is shown in Fig.\ \ref{k310},
where a comparatively large value 
of the anisotropic vacuum coefficient $\kVdjm{3}{10}$
has been chosen to illustrate the effects.
While the local polarization rotations
are similar to those 
from the presence of a nonzero $\kVdjm{3}{00}$,
the anisotropies in this case
cause the correlations to disappear globally.
The effects saturate for large rotations,
and the net result 
is roughly equal amounts of $E$ and $B$.
The overall degree of polarization is unaltered,
and there is little correlation between any two modes.

Birefringent operators that are CPT-even 
are necessarily both frequency dependent and anisotropic,
resulting in similar behavior to the previous case.
However, 
a distinctive feature of CPT-even violations
is the mixing of linear and circular polarizations.
Figure \ref{ke420} illustrates 
the generation of $V$ polarization
in the presence of a nonzero vacuum coefficient $\kEdjm{4}{20}$.
In this case,
we find that the anisotropic mixing 
causes a depletion of $E$ polarization,
which is rotated into roughly comparable amounts 
of $B$ and $V$ polarization.
As in the case of nonzero $\kVdjm{3}{10}$
and indeed in all cases we have studied
involving nonzero coefficients with $j\neq 0$,
the anisotropic effects tend to deplete correlations 
when Lorentz violations are large.

Our survey reveals that similar features  
as those illustrated above 
also recur for other vacuum coefficients for Lorentz violation
that control birefringent operators.
We thereby find that generic signals of Lorentz violation 
in the CMB can incorporate 
one or more of the following basic features:
(a) a depletion in the $EE$ and $TE$ spectra;
(b) the introduction of unconventionally large $B$ polarization;
(c) the appearance of $TB$ or $EB$ correlations;
(d) the development of significant $V$ polarization;
and (e) frequency dependences of the power spectra.

None of these basic features is readily apparent 
in the observed data,
which suggests that any Lorentz violation 
in the CMB must be small.
A study of the high-frequency BOOMERANG data 
\cite{boomerang}
places constraints on several of the vacuum coefficients
with $d\leq 6$,
finding some evidence for nonzero Lorentz violation
\cite{km_cmb}.
A more recent analysis of the $d=3$ coefficients 
using the five-year data from the
Wilkinson Microwave Anisotropy Probe 
\cite{wmap5yr}
finds no evidence for isotropic violations
involving the coefficient $\kVdjm{3}{00}$,
but uncovers some support 
for nonzero anisotropic vacuum coefficients 
$\kVdjm{3}{1m}$ \cite{km_apjl}.
This study constrains the coefficients $\kVdjm{3}{jm}$
at the level of $10^{-43}$ GeV.
For purposes of reporting constraints,
the spherical coefficients with $d=3$ can conveniently 
be separated into the isotropic component $\kVdjm{3}{00}$
and the scalar magnitude
\beq
|\mbf{\kafd{3}}|=
\frac 1{\sqrt{4\pi}}
\big(6|\kVdjm{3}{11}|^2+3|\kVdjm{3}{10}|^2\big)^{1/2}.
\label{scalmag}
\eeq
Table \ref{cmb3bounds}
lists existing constraints on these quantities
obtained from studies of the CMB.
Other limits are given in the data tables
of Ref.\ \cite{datatables}.
Improved sensitivities can be expected from future CMB data 
to be obtained by various experiments and missions
including, 
for example,
the Planck satellite
\cite{planck},
the Q/U Imaging Experiment
(QUIET)
\cite{quiet},
the CMBPol mission
\cite{CMBPol},
the $E$ and $B$ Experiment
(EBEX)
\cite{ebex},
the Experimental Probe of Inflationary Cosmology
(EPIC)
\cite{epic},
and 
the Spider balloon observatory
\cite{spider}.

\section{Cavity experiments}
\label{sec_cavities}

Laboratory experiments provide alternative methods
to search for Lorentz violation in electrodynamics
and have the ability to probe many coefficients 
inaccessible in astrophysical searches.
The most common Earth-based tests
are contemporary versions of the classic
Michelson-Morley \cite{mm}
and Kennedy-Thorndike \cite{kt} 
experiments 
and are based on electromagnetic resonant cavities
\cite{cav1,cav2,cav3,cav4,cav5,cav6}.
In this section,
we discuss possible laboratory studies
of the effects of Lorentz-violating operators 
of arbitrary mass dimension $d$.
We outline a theoretical approach 
for determining the resonance frequency
of a cavity in the presence of Lorentz violation.
The approach is illustrated by applying it 
in the context of the camouflage model,
for which the Lorentz violation
has no leading-order birefringent or dispersive effects
in astrophysical photon propagation.
As a specific example of the techniques,
we derive an explicit result 
for the fractional frequency shift
of the TM$_{010}$ mode in a cylindrical cavity
with circular cross section,
including the time dependence induced in the signal
by the rotation of the Earth.
While we focus specifically on resonant cavities,
many of the ideas discussed here are generic
and can be applied in the context 
of other laboratory-based experiments,
including space-based missions such as
the Atomic Clock Ensemble in Space 
(ACES)
\cite{aces}
or the Space and Time Anisotropy Tests
(STAT) 
combining the former
Space Time Asymmetry Research (STAR)
and OPTIS missions 
\cite{stat}.

\subsection{General theory}
\label{sec_cavity_th}

The strategy behind many laboratory tests 
of Lorentz invariance 
is a search for minute variations in some observable
with changes in the orientation or velocity of the apparatus.
For cavity experiments,
a suitable observable is typically the fractional
frequency shift $\de\nu/\nu$
induced by the Lorentz violation 
\cite{km}.
At leading order in coefficients for Lorentz violation,
the fractional frequency shift takes the generic form
$\de\nu/\nu = \sum_{jm} {\cal M}_{jm}{\cal K}_{jm}$,
where ${\cal M}_{jm}$ 
is an experiment-dependent matrix factor 
and ${\cal K}_{jm}$ represents the relevant
spherical coefficients for Lorentz violation
discussed in Sec.\ \ref{sec_gen_coeffs}.

Normally the experiment-dependent matrix factors
are determined in a cavity frame that
is fixed with respect to the system in question.
In this frame,
the matrices are denoted ${\cal M}^{\rm cav}_{jm}$
and are constant.
However,
since this frame is noninertial,
the cavity-frame coefficients ${\cal K}^{\rm cav}_{jm}$
for Lorentz violation
vary with changes in the orientation and velocity
of the apparatus with respect to the standard Sun-centered frame.
In practice,
boost effects are suppressed by one or more powers
of the typically small velocities $\sim 10^{-4}$ involved,
so for simplicity in what follows
we neglect boost effects and focus 
on violations of rotation invariance.

The rotations relating spherical coefficients
in the cavity, laboratory, and Sun-centered frames
are given in Eqs.\ \rf{k_rot1} and \rf{k_rot2}.
Since the rotation between the cavity frame 
and the laboratory frame is often constant,
it is convenient to define laboratory-frame
matrices ${\cal M}^{\rm lab}_{jm}$ via
\beq
{\cal M}^{\rm lab}_{jm} = \sum_{m'}
{\cal M}^{\rm cav}_{jm'} 
D^{(j)}_{m'm}(-\ga,-\be,-\al) .
\label{Mlab}
\eeq
In terms of these and the spherical coefficients
in the Sun-centered frame,
the fractional frequency shift takes the form
\beq
\frac{\de\nu}{\nu} = \sum_{jmm'}
{\cal M}^{\rm lab}_{jm} 
e^{im\ph}e^{im'\om_\oplus T_\oplus} d^{(j)}_{mm'}(-\ch)\,
{\cal K}_{jm'}^{\rm Sun} ,
\label{dnurot}
\eeq
which explicitly reveals the sidereal dependence.
For experiments involving turntables,
where the cavity and laboratory frames 
rotate relative to each other, 
the variations resulting from the turntable rotation
are incorporated through the phase 
$\ph = \om_{tt} T_{tt}$.

Manipulation of the modified Maxwell equations \rf{maxeq}
leads to a perturbative estimate for the 
fractional frequency shift given by
\cite{km}
\begin{align}
\fr{\de\nu}\nu &\approx
-\fr{1}{4\vev{U}}
\int d^3x 
\big(\mbf E^*\cdot\mbf{\de D} -\mbf B^*\cdot\mbf{\de H}\big) 
\label{dnu}\\
&= \fr{1}{8\vev{U}} \int d^3x F_{\mu\nu}^* (\de G)^{\mu\nu} ,
\notag
\end{align}
where
\beq
\vev{U} = 
\fr14\int d^3x \left( 
\mbf E^*\cdot\mbf D + \mbf B^*\cdot\mbf H 
\right)
\eeq
is the unperturbed energy inside the resonator.
This formulation allows 
for a general linear and lossless medium inside the cavity
in addition to modifications due to Lorentz violation.
The fields $\mbf E$, $\mbf B$, $\mbf D$, and $\mbf H$
are understood to be solutions 
of conventional electrodynamics 
in the absence of Lorentz violation,
while
\begin{align}
\mbf {\de D} &= \kde\cdot\mbf E + \kdb\cdot\mbf B
+ 2\mbf\kaf\times\mbf A , 
\notag\\
\mbf {\de H} &= \khe\cdot\mbf E + \khb\cdot\mbf B
- 2(\kaf)_0 \mbf A+ 2\mbf\kaf A_0 
\label{dDdH}
\end{align}
represent leading-order perturbations
to the $\mbf D$ and $\mbf H$ fields
due to Lorentz violation.
The result \rf{dnu} assumes that the fields vanish 
outside the cavity volume $V$.
For simplicity in what follows,
we suppose that the resonant modes 
under consideration are nondegenerate.
In the case of degenerate resonances,
Eq.\ \rf{dnu} yields a weighted average value 
for the fractional frequency shift.

To express the fractional frequency shift 
in terms of the spherical coefficients 
introduced in Sec.\ \ref{sec_gen_coeffs},
we next convert to momentum space.
However, 
some care is required to avoid divergences 
arising from discontinuities at the boundary of the cavity.
This technical issue stems from the differential nature 
of the operators $\kf$ and $\kaf$.
Although the fields are taken to be continuous 
throughout the interior volume $V$ of the cavity,
they may be discontinuous across the surrounding surface $S$.
As a result,
the fields may not be strictly differentiable over all space,
and the derivatives in Eq.\ \rf{dDdH} may be associated with 
$\de$-function behavior on the surface $S$.
The usual reciprocal nature of position and momentum spaces
then leads to momentum-space representations
that fail to vanish sufficiently rapidly at infinite momentum,
which introduces divergences in the integral.
This issue is absent in the minimal SME with $d=3$ and $d=4$ 
because no derivatives appear in the constitutive relations
for that case,
but it is endemic for Lorentz-violating operators 
with $d\geq 5$ and requires a procedure 
to eliminate the divergences.

One way to address this technical issue 
is to define new everywhere-differentiable fields 
$\mbf\sE$ and $\mbf\sB$ 
that are equal to $\mbf E$ and $\mbf B$ inside $V$
but that may be nonzero in the region outside $V$
where the original fields $\mbf E$ and $\mbf B$ vanish.
We then have two sets of fields,
both satisfying the Maxwell equations inside $V$.
Derivatives of the extended fields $\mbf\sE$ and $\mbf\sB$
remain finite everywhere,
including on the surface $S$ of the cavity,
but $\mbf\sE$ and $\mbf\sB$ need not satisfy 
the Maxwell equations outside the cavity.
Using both sets of fields,
we can construct a finite version of 
the fractional frequency shift \rf{dnu} 
by replacing the fields in Eq.\ \rf{dDdH} 
with their extended versions.
This procedure removes the divergences 
at the cavity boundaries.

Performing a Fourier transform,
we obtain a convergent momentum-space
expression for the fractional frequency shift \rf{dnu}
given by
\begin{align}
\fr{\de\nu}\nu &=
-\fr{1}{4\vev{U}}
\int d^3p
\big(\mbf E^*\cdot\kde\cdot\mbf \sE
-\mbf B^*\cdot\khb\cdot\mbf \sB
\notag \\
&\hspace*{66pt}
+\mbf E^*\cdot\kdb\cdot\mbf \sB
-\mbf B^*\cdot\khe\cdot\mbf \sE
\notag \\
&\hspace*{30pt}
+\fr{2i}{\om^2}(\om\mbf\kaf - \mbf p (\kaf)_0 )
\cdot(\mbf E^*\times\mbf \sE)\big) .
\label{dnu2}
\end{align}
This result is independent of gauge choice,
as expected.
Since the fields and their extensions agree inside $V$,
the cavity energy $\vev{U}$
can be calculated using either of the forms
\begin{align}
\vev{U} 
&= \fr14\int d^3p \left( 
\mbf E^*\cdot\mbf D 
+ \mbf B^*\cdot\mbf H \right)
\notag\\
&= \fr14\int d^3p \left( 
\mbf E^*\cdot\mbf \sD 
+ \mbf B^*\cdot\mbf \sH \right).
\end{align}
Note that both sets of fields are needed 
in the integral \rf{dnu2} for this procedure to work.
The extended fields control divergences in momentum space,
while the unextended fields restrict the integration 
to the volume $V$ of the cavity in position space.

The general procedure for determining
the effects of Lorentz violation 
on the resonant frequency of a given cavity
then involves the following steps.
First,
obtain 
the fields $\mbf E$, $\mbf B$, $\mbf D$, $\mbf H$
in the cavity in the context of conventional electrodynamics,
incorporating in the usual way
any permittivity and permeability of the media involved.
Next,
construct extensions of the fields
that are everywhere smooth beyond the cavity volume.
Perform the Fourier transform
to derive the corresponding momentum-space fields.
Then,
using the spherical-harmonic expansions
of $\kaf$ and the $\hat\ka$ matrices
described in Sec.\ \ref{sec_gen_coeffs},
calculate the cavity-frame matrices ${\cal M}^{\rm cav}_{jm}$ 
from Eq.\ \rf{dnu2}.
Finally,
combine the results using Eq.\ \rf{dnurot}
to extract the desired fractional frequency shift $\de\nu/\nu$ 
displaying the explicit sidereal time dependences.

\subsection{Example: camouflage model}
\label{sec_nb_cav}

A general analysis of cavity experiments
incorporating all spherical coefficients
for Lorentz violation is challenging 
due to the large variety of effects.
Moreover,
although terrestrial tests can provide
valuable independent and fully controlled checks
on vacuum birefringence and dispersion,
the high sensitivities attainable 
in astrophysical searches for birefringence and dispersion
make it reasonable 
to neglect effects from vacuum coefficients 
in the context of resonator experiments.
Laboratory searches using cavities
are therefore well suited to study
the vacuum-orthogonal coefficients
for Lorentz violation,
which play no role in the vacuum propagation of light.

In this subsection,
we illustrate the treatment of the spherical coefficients
in the derivation of the 
cavity-frame matrices ${\cal M}^{\rm cav}_{jm}$.
For simplicity,
we focus primarily on Lorentz-violating operators
that produce no leading-order birefringence 
or vacuum dispersion.
For $d=4$,
this class of operators includes ones corresponding
to the coefficients $\kIdjm{4}{jm}$
that have already been widely studied
in cavity experiments
\cite{cav1,cav2,cav3,cav4,cav5,cav6}.
For general $d$,
it corresponds to the special subset 
of the coefficients $\cfzE$
formed by the camouflage coefficients $\cftzE$
introduced in Sec.\ \ref{sec_nonbire}.
 
We begin by considering a scenario involving  
the coefficients $\cfzE$ 
controlling nonbirefringent effects.
We seek an expression for the
fractional frequency shift of the form
\beq
\fr{\de\nu}{\nu} = \sum_{dnjm} \Mc \cfzE .
\eeq
To find the matrices $\Mc$,
note that the fractional frequency shift 
in the nonbirefringent case may be written as
\begin{align}
\fr{\de\nu}{\nu} &= \frac{1}{4\vev{U}} \int d^3p~ 
F^*_{\mu\nu}{\sF^\mu}_\rh (\cf)^{\nu\rh}
\notag \\
&= \frac{1}{4\vev{U}} \int d^3p~
F^*_{\mu\nu}{\sF^\mu}_\rh \prt^\nu\prt^\rh \hat\Ph_F ,
\end{align}
where $\hat\Ph_F$ is given by Eq.\ \rf{phi}
and the derivatives act in momentum space,
$\prt^\nu=\prt/\prt p_\nu$.
Separating explicitly the temporal and spatial components gives
\begin{align}
\fr{\de\nu}{\nu} &= \frac{1}{4\vev{U}} \int d^3p
\big[ -{\mbf E}^*\cdot \mbf\sE \ \Frac{\prt^2\phantom{\om}}{\prt\om^2}
\notag \\
&\quad\quad
+ \big({\mbf B}^*\times\mbf\sE - {\mbf E}^*\times\mbf\sB\big)^a
\nabla_a \Frac{\prt\phantom{\om}}{\prt\om}
\notag \\
&\quad\quad
+\half \big( (E^*)^{(a}\sE^{b)} + (B^*)^{(a}\sB^{b)} 
\notag\\
&\quad\quad\quad\quad\quad
-2g^{ab} {\mbf B}^*\cdot\mbf\sB \big) \nabla_a \nabla_b
\big]  \hat\Ph_F .
\end{align}
Inserting the spherical-harmonic expansion \rf{phi} of $\hat\Ph_F$ 
and using the identities in Appendix \ref{sec_am},
we can determine the contribution to the
fractional frequency shift from each coefficient $\cfzE$
and hence identify the matrices $\Mc$.

To obtain an explicit integral expression
for the matrices $\Mc$,
we can eliminate the magnetic fields using
the momentum-space Faraday law
$\om \mbf B = \mbf p\times \mbf E$.
This determines the matrices $\Mc$ as integrals 
over the fields $\mbf E$ and $\mbf \sE$.
In terms of the coordinate basis 
described in Appendix \ref{sec_am},
there are six field combinations contributing
to the integrals.
They can be specified as
\begin{align}
\ss^0  &= (E_+)^* \sE_+ + (E_-)^* \sE_- ,
\notag \\
\ss_{(\pm 2)} &= 2 (E_\pm)^* \sE_\mp ,
\notag \\
\tilde\ss_{(\pm 1)} &= E^*_r \sE_\mp + (E_\pm)^* \sE_r ,
\notag \\
\tilde\ss_{(0)} &= (E_r)^* \sE_r ,
\label{cav_stokes}
\end{align}
where 
\beq
E_\pm = \frac 1 {\sqrt{2}}(E_\th \pm i E_\ph) .
\eeq
The combinations $\ss^0$ and $\ss_{(\pm 2)}$
represent smoothed versions of the usual
Stokes parameters $s^0$ and $s_{(\pm 2)}$.
The combinations 
$\tilde\ss_{(\pm 1)}$ and $\tilde\ss_{(0)}$
are helicity-$(\pm 1)$ and helicity-$0$ combinations,
which provide two new transverse Stokes parameters 
vanishing when $E_r = \sE_r = 0$.
In terms of these quantities,
the explicit integral expression for the matrices $\Mc$ is
\begin{widetext}
\begin{align}
\Mc &= \fr{\om^{d-4-n}}{4\vev{U}}\int d^3p\, p^{n-2} 
\bigg[
\Frac14 (\om^2-p^2)\sqrt{\Frac{(j+2)!}{(j-2)!}}
\big(\syjm{+2}{jm}(\mbf{\hat p})\, \ss_{(+2)}(\mbf p)
+ \syjm{-2}{jm}(\mbf{\hat p})\, \ss_{(-2)}(\mbf p)\big)\,
\notag\\
&\hskip 50pt
-\big((n-1)\om^2+(d-2-n)p^2\big)\sqrt{\Frac{j(j+1)}{2}}
\big(\syjm{+1}{jm}(\mbf{\hat p})\, {\tilde \ss}_{(+1)}(\mbf p)
- \syjm{-1}{jm}(\mbf{\hat p})\, {\tilde \ss}_{(-1)}(\mbf p)\big)
\notag\\
&\hskip 50pt
+\big((n-\Frac{j(j+1)}{2})(\om^2-p^2) 
- (d-2-n)(d-3-3n)p^2 - n(n-1)p^2 \big)\,
\syjm{0}{jm}(\mbf{\hat p})\, \ss^0(\mbf p)
\notag\\
&\hskip 50pt
+ \big(n(n-1)\om^2 -(d-2-n)(d-3-n)p^2\big)\,
\syjm{0}{jm}(\mbf{\hat p})\, {\tilde \ss}_{(0)}(\mbf p) 
\bigg] ,
\label{dnu_nb}
\end{align}
\end{widetext}
where $\om$ is the usual
resonant angular frequency.

The result \rf{dnu_nb} applies both to empty cavities
and to ones containing a material medium,
provided the medium is lossless.
The form of the integral reveals that
the sensitivity to Lorentz violation 
depends both on the shape of the cavity 
and on the properties of any material medium it contains.
This means that different geometries or media
can be adopted to access different combinations 
of coefficients for Lorentz violation.
For example,
parity-breaking resonators may be used 
to access parity-odd Lorentz violations 
that cannot be detected at unsuppressed levels
using parity-symmetric systems
\cite{mp}.
The properties of the resonator enter implicitly 
through the conventional fields $\mbf E$,
which are determined by solving 
the conventional Maxwell equations inside the cavity
in the presence of the medium,
if any.
We emphasize that the frequency $\om$
is fixed for a given mode,
but any momentum $\mbf p$ can contribute
to the integral and hence to the matrices $\Mc$.

Next,
we focus attention specifically on the combinations
of the coefficients $\cfzE$
that govern nondispersive effects.
For $d=4$,
the $\cfzEdnjm{4}{njm}$ coefficients correspond directly 
to the vacuum coefficients $\kIdjm{4}{jm}$,
as described in Sec.\ \ref{sec_renorm}.
The matrices $\MIdjm{4}{jm}$ can therefore be expressed
in terms of the matrices $\Mcdnjm{4}{njm}$.
We find
\begin{align}
\MIdjm{4}{00} &= \fr34\Mcdnjm{4}{000}+\fr14\Mcdnjm{4}{000} ,
\notag \\
\MIdjm{4}{1m} &= -2\Mcdnjm{4}{11m} ,
\notag \\
\MIdjm{4}{2m} &= \Mcdnjm{4}{22m} .
\label{MI}
\end{align}
For $d>4$,
the relevant matrices are associated 
with the camouflage coefficients $\cftzE$ instead.
Using the relation \rf{ct},
we find
\beq
\Mct=\Mc-\Mcdnjm{d}{(n+2)jm} .
\label{Mct}
\eeq
With these results and Eq.\ \rf{dnu_nb},
the sensitivity of any given cavity 
to camouflage operators for Lorentz violation
can be determined.

Typically,
the above analysis is performed in the cavity frame,
although many of the equations hold 
for an arbitrary inertial frame.
Given the results for the
cavity-frame matrices $\cal M$,
the corresponding expressions 
for the laboratory-frame matrices 
follow from Eq.\ \rf{Mlab} and
from the orientation of the cavity in the laboratory.
Applying Eq.\ \rf{dnurot},
we finally obtain the fractional frequency shift
\begin{align}
\frac{\de\nu}{\nu} & = 
\sum_{dnjmm'} \Mctlab
e^{im\ph}e^{im'\om_\oplus T_\oplus}
d^{(j)}_{mm'}(-\ch)\, \cftzEdnjm{d}{njm'}
\notag \\
& \quad
+ \sum_{jmm'} \MIlabdjm{4}{jm}
e^{im\ph}e^{im'\om_\oplus T_\oplus}
d^{(j)}_{mm'}(-\ch)\, \kIdjm{4}{jm'} 
\label{fffcc}
\end{align}
which exhibits the variation with sidereal time.

\subsection{Example: circular-cylindrical cavity}
\label{sec_cav_example}

As an explicit illustration,
we calculate in this subsection
a number of elements of the matrix $\cal M$
for the TM$_{010}$ mode of a cylindrical cavity 
with circular cross section.
The cavity is centered at the origin 
of a cavity frame,
with the $z'$ direction aligned with the symmetry axis.
For definiteness,
let $R$ be the cavity radius and $2R$ be its length,
so that its ends lie at $z'=\pm R$.
Here,
we consider the case of a vacuum cavity for simplicity.
The solutions for the TM$_{010}$ mode 
in the absence of Lorentz violation are given by
\beq
\mbf E =
\begin{cases}
J_0(\rh' x_{01}/R) \mbf{\hat \rh'} & \mbox{inside $V$,}\\
0 & \mbox{outside $V$,}
\end{cases}
\eeq
where $x_{01}$ is the first zero of the Bessel function $J_0$,
and where $\mbf{\hat \rh'}$ represents the radial unit vector.

The electric field $\mbf E$ 
is discontinuous at the ends of the cavity,
and its derivatives are discontinuous
at the sides $\rh'=R$.
We therefore must seek a differentiable extension $\mbf\sE$
of the electric field
of the type described in Sec.\ \ref{sec_cavity_th}.
This is relatively straightforward
to find in the present case 
because we have an analytic solution 
that could be extended to infinity.
However,
for the purposes of numerical calculation
it is beneficial to construct 
instead a field $\mbf\sE$ 
that vanishes outside a larger volume $V'$
containing the cavity volume $V$.
With a field of this type,
the Fourier transforms can be accurately determined 
by numerical integration over only a finite region $V'$ of space.

An extension suitable for numerical work
can be constructed 
with the aid of the $C^q$ smoothing function $g_q(x;a,b)$,
defined by
$g_q(x;a,b)\equiv 0$ for $x<a<b$ or $b<a<x$,
$g_q(x;a,b)\equiv 1$ for $a<b<x$ or $x<b<a$,
and 
\beq
g_q(x;a,b) =
\frac{(2q+1)!}{(b-a)^{2q+1}}\sum_{n=0}^q
\frac{(x-a)^{q+1+n}(b-x)^{q-n}}{(q+1+n)!(q+n)!}
\eeq
otherwise.
This function continuously interpolates 
from $0$ at $x=a$ to $1$ at $x=b$,
is constant outside that interval,
and is $q$-times differentiable everywhere.
Note that $C^\infty$ functions of this type also exist, 
but for numerical calculations the above polynomial 
is easier to handle and suffices for our purposes.

Using this smoothing function,
we can define an extended electric field by
\begin{align}
\mbf\sE &=
g_q(\rh';2R,R)g_q(z';2R,R) g_q(z';-2R,-R)
\notag \\
&\quad 
\times J_0(\rh' x_{01}/R) \mbf{\hat \rh'} .
\end{align}
Taking the integration volume $V'$
as a cube of side length $4R$,
we see that the extended field $\mbf\sE$ 
matches $\mbf E$ inside cavity volume $V$, 
vanishes outside $V'$,
and is $q$-times differentiable.
The value of $q$ that is required to ensure 
the finiteness of the integrals
depends on the mass dimension $d$ of the
Lorentz-violating operators being considered.
For a given dimension $d$,
the matrices $\kde$, $\khb$, and $\kdb$ involve
$d-4$ derivatives.
Use of the Faraday law to eliminate 
the magnetic field $B$ introduces another derivative.
Consequently,
the extended field $\mbf\sE$ must be at least $C^{d-3}$,
so choosing $q>d-4$ should be sufficient.

\begin{table}
\renewcommand{\arraystretch}{1.1}
\begin{tabular}[t]{cc|c}
$j$&$m$&$\MIcavdjm{4}{jm}$\\
\hline
0 & 0 & -0.28 \\
2 & 0 & 0.32 \\[4pt]
\end{tabular}
\hfill
\begin{tabular}[t]{ccc|c}
$n$&$j$&$m$&$\Mctcavdnjm{6}{njm}$\\
\hline
0 & 0 & 0 & -13 \\
2 & 0 & 0 & 13 \\
2 & 2 & 0 & -12
\end{tabular}
\hfill
\begin{tabular}[t]{ccc|c}
$n$&$j$&$m$&$\Mctcavdnjm{8}{njm}$\\
\hline
0 & 0 & 0 & -150 \\
2 & 2 & 0 & 26 \\
4 & 0 & 0 & 150 \\
4 & 2 & 0 & -170 \\
4 & 4 & 0 & 56
\end{tabular}
\caption{\label{Mcav_values}
Nonzero matrix elements $\MIcavf$ and $\Mctcav$
for the fundamental mode of a circular cylindrical cavity.
Values for the cases $d=4,6,8$ are displayed.
The units are $R^{4-d}$,
where $R$ is the radius and half the length of the cavity. }
\end{table}

The next step involves obtaining the Fourier transforms
of the field components 
$E_{x'}$, $E_{y'}$, $E_{z'}$,
$\sE_{x'}$, $\sE_{y'}$, $\sE_{z'}$.
We implement these via fast Fourier transform
over an $N\times N\times N$ grid in the volume $V'$.
We use the transformed fields 
to determine the momentum-space Stokes parameters
from Eq.\ \rf{cav_stokes},
and we then perform numerical integration 
to evaluate the integral \rf{dnu_nb}
and obtain values for $\Mccav$.
Lastly, 
we use Eqs.\ \rf{MI} and \rf{Mct} 
to determine the components of interest
for the matrices $\MIcavf$ and $\Mctcav$.
As expected,
the results converge to stable 
$q$-independent values for large $N$,
provided $q$ is sufficiently large.
Table \ref{Mcav_values}
summarizes the results obtained 
through this procedure for $d=4,6,8$.
Note that the symmetry of the mode
implies no contribution from
parity-breaking coefficients with odd $j$ 
or from coefficients with $m\neq 0$,
a result that is recovered numerically.

Cavity experiments searching for Lorentz violation
typically compare two identical cavities
with different orientations.
In the present example,
the cavity-frame matrices $\cal M$ 
are then identical for the two cavities.
However,
their differing orientation implies
their laboratory-frame values would differ.
This leads to the slight frequency difference
that constitutes the signal for Lorentz violation.

\section{Summary and discussion}
\label{Summary and discussion}

In this paper,
we derive and study
gauge-invariant Lorentz- and CPT-violating terms 
associated with the effective photon propagator
in the Lagrange density of the SME,
allowing for operators of arbitrary mass dimension $d$.
We begin by showing that Lorentz violation at mass dimension $d$ 
is characterized by a set of 
$(d+1)(d-1)(d-2)/2$
independent coefficients 
for the CPT-odd case
and another set of $(d+1)d(d-3)$
independent coefficients
for the CPT-even case.
The compact Lagrange density \rf{lagrangian}
incorporates these effects for all $d$.
It includes and extends 
the pure-photon sector of the minimal SME
\cite{ck,km}.

The Lagrange density \rf{lagrangian} implies
the equations of motion \rf{eqnmot} for the photon field.
An interpretation of these equations is elaborated 
in terms of electrodynamics in macroscopic media
via the introduction of the operator constitutive tensors
\rf{constit}.
We derive the covariant scalar dispersion relation \rf{dr},
which must be satisfied 
by nontrivial plane-wave solutions.
For the CPT-even violations,
we extend the widely used $\kappa$ matrices
\cite{km}
of the minimal SME
to general $\hat\ka$ operators 
\rf{kappastwo}
that are relevant for studies at arbitrary $d$.

The unconventional properties 
of the Lorentz-violating eigenmodes 
can be characterized in terms of 
birefringence, dispersion, and anisotropy.
All CPT-odd operators lead to birefringence.
We study the conditions for leading-order birefringence
from CPT-even operators
via a Weyl decomposition of the constitutive tensor.
For this case,
we conjecture that nonbirefringent terms 
are uniquely associated with the non-Weyl part
of the constitutive tensor
when the decomposition is expressed 
in terms of a suitable effective metric.
For arbitrary Lorentz violation,
we introduce a duality symmetry
determined by the effective metric 
and argue that birefringence is a consequence
of the breaking of this symmetry.

\begin{table*}
\renewcommand{\arraystretch}{1.6}
\begin{tabular}{c|c|c|c|c|c}
& \ coefficient\ & $d$ & $n$ & $j$ & number \\
\hline
\hline
CPT even, $d$ even
& $\cfzE$  & $\geq 4$ & $0,1,\ldots, d-2$ 
& $n,n-2,n-4\ldots, \geq 0$    
& $\frac{(d+1)d(d-1)}{6}$ \\
$(d+1)d(d-3)$
& $\kfzE$  & $\geq 4$ & $0,1,\ldots, d-4$ & 
2 for $n=0$,
& $\frac{d^3-d-30}{6}$ \\
& & & & $n+2,n,n-2\ldots, \geq 0$ for $n\neq 0$ & \\
& $\kfoE$  & $\geq 6$ & $1,2,\ldots, d-4$ 
& $n+1,n-1,n-3\ldots, \geq 1$    
& $\frac{(d-4)(d^2+d+3)}{6}$ \\
& $\kftE$  & $\geq 6$ & $2,3,\ldots, d-4$ 
& $n,n-2,n-4\ldots, \geq 2$    
& $\frac{(d-4)(d^2-2d-9)}{6}$ \\
& $\kfoB$  & $\geq 4$ & $0,1,\ldots, d-4$ 
& $n+2,n,n-1\ldots, \geq 1$    
& $\frac{d^3-4d-18}{6}$ \\
& $\kftB$  & $\geq 6$ & $1,2,\ldots, d-4$ 
& $n+1,n-1,n-3\ldots, \geq 2$    
& $\frac{(d+3)(d-2)(d-4)}{6}$ \\
\hline
CPT odd, $d$ odd
& $\kafzB$ & $\geq 3$ & $0,1,\ldots, d-3$ 
& $n,n-2,n-4\ldots, \geq 0$    
& $\frac{d(d-1)(d-2)}{6}$ \\
$\Frac{(d+1)(d-1)(d-2)}{2}$
& $\kafoB$ & $\geq 3$ & $0,1,\ldots, d-3$ 
& $n+1,n-1,n-3\ldots, \geq 1$    
& $\frac{(d-1)(d^2+d-3)}{6}$ \\
& $\kafoE$ & $\geq 5$ & $1,2,\ldots, d-3$ 
& $n,n-2,n-4\ldots, \geq 1$    
& $\frac{(d+1)(d-1)(d-3)}{6}$ \\
\hline
\hline
\end{tabular}
\caption{\label{summary_table}
Summary of spherical coefficients 
for Lorentz-violating operators of arbitrary mass dimension.
The first column specifies 
the CPT property
and the total number of independent operators at each $d$.
The corresponding spherical coefficient sets 
are listed in the second column.
The remainder of the table
provides the allowed ranges for the indices $d$, $n$, $j$ 
and the number of independent components 
for each coefficient set.}
\end{table*}

\begin{table*}
\renewcommand{\arraystretch}{1.25}
\begin{tabular}{c|c|c|c|c|c}
limit & coeff.\ & $d$ & $n$ & $j$ & number \\
\hline
\hline
vacuum
& $\kI$ & even, $\geq 4$ &--& $0,1,\ldots, d-2$ & $(d-1)^2$   \\
& $\kE$ & even, $\geq 4$ &--& $2,3,\ldots, d-2$ & $(d-1)^2-4$ \\
& $\kB$ & even, $\geq 4$ &--& $2,3,\ldots, d-2$ & $(d-1)^2-4$ \\
& $\kV$ & odd,  $\geq 3$ &--& $0,1,\ldots, d-2$ & $(d-1)^2$   \\
\hline
vacuum orthogonal 
& $\cftzE$  & even, $\geq 4$ 
& $0,1,\ldots, d-4$ & $n,n-2,n-4\ldots, \geq 0$    
& $\frac{(d-1)(d-2)(d-3)}{6}$ \\
& $\kftzE$  & even, $\geq 6$ & $1,2,\ldots, d-4$   
& $n,n-2,n-4\ldots, \geq 0$    
& $\frac{(d-1)(d-2)(d-3)}{6}\scriptstyle{-1}$ \\    
& $\kfoE$  & even, $\geq 6$ & $1,2,\ldots, d-4$ 
& $n+1,n-1,n-3\ldots, \geq 1$    
& $\frac{(d-4)(d^2+d+3)}{6}$ \\
& $\kftE$  & even, $\geq 6$ & $2,3,\ldots, d-4$ 
& $n,n-2,n-4\ldots, \geq 2$    
& $\frac{(d-4)(d^2-2d-9)}{6}$ \\
& $\kftoB$  & even, $\geq 6$ & $1,2,\ldots, d-4$   
& $n,n-2,n-4\ldots, \geq 1$    
& $\frac{d(d-2)(d-4)}{6}$ \\    
& $\kftB$  & even, $\geq 6$ & $1,2,\ldots, d-4$ 
& $n+1,n-1,n-3\ldots, \geq 2$    
& $\frac{(d+3)(d-2)(d-4)}{6}$ \\
& $\kaftzB$  & odd, $\geq 5$ & $0,1,\ldots, d-4$   
& $n,n-2,n-4\ldots, \geq 0$    
& $\frac{(d-1)(d-2)(d-3)}{6}$ \\    
& $\kaftoB$  & odd, $\geq 5$ & $0,1,\ldots, d-4$   
& $n+1,n-1,n-3\ldots, \geq 1$    
& $\frac{(d+1)(d-1)(d-3)}{6}$ \\    
& $\kafoE$ & odd,  $\geq 5$ & $1,2,\ldots, d-3$ 
& $n,n-2,n-4\ldots, \geq 1$    
& $\frac{(d+1)(d-1)(d-3)}{6}$ \\
\hline
\hline
camouflage 
& $\cftzE$  & even, $\geq 4$ & $0,1,\ldots, d-4$ 
& $n,n-2,n-4\ldots, \geq 0$ 
& $\frac{(d-1)(d-2)(d-3)}{6}$ \\
\hline
isotropic 
& $\cffc$  & even, $\geq 4$ & $0,2,\ldots, d-2$ 
& $0$ & $d/2$ \\
& $\kffc$  & even, $\geq 4$ & $2,4,\ldots, d-4$ 
& $0$ & $(d-4)/2$ \\
& $\kaffc$ & odd,  $\geq 3$ & $0,2,\ldots, d-3$ 
& $0$ & $(d-1)/2$ \\
\hline
minimal SME
& $\kIdjm{4}{jm}$ & $4$ &--& $0,1,2$ & 9 \\
& $\kEdjm{4}{jm}$ & $4$ &--& $2$     & 5 \\
& $\kBdjm{4}{jm}$ & $4$ &--& $2$     & 5 \\
& $\kVdjm{3}{jm}$ & $3$ &--& $0,1$   & 4 \\
\hline
\hline
\end{tabular}
\caption{\label{limiting_table}
Summary of limiting cases.
The first column specifies the limit,
while the corresponding spherical coefficient sets 
are listed in the second column.
The remainder of the table
provides the allowed ranges for the indices $d$, $n$, $j$ 
and the number of independent components 
for each coefficient set.}
\end{table*}

In Sec.\ \ref{sec_gen_coeffs},
we obtain a complete classification
of the coefficients for Lorentz violation at arbitrary $d$,
using an SO(3) decomposition 
in terms of spin-weighted spherical harmonics.
A review of the spin-weighted spherical harmonics
and derivations of some useful mathematical results
are provided in Appendix \ref{sec_decomp},
along with a discussion of the relationships
to angular momentum, helicity, and parity.
The SO(3) decomposition reveals a total of nine independent sets 
of spherical coefficients for Lorentz violation
that control birefringence, dispersion, and anisotropy
in the photon propagator. 
Table \ref{summary_table}
lists these spherical coefficients 
and displays the ranges of their indices
and their counting.
The labels and indices on a given spherical coefficient
identify the key properties of the corresponding 
Lorentz-violating operator,
and their interpretation is summarized in the paragraph
containing Eq. \rf{ccrel}.
More detailed properties of these nine sets 
are compiled in Tables
\ref{kaf_ranges}, \ref{kf_B_ranges}, 
\ref{cf_ranges}, and \ref{kf_E_ranges}.

Table \ref{summary_table} reveals
the existence of several classes of operators
that are absent in the minimal SME,
which is restricted to $d=3$ and $d=4$.
The CPT-odd case at $d=3$
involves two sets containing four independent spherical coefficients,
while the CPT-even case at $d=4$ involves three sets
with 20 coefficients,
one of which is an unobservable constant.
However,
the CPT-odd case at $d=5$ has 35 coefficients
distributed among three sets rather than two.
The additional set $\kafoEdnjm{5}{njm}$ 
controls $E$-parity CPT-odd effects,
which are absent in the minimal SME. 
Similarly,
the CPT-even case at $d=6$ has 126 coefficients
distributed among six sets rather than three.
The additional sets 
$\kfoEdnjm{6}{njm}$, $\kftEdnjm{6}{njm}$, $\kftBdnjm{6}{njm}$ 
govern $E$-parity spin-one operators 
and also spin-two operators with both $E$- and $B$-type parities,
all associated with qualitatively new effects.

For applications to observation and experiment,
it is valuable to define various limiting cases 
of the general theory.
Sec.\ \ref{sec_models}
presents several of these limits,
while Table \ref{limiting_table}
summarizes their specific content
in terms of spherical coefficients for Lorentz violation.
The most widely studied limit to date is the minimal SME,
which involves a total of 23 independent 
nontrivial spherical coefficients.
The explicit connections between 
the spherical coefficients 
and the usual cartesian coefficients
for the minimal SME is provided via Tables
\ref{renorm-odd}, \ref{renorm-even-even},
and \ref{renorm-even-odd}.

Another useful limit involves the specification
of a preferred frame in which all Lorentz violation is isotropic.
The corresponding isotropic or `fried-chicken' models 
are discussed in Sec.\ \ref{sec_fc}.
The isotropic requirement eliminates 
all but three sets of spherical coefficients
and reduces the growth of coefficient numbers 
to be linear instead of cubic at large $d$.
For CPT-odd isotropic operators with $d=3,5,7\ldots$
there are only $1,2,3\ldots$ types of Lorentz violation,
all of which are birefringent.
Similarly,
for CPT-even isotropic operators with $d=4,6,8\ldots$
only $2,4,6\ldots$ coefficients arise,
of which $0,1,2\ldots$ are associated 
with leading-order birefringence.
Note that isotropic CPT-even birefringence
is a physical feature only for $d\geq 6$.

A third useful limiting subset of the general theory
is obtained by restricting attention
to Lorentz-violating operators that
are nonbirefringent and also 
are nondispersive in the vacuum at leading order.
Sec.\ \ref{sec_nonbire} constructs  
the corresponding camouflage models.
These models involve effects that are challenging
to detect in astrophysical studies
of birefringence and dispersion.
They are described by the single subset 
of coefficients $\cftzE$ appearing at even $d$
and hence associated with CPT-even Lorentz violation.
Table \ref{cft_ranges}
summarizes some properties of these coefficients.
For $d=4$ only a Lorentz-invariant trace appears,
so for most purposes it suffices to take $d>4$.
The camouflage coefficients then govern effects 
outside the minimal SME.
For operators of larger mass dimension
$d = 6,8,10\ldots$ there are $10, 35, 84\ldots$
independent effects,
with the number of independent coefficients growing rapidly 
as the cube of $d$ for large $d$.
Also,
for $d=4,6,8\ldots$ there are $1,2,3\ldots$
isotropic camouflage coefficients,
which therefore also belong to the general isotropic model.

The extreme sensitivities to Lorentz violation
available via studies of birefringence and dispersion 
of astrophysical sources
provides motivation for a further refinement
in the classification of spherical coefficients,
based on separating 
those coefficients that affect
the propagation of light in the vacuum 
from all others.
We refer to the former as vacuum coefficients
and to the complement
as vacuum-orthogonal coefficients,
and we identify the latter by a negation diacritic $\neg$. 
This classification is summarized 
in Table \ref{limiting_table}.
The vacuum-orthogonal coefficients
appear only for $d\geq 5$,
so they represent qualitatively new effects 
that are absent in the minimal SME.
Moreover,
the numbers of vacuum and vacuum-orthogonal coefficients
grow as $d^2$ and $d^3$ for large $d$, 
respectively,
so the vacuum-orthogonal coefficients
represent most of the coefficient space for large $d$. 
For example,
the vacuum coefficents govern all the physical effects
in the minimal SME for which $d=3$ and $d=4$,
but for $d=5$ they span only 16 of the 36 possibilities
and for $d=6$ only 67 of 126. 

The vacuum coefficients are constructed in Sec.\ \ref{sec_vac}.
At leading order,
they are identified by requiring 
that the radiation fields are plane waves.
This restricts attention to four sets of coefficients 
$\kI$, $\kE$, $\kB$, $\kV$.
The coefficients $\kI$ are associated 
with CPT-even Lorentz violation 
that is nonbirefringent at leading order
but dispersive for $d>4$.
The coefficients $\kE$ and $\kB$ control
CPT-even birefringent effects 
that are also dispersive for $d>4$.
Only $\kV$ is associated with CPT-odd effects,
which are also birefringent and dispersive.
For any given $d$,
the numbers of each type of vacuum coefficients
are roughly the same.
For example,
at $d=6$ there are 
25 coefficients in $\kIdjm{6}{jm}$
and 21 each in $\kEdjm{6}{jm}$ and $\kBdjm{6}{jm}$.
 
The vacuum-orthogonal models are
introduced in Sec.\ \ref{vacorthog}.
There are nine sets of vacuum-orthogonal coefficients,
of which four are identical to 
and five are reduced subsets of 
the nine sets in the general analysis.
One of the latter consists of the camouflage coefficients.
Some properties of the remainder are provided in
Tables \ref{kftE_ranges}, \ref{kftB_ranges}, 
and \ref{kaft_ranges}.
Except for a single constant scale factor in $d=4$,
the vacuum-orthogonal coefficients appear only for $d>4$.
They govern Lorentz-violating operators that produce
no leading-order birefringence or dispersion
in vacuum propagation,
although birefringent or dispersive effects
can appear in other physical contexts.
For example,
all 20 vacuum-orthogonal coefficients for $d=5$
and 49 of the 59 for $d=6$
are associated with birefringent and dispersive effects
in suitable circumstances.
The remaining 10 for $d=6$ are camouflage coefficients,
which govern nonbirefringent effects.

The remainder of the paper is concerned 
with applications of the theory 
to observation and experiment.
Results of measurements are conventionally reported
in the canonical Sun-centered frame, 
so for some applications
it is necessary to perform transformations
relating spherical coefficients in the Sun-centered frame 
to a laboratory or other frame.
These transformations are provided explicitly
in Sec.\ \ref{sec_rotations}
for the case where the boost component is negligible.

Section \ref{sec_astro} describes some applications
in the astrophysical context.
We focus on astrophysical studies 
of vacuum birefringence and dispersion.
Birefringence studies offer extreme sensitivity
to many vacuum coefficients,
while dispersion tests access the remainder
at lesser but nonetheless impressive sensitivities.
Dispersion constraints are discussed in Sec.\ \ref{sec_disp}.
We obtain expressions applicable 
to both isotropic and anisotropic dispersion,
and we use the recent measurements of GRB 080916C 
made by the Fermi Observatory
to obtain new constraints on Lorentz violation
involving operators of mass dimension six and eight.
A summary of existing constraints
from astrophysical dispersion tests 
is provided in Table \ref{dispersionbounds}.

Birefringence constraints are considered in Sec.\ \ref{sec_bire}.
The general theory is outlined,
and the net rotation induced by arbitrary spherical coefficients
is described quantitatively in terms of Stokes parameters.
We use polarimetric data from the gamma-ray bursts
GRB 930131 and GRB 960924
to set tight constraints on spherical coefficients 
associated with Lorentz-violating operators
of mass dimensions four through nine.
A summary of current limits from GRB polarimetry
is provided in Table \ref{pointbounds}.
We also consider constraints obtained 
from polarimetric studies of the CMB.
The maximal photon-propagation distances 
and the photon frequencies
make the CMB particularly well-suited
for measurements of spherical coefficients with $d=3$.
A survey is performed to categorize the mixing 
of polarizations in the CMB
induced by Lorentz violation.
Table \ref{cmb3bounds} compiles
some existing limits from the CMB 
on both isotropic and anisotropic Lorentz violation.

Section \ref{sec_cavities} discusses applications
in the laboratory context,
focusing on the use of resonant cavities 
to search for Lorentz violation.
These systems offer sensitivities to spherical coefficients
that are challenging to access in studies
of vacuum birefringence and vacuum dispersion.
A general theoretical procedure
for deriving the fractional frequency shift
in a cavity is presented in Sec.\ \ref{sec_cavity_th}.
In the following subsections,
the results are applied to nonbirefringent Lorentz violation
and in particular to the camouflage coefficients,
for which some relevant experiment-dependent factors
are explicitly obtained.
For the specific case of a circular-cylindrical cavity,
we provide in Eq.\ \rf{fffcc}
the fractional frequency shift
including the explicit time dependence.

The analysis in this work demonstrates
that a comprehensive search for Lorentz violation 
is best performed with multiple types of measurements. 
The most sensitive tests use astrophysical birefringence,
which provides access to some vacuum coefficients.
Astrophysical dispersion offers high sensitivity
to the remaining vacuum coefficients.
However,
the bulk of effects involves vacuum-orthogonal coefficients. 
Direct sensitivity to these requires non-vacuum
boundary conditions and hence laboratory studies. 
Disentangling the various birefringence, dispersion,
and anisotropy effects requires
a variety of laboratory experiments
involving different boundary conditions and different media,
and also the exploitation of
signals from different rotations and boosts.
Even if attention is limited to coefficients 
with comparatively low values of $d$,
considerable room remains for investigations
via both astrophysical observations and laboratory experiments.

\section*{Acknowledgments}

This work was supported in part
by the United States Department of Energy
under grant DE-FG02-91ER40661.

\appendix

\section{Spherical harmonics}
\label{sec_decomp}

Angular-momentum eigenstates are irreducible representations
of rotations,
so tensors in three dimensions
can be decomposed into components 
with definite orbital angular momentum and spin.
For spin-zero scalars,
the spherical harmonics $Y_{jm}$ provide
a basis for this decomposition.
For nontrivial tensors,
more general tensor spherical harmonics 
are needed to incorporate the spin.
Among the most widely used
are the spin-weighted spherical harmonics
\cite{sYjm1,sYjm2},
denoted $\syjm{s}{jm}$.
In this Appendix,
we briefly review the definitions and basic features 
of these spin-weighted spherical harmonics.
We also obtain some mathematical properties 
that are used in Sec.\ \ref{sec_gen_coeffs}.

\subsection{Spin-weighted spherical harmonics}
\label{sec_swsh}

A key concept underlying the spin-weighted spherical harmonics
is the notion of spin weight.
To introduce this idea,
consider the problem of characterizing radiation 
propagating inward toward the Earth 
from a distant point source in the sky.
The electric-field vector $\mbf E$
is oriented perpendicular to the line of sight,
so it lies in the tangent space of a sphere
surrounding the Earth.
In spherical polar coordinates,
the angular components of $\mbf E$ are $E_\th$ and $E_\ph$.
However,
the alternative components
$E_\pm \propto E_\th \mp i E_\ph$
can be considered instead.
These have the advantage of transforming elegantly as 
$E_\pm \rightarrow e^{\mp i\de} E_\pm$
under a rotation of the local coordinates
by an angle $\de$ about the line of sight.
The irreducible combinations $E_\pm$
are said to be spin-weighted functions
of spin weight $\pm 1$.
More generally,
a function $_sf$ is said to have spin weight $s$ 
if it transforms according to
$_sf\rightarrow e^{-is\de} \phantom{}_sf$
under a local rotation about the line of sight.

The generator of rotations is the
the angular-momentum operator $\mbf J$.
Denoting the radial unit vector as $\mbf{\hat n}$,
the generator of local rotations
about the line of sight is the operator
$\mbf{\hat n}\cdot\mbf J$.
This is the helicity operator
with respect to the line of sight.
We see that the spin weight 
can be understood as the eigenvalue
of the helicity operator.
For instance,
the irreducible combinations $E_\pm$
in the above example
are components of definite helicity 
with respect to the line of sight.
In this particular example,
the light propagates in the 3-momentum direction 
$\mbf{\hat p} = -\mbf{\hat n}$,
so the helicity operator $\mbf{\hat n}\cdot\mbf J$
is equivalent up to a sign to
the helicity operator with respect to the momentum,
$\mbf{\hat p}\cdot\mbf J$.
However,
this equivalence fails for non-radial propagation.
So in general there are two basic options for defining spin weight,
as the eigenvalue up to a sign of either
$\mbf{\hat n}\cdot\mbf J$ or $\mbf{\hat p}\cdot\mbf J$.
The choice of definition can be made based on suitability
for the problem at hand.

The helicity operator
$\mbf{\hat p}\cdot\mbf J$
or $\mbf{\hat n}\cdot\mbf J$
commutes with both
the squared total angular momentum ${\mbf{J}}^2$
and the component $J_z$, 
where by convention we choose the projection axis 
to be the $z$ direction.
It is therefore possible to introduce
simultaneous eigenfunctions for all three operators.
We show in the next subsection that
these simultaneous eigenfunctions are 
the spin-weighted spherical harmonics
$\syjm{s}{jm}$.
However,
typical discussions 
\cite{sYjm1,sYjm2,zs}
of $\syjm{s}{jm}$
are based on raising and lowering operators
for the spin weight,
which connect different harmonics.
In this subsection,
we define these operators
and provide explicit expressions for 
$\syjm{s}{jm}$.

Acting on a function $\phantom{}_sf$ of spin weight $s$,
the raising and lowering operators are given by
\beq
\left\{
\begin{array}{c}
\eth \\
\bar\eth
\end{array}
\right\} \phantom{}_sf = 
-\sin^{\pm s}\th(\prt_\th \pm i \csc\th\prt_\ph)\sin^{\mp s}\th\
\phantom{}_sf .
\label{rl_ops}
\eeq
The raising operator $\eth$ 
acts on a function of spin weight $s$
to yield a function of spin weight $s+1$.
Similarly,
the lowering operator $\bar\eth$
decreases spin weight by one.
The spin-weighted spherical harmonics $\syjm{s}{jm}$ 
of spin weight $s$
can be generated from the usual scalar spherical harmonics
$Y_{jm} = \syjm{0}{jm}$ 
by raising or lowering the spin weight $|s|$ times:
\beq
\syjm{s}{jm} = 
\begin{cases}
\hfill \sqrt\frac{(j-s)!}{(j+s)!}\ 
\eth^s \syjm{0}{jm}, & \hfill 0< s\leq j , 
\\[10pt]
(-1)^s \sqrt\frac{(j+s)!}{(j-s)!}\ 
\bar\eth^s \syjm{0}{jm}, & -j\leq s< 0 .
\label{syjm_gen}
\end{cases}
\eeq
As discussed in the following subsection,
the index $j$ corresponds to the
eigenvalue of the squared total angular momentum 
$\mbf J^2 = j(j+1)$
and $m$ to the eigenvalue of $J_z$. 

The functions $\syjm{s}{jm}$
are nonvanishing for index values $|m|\leq j$,
as usual.
Since the helicity is limited by $j$,
they are also nonvanishing for $j\geq |s|$.
They satisfy a relation analogous 
to that obeyed by the usual scalar spherical harmonics,
\begin{align}
\syjm{s_1}{j_1m_1}~\syjm{s_2}{j_2m_2} =
\ & \sum_{s_3j_3m_3} 
\sqrt{\Frac{(2j_1+1)(2j_2+1)}{4\pi(2j_3+1)}}
\notag\\
&\times \cg{j_1j_2(-s_1)(-s_2)}{j_3 (-s_3)}
\notag\\
&\times \cg{j_1j_2m_1m_2}{j_3 m_3}
\ \syjm{s_3}{j_3m_3},
\label{orth}
\end{align}
where the symbols $\cg{j_1j_2m_1m_2}{j_3 m_3}$
represent the Clebsch-Gordan coefficients.
This implies orthogonality of harmonics of equal spin weight,
\beq
\int \syjm{s}{jm}^*(\mbf{\hat n})\, 
\syjm{s}{j'm'}(\mbf{\hat n})\, 
\sin\th d\th d\ph = 
\de_{jj'}\de_{mm'}.
\eeq
Note,
however,
that this orthogonality does not extend to
$\syjm{s}{jm}$ of different spin weight.
The harmonics $\syjm{s}{jm}$ also satisfy
the completeness relations
\beq
\sum_{jm} 
\syjm{s}{jm}^*(\mbf{\hat n})\, 
\syjm{s}{jm}(\mbf{\hat n}')
= \de(\mbf{\hat n} - \mbf{\hat n}')\ .
\label{compl}
\eeq
Here,
we adopt phase conventions ensuring 
\beq
\syjm{s}{jm}^*=(-1)^{s+m}\, \syjm{-s}{j(-m)}
\label{conj}
\eeq
under complex conjugation
and  
\beq
\syjm{s}{jm}(-\mbf{\hat n}) =(-1)^j\, \syjm{-s}{jm}(\mbf{\hat n})
\label{parity}
\eeq
under parity.

Although Eq.\ \rf{syjm_gen} can be used
to generate the spin-weighted spherical harmonics,
in practice it is often easier to
use the explicit expression
\begin{align}
\syjm{s}{jm}(\th,\ph) =\ & \left[
\Frac{2j+1}{4\pi}
\Frac{(j+m)!(j-m)!}{(j+s)!(j-s)!}
\right]^{\frac12}
e^{im\ph}\sin^{2j}\Frac{\th}{2}
\notag \\
&\hspace{-55pt} \times \sum_r
(-1)^{j+m+s+r}
\bc{j-s}{r} \bc{j+s}{r+s-m}
\cot^{2r+s-m}\Frac{\th}{2} ,
\label{swh}
\end{align}
where $\bc{m}{n}$
denotes the binomial coefficients.
However,
this expression can be numerically troublesome
for high values of $j$.
One strategy for numerical applictions
is to use Eq.\ \rf{swh} to generate the low-$j$ harmonics 
but instead to use recursion relations
derived from Eq.\ \rf{orth}
to extract the higher-$j$ ones.
We find that two recursion relations are useful
for this purpose.
Taking $s_1=0, j_1=1, m_1=1$,
and $m_2=\pm j_2$ in Eq.\ \rf{orth},
we obtain the recursion
\beq
\syjm{s}{j(\pm j)} = \mp\sqrt{\Frac{j(2j+1)}{2(j^2-s^2)}}
e^{\pm i \ph}\sin\th
\syjm{s}{(j-1)(\pm j \mp 1)} ,
\label{rec1}
\eeq
which relates harmonics with $j=|m|$.
Also,
taking $s_1=0, j_1=1, m_1=0$ in Eq.\ \rf{orth} 
leads to the recursive formula
\begin{align}
\syjm{s}{jm} =\ &
\sqrt{\Frac{j^2((2j)^2-1)}{(j^2-m^2)(j^2-s^2)}}
\bigg[ (\cos\th+\Frac{ms}{j(j-1)})\ \syjm{s}{(j-1)m}
\notag \\
&\quad 
-\sqrt{\Frac{((j-1)^2-m^2)((j-1)^2-s^2)}{(j-1)^2(2j-1)(2j-3)}}\
\syjm{s}{(j-2)m}
\bigg] ,
\label{rec2}
\end{align}
which permits the calculation of higher-$j$ harmonics 
from lower-$j$ ones with the same $m$.
A practical procedure for determining numerical values
of spin-weighted spherical harmonics
then involves first using Eq.\ \rf{swh}
to find values for $j=|s|$,
followed by using Eq.\ \rf{rec1} to obtain harmonics with $j=|m|$,
and finally using Eq.\ \rf{rec2} to
fill in all remaining values up to the desired maximum $j$.

\subsection{Covariant angular momentum}
\label{sec_am}

In this subsection,
we further explore the relationships between 
angular momentum, helicity,
and the spin-weighted spherical harmonics.
For definiteness,
we consider helicity 
with respect to the momentum direction 
$\mbf{\hat p}$,
but the following discussion remains valid 
if $\mbf{\hat n}$ is substituted for $\mbf{\hat p}$ throughout.
Note also that we adopt a metric with positive signature
when working with 3-dimensional tensors.

We introduce standard angles in spherical polar coordinates,
so that the momentum direction can be written as 
\beq
\mbf{\hat p}= \sin\th\cos\ph\, \mbf{\hat e}_x 
+\sin\th\sin\ph\, \mbf{\hat e}_y
+\cos\th\, \mbf{\hat e}_z,
\eeq
where $\mbf{\hat e}_x$, $\mbf{\hat e}_y$, $\mbf{\hat e}_z$
form the cartesian basis vectors.
The orthonormal basis vectors in the spherical polar coordinates
are taken as
\beq
\mbf{\hat e}_r=\mbf{\hat e}^r=\mbf{\hat p},
\quad
\mbf{\hat e}_\th=\mbf{\hat e}^\th,
\quad
\mbf{\hat e}_\ph=\mbf{\hat e}^\ph,
\eeq
where $\mbf{\hat e}_\th$ and $\mbf{\hat e}_\ph$
are the coordinate unit vectors associated 
with the spherical coordinates $\th$ and $\ph$.
We also define complex helicity-basis vectors
\beq
\mbf{\hat e}_r=\mbf{\hat e}^r=\mbf{\hat p},
\quad
\mbf{\hat e}_\pm ={\mbf{\hat e}}^\mp = 
\Frac{1}{\sqrt{2}}(\mbf{\hat e}_\th 
\pm i \mbf{\hat e}_\ph) ,
\eeq
where an explicit choice of phase has been made.
Other phase choices would change some of the relations below.
With this definition, 
a rotation of the local coordinates about $\mbf{\hat p}$ 
by an angle $\de$ generates a phase shift 
in the helicity-basis vectors, 
$\mbf{\hat e}_\pm \rightarrow e^{\mp i\de}\mbf{\hat e}_\pm$.

Using the helicity basis,
we can readily decompose any tensor 
into components of definite helicity and spin weight.
For example, a vector $\mbf V$ has
0-helicity component $V_r=\mbf{V}\cdot\mbf{\hat e}_r$
and $\pm 1$-helicity components 
$V^\pm=V_\mp=\mbf{V}\cdot\mbf{\hat e}^\pm$,
the latter corresponding to spin-weight = $\mp 1$. 
The spin weight and helicity of any tensor component
is readily obtained by simple counting,
since each $\mp$ contravariant or $\pm$ covariant index 
adds $\pm 1$ to the spin weight.

To construct operators acting to raise or lower
the spin weight,
it is convenient first to introduce 
directional derivatives in momentum space
with respect to arbitrary basis vectors ${\hat e}_a$,
according to
\beq
\prt_a = {\mbf{\hat e}}_a\cdot\mbf\prt ,
\qquad
\prt_a \mbf{ p} = \mbf{\hat e}_a.
\eeq
The metric is 
\beq
g_{ab} = {\hat e}_a\cdot {\hat e}_b,
\eeq
and we define covariant directional derivatives 
$\nabla_a$ with connection 
\beq
\Ga^c_{\pt{c}ab}=(\prt_a\mbf{\hat e}_b)\cdot\mbf{\hat e}^c,
\eeq
as usual.
For example,
acting with $\nabla_a$ 
on the components of a vector $\mbf V$ gives
\beq
\nabla_a V^b = \prt_a V^b + \Ga^b_{\pt{b}ac} V^c,
\quad
\nabla_a V_b = \prt_a V_b - \Ga^c_{\pt{c}ab} V_c.
\eeq
In the helicity basis 
$\{\mbf{\hat e}_+,\mbf{\hat e}_r,\mbf{\hat e}_-\}$,
we find the metric takes the explicit form
\beq
g_{ab}=g^{ab}=
\left(
\begin{array}{ccc}
0 & 0 & 1 
\\ 
0 & 1 & 0 
\\ 
1 & 0 & 0 
\end{array}
\right),
\eeq
while the nonzero connection elements are
\begin{align}
\Ga^\pm_{\pt{\pm}\pm r} &= 
-\Ga^r_{\pt{r}\pm\mp} = p^{-1} , 
\notag \\
\Ga^\pm_{\pt{\pm}\pm\pm} &= 
-\Ga^\pm_{\pt{\pm}\mp\pm} = (\sqrt{2}~p~\tan\th)^{-1} ,
\end{align}
where $p=|\mbf p|$.
The point is that the derivative $\nabla_+$ 
operating on an arbitrary tensor 
acts to create tensor components 
whose spin weight is increased by one,
while $\nabla_-$ lowers the spin weight by one.

Our goal is a decomposition in angular momentum, 
so we seek a covariant description of the angular-momentum operators.
In momentum space,
the orbital angular momentum $\mbf L$ has components
\beq
L^a=-i\ve^{abc}p_b\nabla_c
\eeq
and obeys the algebra
\beq
[L_a,L_b]=-i\ve_{abc}L^c.
\eeq
Here,
$\ve^{abc}$ is the 
totally antisymmetric tensor,
which in the helicity basis satisfies
\beq
\ve_{+r-} = -\ve^{+r-} = i.
\eeq
In particular,
we have 
\beq
L_r = 0, 
\qquad
L_\pm = \pm p\nabla_\pm,
\eeq
which implies that orbital angular momentum is
perpendicular to $\mbf{\hat p}$,
as expected.

To complete the description of angular momentum,
it is useful to define a covariant spin operator $\mbf S$
with components $S_a$.
Acting on a vector $V_a$,
we define 
\beq
S_a V_b = -S^c_{\pt{c}ab} V_c,
\qquad
S_a V^b = S^b_{\pt{b}ac} V^c,
\eeq
where
\beq 
S^c_{\pt{c}ab}=i{\ve^c}_{ab} .
\eeq
This provides a covariant formulation of the general spin relation
\beq
({\mbf V}_1\cdot\mbf S) {\mbf V}_2 = i{\mbf V}_1 \times {\mbf V}_2
=  (V_1)^a \mbf{\hat e}^{\, b} (S_a (V_2)_b) .
\eeq
The spin operator acting on more general tensors 
obeys rules like those of the connection.
For example,
we obtain
\beq
S_a {T^b}_c= S^b_{\pt{b}ad}{T^d}_c - S^d_{\pt{d}ac} {T^b}_d.
\eeq
Since the $S_a$ are covariant operators
rather than matrix operators,
they obey slightly modified commutation relations
\beq
[S_a,S_b]=-i\ve_{abc}S^c.
\eeq

With the above definitions,
we can introduce a covariant operator 
for the total angular momentum as 
\beq
J_a=L_a+S_a.
\eeq
Like the covariant derivative,
this operator has the advantage
of maintaining the explicit tensor nature of expressions
when operating on tensor components.
It obeys the modified commutation relations
\beq
[J_a,J_b]=-i\ve_{abc}J^c.
\eeq
An interpretation of these operators is as follows.
The radial angular momentum $J_r = S_r$
is the helicity,
which is the negative of the spin weight.
The operator $J^+$ raises the helicity
and $J_+$ raises the spin weight,
while $J^-$ lowers the helicity 
and $J_-$ lowers the spin weight.
These operators are equivalent 
to the raising and lowering operators Eq.\ \rf{rl_ops}
via the correspondence
\beq
J_+=-\eth/\sqrt2, 
\qquad 
J_-=\bar\eth/\sqrt2. 
\eeq

The spin-weighted spherical harmonics $\syjm{s}{jm}$
are eigenfunctions of the commuting operators
\beq
-J_r=s, 
\qquad
J^2=j(j+1), 
\qquad
J_z=m.
\eeq
The ladder operators $J_\pm$ 
can be shown to commute with $\mbf J^2$ and $J_z$.
In fact, 
$J_\pm$ commutes with $\mbf V\cdot \mbf J$
for any covariantly constant vector $\mbf V$, 
so that $\nabla_a V_b = 0 $.
Also,
the commutator
\beq
[J_\pm,-J_r]=i\ve_{\pm r\mp} J^\mp = \mp J_\pm,
\eeq
implies that $J_+$ and $J_-$
raise and lower the spin weight, 
respectively,
while leaving $j$ and $m$ unaltered.
The ladder nature of the $J_\pm$ can thus be seen 
either from the differential geometry
or from the operator algebra.

Since the spin-weight ladder operators $J_\pm$ 
are analogous to the conventional ladder operators
that raise and lower the eigenvalue of the
$z$ component of the angular momentum,
many of the usual techniques and results apply.
For example, 
the standard normalization scheme gives
\beq
J_\pm\ \syjm{s}{jm}= 
-\sqrt{\half\big(j(j+1)-s(s\pm 1)\big)}~ \syjm{s\pm1}{jm} ,
\eeq
where the phase is chosen to match the conventional definitions 
when expressed in terms of $\eth$ and $\bar\eth$.
Repeated application gives the useful expressions
\beq
\syjm{0}{jm} =  (-1)^s 2^{|s|/2}\sqrt\Frac{(j-|s|)!}{(j+|s|)!}
(J_\pm)^{|s|} ~\syjm{\mp|s|}{jm} , 
\label{J-ident3}
\eeq
and
\beq
\syjm{\pm|s|}{jm} = (-1)^s 2^{|s|/2}\sqrt\Frac{(j-|s|)!}{(j+|s|)!}
(J_\pm)^{|s|} ~\syjm{0}{jm} , 
\label{J-ident4}
\eeq
which are equivalent to Eq.\ \rf{syjm_gen}.

In practical terms,
the above formalism leads to the simple results
\begin{align}
\nabla_r &= \prt/\prt p ,   
\notag \\
\nabla_\pm &= \pm(J_\pm - S_\pm)/p , 
\label{J-ident}
\end{align}
which will be used extensively in Section \ref{sec_gen_coeffs}.
Note, 
however, 
that some care is needed
when using the results \rf{J-ident} 
since they are non-tensorial relations,
valid only in the helicity basis.
For example, 
the $\pm$ indices on the right-hand side 
are noncovariant once applied,
so further manipulations with covariant operators
such as $\nabla_a$, $S_a$, $J_a$
are inappropriate.

As a simple illustration of the above formalism,
consider the covariant momentum-space laplacian 
acting on a scalar $\psi$:
\begin{align}
\nabla_a\nabla^a \psi &= 
\big(\nabla_r\nabla^r +\nabla_+\nabla^+ +\nabla_-\nabla^- \big)\psi
\notag\\
&= \big(\Frac{\prt^2 \phantom{p}}{\prt p^2} 
+\Frac{1}{p}(J_+ - S_+) \nabla^+
-\Frac{1}{p}(J_- - S_-)\nabla^- \big) \psi .
\end{align}
Using the results 
\begin{align}
S_a\nabla^b\psi & = 
S^b_{\pt{b}ac} \nabla^c\psi = -i{\ve_a}^{bc}\nabla_c\psi,
\notag\\
\nabla^\pm \psi & = \mp\Frac{1}{p} (J^\pm - S^\pm) \psi
= \mp\Frac{1}{p} J^\pm \psi,
\notag\\
(J_+J^+ +J_-J^-)\psi & = J^2\psi,
\end{align}
we can reexpress $\nabla_a\nabla^a \psi$ as follows:
\begin{align}
\nabla_a\nabla^a \psi 
&= \big(\Frac{\prt^2 \phantom{p}}{\prt p^2} 
+\Frac{1}{p}( J_+ \nabla^+ + \nabla_r)
-\Frac{1}{p} (J_- \nabla^- - \nabla_r ) \big) \psi
\notag\\
&= \big(\Frac{\prt^2 \phantom{p}}{\prt p^2} 
+\Frac{2}{p} \Frac{\prt \phantom{p}}{\prt p}
- \Frac{1}{p^2}( J_+ J^+ +J_- J^- ) \big) \psi
\notag\\
&= \big(\Frac{\prt^2 \phantom{p}}{\prt p^2} 
+\Frac{2}{p} \Frac{\prt \phantom{p}}{\prt p}
- \Frac{1}{p^2}J^2 \big) \psi .
\end{align}
We thereby recover the familiar expression 
for $\nabla_a\nabla^a \psi$ involving the angular momentum.

It has been noted in the literature that
the operators $\eth$ and $\bar\eth$
are equivalent to covariant derivatives 
in the two-dimensional tangent space of the sphere
\cite{sYjm1}.
In our present language,
this can be made apparent by defining 
a new covariant derivative as 
\beq
\widetilde\nabla_a=\nabla_a-i\ve_{arb}S^b/p,
\eeq
which in the helicity basis
gives nonzero connection elements
\beq
\widetilde\Ga^\pm_{\pt{\pm}\pm\pm} 
= -\widetilde\Ga^\pm_{\pt{\pm}\mp\pm} 
= (\sqrt{2}~p~\tan\th)^{-1}.
\eeq
The derivative $\widetilde\nabla_a$
corresponds to the projection
of the usual three-dimensional geometry
onto the embedded 2-sphere.
In terms of this derivative,
we can express the ladder operators as 
\beq
J_\pm = \pm \widetilde\nabla_\pm/p.
\eeq
This relates our three-dimensional picture
to the two-dimensional one.

\subsection{Parity}
\label{sec_parity}

In a typical application of the above formalism,
a tensor is decomposed
by considering its components in the helicity basis,
which are spin-weighted functions,
and expanding them using spin-weighted spherical harmonics.
In many cases, 
it is useful to decompose further the results
in terms of parity properties
\cite{sYjm1}.
This subsection contains a brief summary 
of the latter procedure,
along with some figures providing insight
into the structure of the resulting modes.

Let $t$ represent an arbitrary tensor component
with spin weight $s$.
Its expansion then takes the form
\beq
t(\mbf{\hat p}) = \sum_{jm} 
t_{jm}\ \syjm{s}{jm}(\mbf{\hat p}).
\eeq
By exchanging all the $+$ and $-$ indices on $t$,
we obtain another tensor component of spin weight $-s$.
Denoting this component as $\bar t$,
we can expand it as 
\beq
\bar t(\mbf{\hat p}) = \sum_{jm} 
\bar t_{jm}\ \syjm{-s}{jm}(\mbf{\hat p}).
\eeq
The two components $t$ and $\bar t$ are parity conjugates,
interchanging under parity according to
\beq
t(\mbf{\hat p}) \leftrightarrow 
(-1)^s\, \bar t(\mbf{-\hat p}).
\eeq
In terms of spherical coefficients,
the parity transformation gives
\beq
t_{jm} \leftrightarrow (-1)^{s+j}\, \bar t_{jm}.
\eeq
It is then useful to define
\begin{align}
t_{jm} &= E_{jm} + i B_{jm} \ ,
\notag \\
\bar t_{jm} &=(-1)^s( E_{jm} - i B_{jm}) .
\end{align}
The result is a splitting of $t$ and $\bar t$
into modes with so-called electric-type parity
or $E$-type parity,
\beq
E_{jm}\rightarrow (-1)^j E_{jm}, 
\eeq
and ones with magnetic-type parity or $B$-type parity,
\beq
B_{jm}\rightarrow (-1)^{j+1} B_{jm},
\eeq
where the nomenclature is borrowed from radiation theory.

\begin{figure}
\begin{center}
\centerline{\psfig{figure=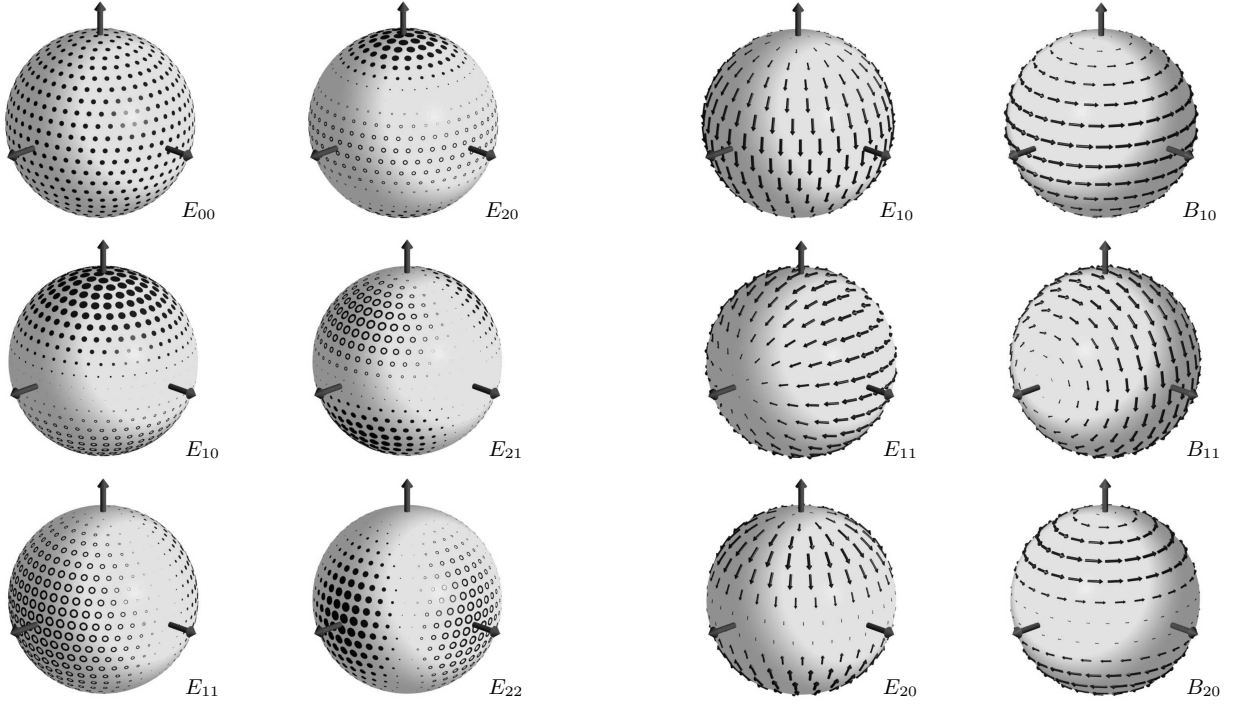,width=0.8\hsize}}
\caption{ \label{0yjm}
Angular distributions for electric-type
scalar components. }    
\end{center}
\end{figure}

As a simple example,
consider a scalar function
\beq
S(\mbf{\hat p}) =\sum_{jm} E_{jm}\, \syjm{0}{jm}(\mbf{\hat p}).
\label{scalarE}
\eeq
Under parity,
$S$ transforms according to  
\beq
S(\mbf{\hat p})\rightarrow S(-\mbf{\hat p}),
\eeq
which in conjunction with Eq.\ \rf{parity}
confirms that scalar functions
contain only $E$-type components.
Figure \ref{0yjm} illustrates 
the angular distribution obtained by considering each mode 
$E_{00}$, $E_{10}$, $E_{11}$, $E_{20}$, $E_{21}$, $E_{22}$
of a real scalar field in turn.
For example,
the real scalar field $S(E_{11})$
associated with real $E_{11}$ is 
\beq
S(E_{11}) = E_{11}(\syjm{0}{11}-\syjm{0}{1(-1)}).
\eeq
For each of the six images in Fig.\ \ref{0yjm},
the three arrows perpendicular to the spherical surface 
represent rectangular right-handed coordinate axes
with $\mbf{\hat z}$ vertical.
The solid disks represent positive values 
of the scalar,
while the rings represent negative ones.
The radius of each disk or ring
is proportional to the magnitude of the scalar
at that point.
The images reveal the angular and parity symmetries
of each scalar mode.
Consider,
for example,
the $E_{10}$ scalar mode.
The figure shows that this mode has 
extremal magnitudes at the poles and equator
and is symmetric under rotations about $\mbf{\hat z}$,
all of which are features of a distribution
with $j=1$ and $m=0$.
Also, the sign of the mode changes 
on interchanging any two antipodal points,
matching its negative parity.
Note that the distributions are plotted for real amplitudes
$E_{jm}$.
Inclusion of a phase in an amplitude
rotates the corresponding figure about $\mbf{\hat z}$.

\begin{figure}
\begin{center}
\centerline{\psfig{figure=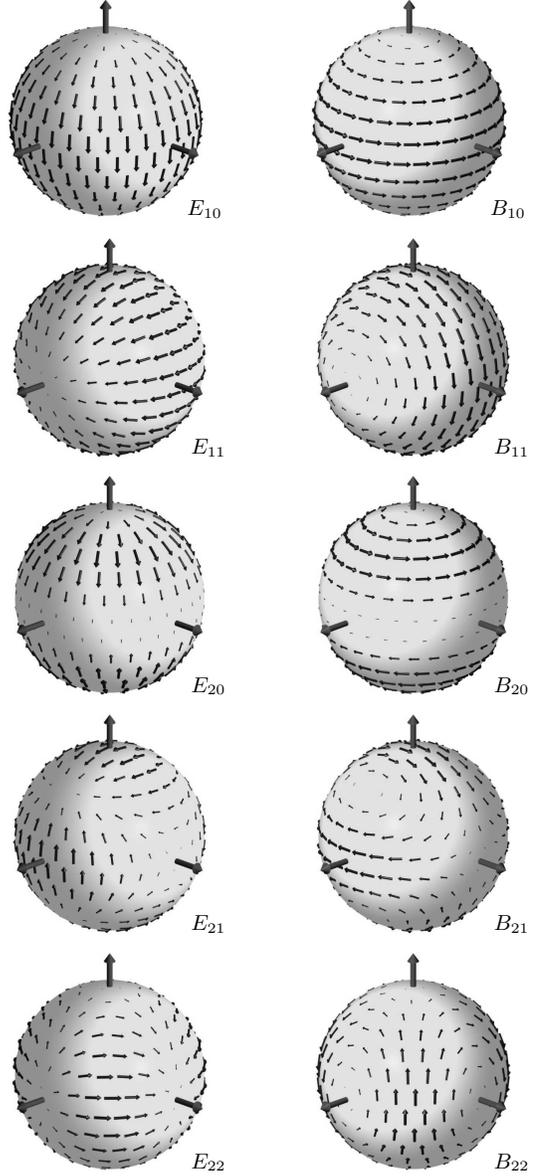,width=0.8\hsize}}
\caption{ \label{1yjm}
Angular distributions for electric- and magnetic-type 
vector components.}    
\end{center}
\end{figure}

An example with both $E$- and $B$-type content 
is provided by the components 
\beq
V_\pm = \sum_{jm} (\pm E_{jm} +i B_{jm})\, \syjm{\pm 1}{jm}
\label{vectorEB}
\eeq
of a vector field $\mbf V$ in the tangent bundle of the sphere.
Figure \ref{1yjm} displays 
the angular distributions for a real vector field
for each separate mode
$E_{00}$, $E_{10}$, $E_{11}$, $E_{20}$, $E_{21}$, $E_{22}$.
For example,
the components of the real vector field $\mbf V (E_{11})$
associated with real $E_{11}$ are 
\begin{align}
V_+(E_{11}) &= 
E_{11}(\syjm{1}{11} - \syjm{1}{1(-1)}),
\notag\\
V_-(E_{11}) &= 
-E_{11}(\syjm{-1}{11} - \syjm{-1}{1(-1)}).
\end{align}
In the figure,
the vector tangent to a sphere at a given point
represents the corresponding mode.
The angular and parity symmetries of each vector mode
can be seen by inspection.
For instance,
the $E_{10}$ vector mode has extremal magnitudes
at the poles and equator
and is symmetric under rotations about $\mbf {\hat z}$,
as expected for a mode with $j=1$ and $m=0$. 
Also, 
inspection reveals that applying the parity operation 
reverses the flow of vectors on the sphere,
as is appropriate for a mode with $E$-type parity and $j=1$.
Note that the $E$-type modes are curl free
while the $B$-type modes are divergence free,
matching the usual properties in electrodynamics.
Note also that for given values of $j$ and $m$
the $E$- and $B$-mode field lines are perpendicular everywhere,
reflecting the orthogonality of the modes. 

\begin{figure}
\begin{center}
\centerline{\psfig{figure=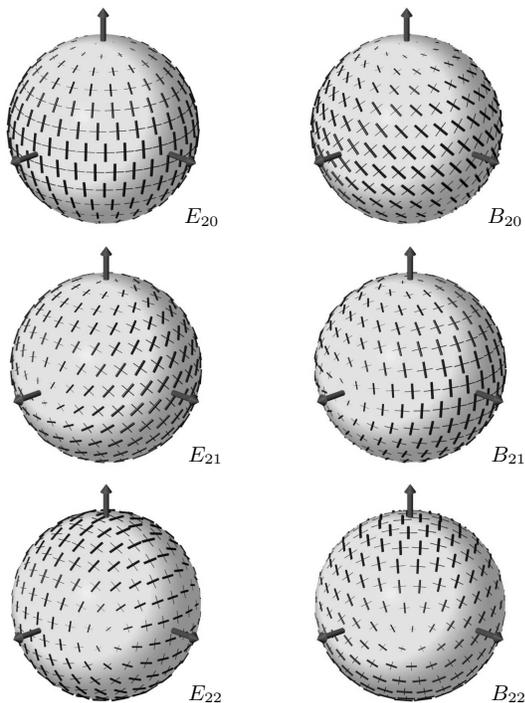,width=0.8\hsize}}
\caption{ \label{2yjm}
Angular distributions for electric- and magnetic-type 
tensor components. }    
\end{center}
\end{figure}

For a symmetric traceless 2-tensor,
we can consider the components 
\beq
T_{\pm\pm} = \sum_{jm} (E_{jm} \pm i B_{jm})\, \syjm{\pm 2}{jm}.
\label{tensorEB}
\eeq
Figure \ref{2yjm} provides representations of
each the six modes
$E_{20}$, $E_{21}$, $E_{22}$,
$B_{20}$, $B_{21}$, $B_{22}$
for a real tensor.
The figure takes advantage of the spectral theorem
applied to the decomposition 
of a real symmetric traceless tensor $T$ 
that lies in in the tangent bundle of the sphere
and has nonzero components $T_{\pm\pm}$.
In terms of its two orthogonal eigenvectors $\mbf v$ and $\mbf w$
lying tangent to the sphere,
we can write
\beq
T = 
\mbf v \otimes \mbf v - \mbf w \otimes \mbf w,
\label{Tdec}
\eeq
where the eigenvalues of $T$ 
are $|\mbf v|^2$ and $-|\mbf w|^2= -|\mbf v|^2$. 
The tensor $T$ can therefore be represented
at each point on the sphere
by the two vectors $\mbf v$ and $\mbf w$.
We denote the vector $\mbf v$ by a thick line
and the vector $\mbf w$ by a thin one,
noting that the vector orientations are unspecified
by Eq.\ \rf{Tdec}
and so the vectors are best represented 
by unoriented line segments.
The angular and parity symmetries of each tensor mode
are visible in the figure.
For example,
the extremal magnitudes of the $E_{20}$ tensor mode 
occur at the poles and equator,
and the mode is symmetric under rotations about $\mbf {\hat z}$,
as expected for a mode with $j=2$ and $m=0$. 
The symmetry under interchange of two antipodal points
is a consequence of the positive parity of this mode.
The orthogonality of the $E$ and $B$ modes 
for given values of $j$ and $m$
implies that each $E$-$B$ pair of modes is related 
by interchanging a plus with a cross at each point.
As an aside,
we remark that this visualization has a parallel
in general relativity,
where the orthogonality
of the plus and cross modes for gravitational radiation
also arises from a symmetric real 2-tensor
tangential to the direction of propagation. 

We can also consider various types of pseudotensors.
All these acquire an additional sign under parity,
which implies that the roles of the $E$- and $B$-type coefficients 
are interchanged in the parity decomposition.
For example,
pseudoscalars contain only $B$-type components.
Also,
the parity decomposition of a pseudovector $\mbf V'$
leads to components with spin weight $\pm 1$ 
of the form
\beq
V'_\pm = \sum_{jm} (\pm B_{jm} +i E_{jm})\, \syjm{\pm 1}{jm}
\eeq
instead.
Similarly for a pseudotensor $T'$,
we have 
\beq
T'_{\pm\pm} = \sum_{jm} (B_{jm} \pm i E_{jm})\, \syjm{\pm 2}{jm}.
\eeq


\begin{thebibliography}{99}
\bibitem{fn1}
For $d=4$,
these figures differ slightly from those in Ref.\ \cite{km_grb},
where the effective degree of polarization
$\Pi_{\rm eff}$ was taken as $\vev{s^1}^2+\vev{s^2}^2$ instead.
\end{thebibliography}
\end{document}